%% file: HaPro.tex
\documentclass[onecolumn,sort,compress,numbers]{els-mrw} 
\usepackage{fix-cm}
\usepackage{xcolor}
\usepackage{lmodern}
\usepackage{amsmath,amssymb,amsfonts,amsthm,makeidx,graphicx}
\usepackage{helvet}
\usepackage{bm}
\usepackage{bbm}
\usepackage{dsfont}
\usepackage{bbold}
\usepackage{placeins}
\usepackage{wasysym}
\usepackage{anyfontsize}
\usepackage{slashed}
\usepackage{microtype}
\usepackage{float}
\begin{document}
\include{definitions}


\chapter{Hadron Production Processes}\label{chap1}

\author[1]{Horst Lenske}%
\author[2]{Igor Strakovsky}%


\address[1]{\orgname{Justus-Liebig-Universit\"at Giessen}, \orgdiv{Institut f\"{u}r Theoretische Physik}, \orgaddress{D-35392 Giessen, Germany}}
\address[2]{\orgname{The George Washington University}, \orgdiv{Institute for Nuclear Studies, Department of Physics}, \orgaddress{Washington, D.C. 20052, USA}}

\maketitle

\vspace{1cm}
\begin{abstract}[Abstract]The experimental search for the pion --  Hideki Yukawa's meson proposed in 1935 as the force carrier of the strong nucleon-nucleon interaction -- was rewarded in 1947 when in cosmic ray photographic emulsion data a charged particle was identified with the proper mass of about 300 times the electron mass, completed three years later by the discovery of the neutral pion. Since then, accelerator-driven pion and meson (photo-) production on the nucleon, along with the associated production of new baryons, has become a key element for ground-breaking discoveries in numerous areas of particle and nuclear physics, ranging from fundamental symmetries and their breaking to low-energy QCD dynamics to laying the foundations for modern elementary particle physics and the Standard Model. This article provides an overview of almost a century of experimental and theoretical research on meson production, from isospin to charm and beyond, shaping our understanding of hadrons, their spectra, structure, and interactions. 
\end{abstract}

\begin{keywords}
 	meson photoproduction\sep baryon spectroscopy\sep coupled channel methods and
    analysis\sep QCD\sep LQCD
\end{keywords}

\begin{figure}[ht]
	\centering
	\includegraphics[width=3.5cm]{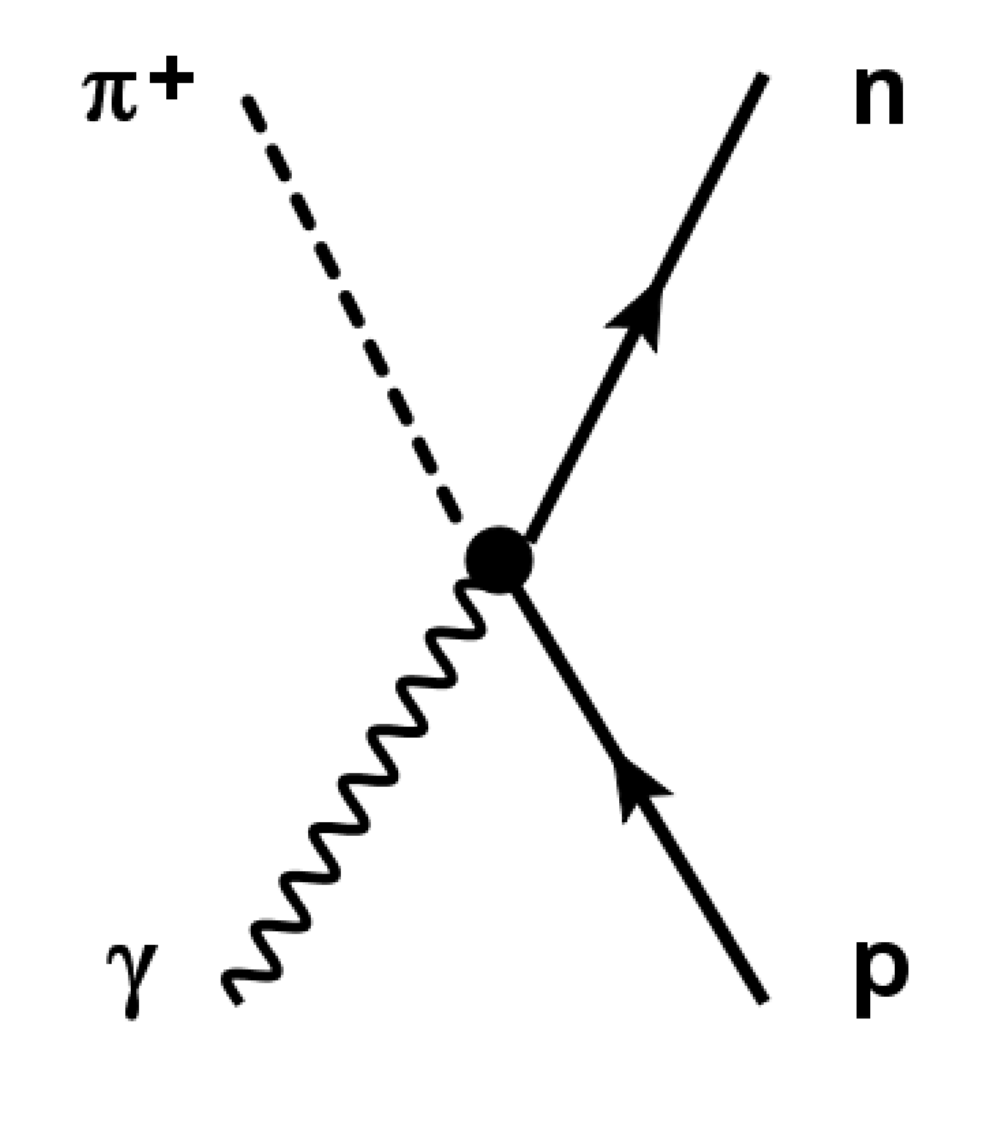}
	\caption{Feynman diagram describing the production of a charged pion (dashed line) and a neutron (full line) by an incident photon $\gamma$ (wavy line), incident on a proton (full line) .}
	\label{fig:titlepage}
\end{figure}

\begin{glossary}[Nomenclature]
\begin{tabular}{@{}lp{34pc}@{}}
        AGS   & Alternating Gradient Synchrotron at BNL\\
        ANL   & Argonne National Laboratory\\
        ALICE & A Large Ion Collider Experiment at CERN\\
        Bevatron & Billions of eV synchrotron at LBL \\
        BEVALAC  & Bevatron plus HILAC linear accelerator at LBL\\
        BIC   & Bound States in the Continuum\\
        BNL   & Brookhaven National Laboratory\\
        BnGa  & Bonn-Gatchina Model\\
        BSE   & Bethe-Salpeter Equation\\
        CERN  & European Organization for Nuclear Research (Conseil Europ\'{e}en pour la Recherche Nucléaire)\\
        CC     & Coupled Channel\\
        CC Model & Coupled Channel Model \\ 
        CBELSA & Crystal-Barrel-Detector at ELSA\\
        CGLN amplitudes & Chew, Goldberger, Low, and Nambu amplitudes \\
        CLAS  & CEBAF Large Acceptance Spectrometer at JLab\\
        c.m.  & Center-of-Mass Frame \\
        COSY  & Cooler Synchrotron \\
        CQM   & Constituent Quark Model\\
        CMS   & Compact Muon Solenoid \\
        ELSA  & Elektronen-Stretcher-Anlage (Electron-Stretcher-Facility)\\
        EM    & Electromagnetic \\
        FAIR  & Facility for Antiproton and Ion Research\\
        FNAL  & Fermi National Laboratory\\
        GANIL & Grand Accélérateur National d'Ions Lourds\\
        GiM   & Giessen Model\\
        GlueX & Gluonic Excitation Experiment \\
        GMO   & Gell-Mann-Okubo mass formula\\
        GRAAL & Grenoble Accelerator LASer \\
        GSI   & Gesellschaft f\"ur Schwerionenforschung\\
        GW    & The George Washington University \\
        HADES & High Acceptance DiElectron Spectrometer \\
        HILAC & Heavy Ion Linear Accelerator at LBL\\
        ITEP  & Institute of Theoretical and Experimental Physics \\
        JINR  & Joint Institute for Nuclear Research \\
        JLab  & Jefferson Laboratory or Thomas Jefferson National Accelerator Facility (TJNAF) \\
        J-PARC& Japan Proton Accelerator Research Complex\\
        LAMPF & Los Alamos Meson Physics Facility \\
        LBL   & Lawrence Berkeley Laboratory\\
        LEPS  & Laser Electron Photon Experiment\\
        LET   & Low Energy Theorem\\
        LHC   & Large Hadron Collider at CERN\\
        LQCD  & Lattice Quantum Chromo-Dynamics\\
        MAMI  & Mainz Microtron\\
        OZI Rule & Okubo–Zweig–Iizuka rule\\
        PDG   & Particle Data Group\\
        PSI   & Paul Scherrer Institute, previously SIN - Swiss Institute for Nuclear Research\\
        PWA   & Partial-Wave Analysis \\
        QCD   & Quantum Chromo-Dynamics\\
        QED   & Quantum Electrodynamics\\ 
        QFT   & Quantum Field Theory \\
        QGP   & Quark–Gluon Plasma\\
        RICH  & Relativistic Heavy Ion Collider at BNL\\
        RIKEN & National Research and Development Agency, Japan\\
        SAID  & Scattering Analyses Interactive Data\\
        SIN   & see PSI\\
        SM    & Standard Model\\
        TAPS  & Two-Arm Photon Spectrometer\\
        TRIUMF& Tri-University Meson Facility  \\
        VMD   & Vector Meson Dominance Model
\end{tabular}
\end{glossary}

\section*{Objectives}
\begin{itemize}
    \item A short story of hadron physics.
	\item Production of short-lived particles underlying strong interactions.
	\item Spectrum of hadrons as asymptotic mass eigenstates of QCD.
	\item Probing the structure and dynamics of hadrons.
	\item Experimental methods and facilities.
    \item Theoretical concepts and results.
\end{itemize}

\input{Introduction.tex}        
\input{MethodsCC.tex}           
\input{PionPhotoproduction.tex} 
\input{DoublePion.tex}
\input{GimStrangeEta.tex}       


\section*{Acknowledgments}
This work was supported in part by the U.~S.~Department of Energy,  Office of Science, Office of Nuclear Physics, under Award No.~DE--SC0016583 (I.S.) and Deutsche Forschungsgemeinschaft (DFG), grants Le439/16 and Le439/17 (H.L.).


\input{biblio.tex}
\end{document}

%% file: Definitions.tex
\newcommand{\bfigh}{\begin{figure}[H]}					

\newcommand{\bfig}{\begin{figure}}		

\newcommand{\efig}{\end{figure}}

\newcommand{\refe}[1]{(\ref{#1})}
\newcommand{\DDelta}{\Delta(1232)}
\newcommand{\NN}{N^*}
\newcommand{\foh}{\frac{1}{2}}
\newcommand{\fot}{\frac{1}{3}}
\newcommand{\fof}{\frac{1}{4}}
\newcommand{\ftt}{\frac{2}{3}}
\newcommand{\fth}{\frac{3}{2}}
\newcommand{\ffh}{\frac{5}{2}}
\newcommand{\mt}{{\rm g}}
\newcommand{\mca}{{\mathcal A}}
\newcommand{\mcd}{{\mathcal D}}
\newcommand{\mcf}{{\mathcal F}}
\newcommand{\mci}{{\mathcal I}}
\newcommand{\mck}{{\mathcal K}}
\newcommand{\mcl}{{\mathcal L}}
\newcommand{\mcm}{{\mathcal M}}
\newcommand{\mcp}{{\mathcal P}}
\newcommand{\mcs}{{\mathcal S}}
\newcommand{\mct}{{\mathcal T}}
\newcommand{\mcv}{{\mathcal V}}
\newcommand{\difd}{\mathrm d}
\newcommand{\mi}{\mbox i}
\newcommand{\ibid}[3]{\textit{ibid.} {\bf #1}, #2 (#3)}
\renewcommand{\slash}{/ \!\!\!\!\,}
\newcommand{\UNIT}[1]{\mbox{$\,{\rm #1}$}}
\newcommand{\MeV}{\UNIT{MeV}}
\newcommand{\GeV}{\UNIT{GeV}}
\newcommand{\GeVc}{\UNIT{GeV/c}}
\newcommand{\TeV}{\UNIT{TeV}}
\newcommand{\AMeV}{\UNIT{AMeV}}
\newcommand{\AGeV}{\UNIT{AGeV}}
\newcommand{\ATeV}{\UNIT{ATeV}}
\newcommand{\fm}{\UNIT{fm}}
\newcommand{\mb}{\UNIT{mb}}
\newcommand{\nb}{\UNIT{nb}}
\newcommand{\fmc}{\UNIT{fm/c}}
\newcommand{\proz}{\UNIT{\%}}
\newcommand{\Pythia}{\textsc{pythia}}
\newcommand{\Jetset}{\textsc{jetset}}
\newcommand{\Fritiof}{\textsc{Fritiof}}
\newcommand{\ds}{\displaystyle}
\newcommand{\qbq}{\mbox{$q\bar{q}$}}
\newcommand{\zerovec}{\vec{0}\,}
\newcommand{\qvec}{\vec{q}\,}
\newcommand{\kvec}{\vec{k}\,}
\newcommand{\lvec}{\vec{l}\,}
\newcommand{\nvec}{\vec{n}\,}
\newcommand{\pvec}{\vec{p}\,}
\newcommand{\rvec}{\vec{r}\,}
\newcommand{\jvec}{\vec{j}\,}
\newcommand{\Gcapvec}{\vec{G}\,}
\newcommand{\Pcapvec}{\vec{P}\,}
\newcommand{\Picapvec}{\vec{\Pi}\,}
\newcommand{\Scapvec}{\vec{S}\,}
\newcommand{\Tcapvec}{\vec{T}\,}
\newcommand{\lambdavec}{\vec{\lambda}\,}
\newcommand{\sigmavec}{\vec{\sigma}\,}
\newcommand{\omegavec}{\vec{\omega}\,}
\newcommand{\sigmacapvec}{\vec{\Sigma}\,}
\newcommand{\thetacapvec}{\vec{\Theta}\,}
\newcommand{\tauvec}{\vec{\tau}\,}
\newcommand{\rhovec}{\vec{\rho}\,}
\newcommand{\deltavec}{\vec{\delta}\,}
\newcommand{\pivec}{\vec{\pi}\,}
\newcommand{\xivec}{\vec{\xi}\,}
\newcommand{\xvec}{\vec{x}\,}
\newcommand{\nrmsq}{\mbox{$\langle r^2\rangle_n$}}

\newcommand{\la}{\langle}
\newcommand{\ra}{\rangle}
\newcommand{\lan}{\langle}
\newcommand{\ran}{\rangle}
\newcommand{\hF}{\hat{F}}
\newcommand{\hr}{\hat{\rho}}
\newcommand{\al}{\alpha}
\newcommand{\Tr}{\text{Tr}}
\newcommand{\beqn}{\begin{eqnarray*}}
\newcommand{\eeqn}{\end{eqnarray*}}
\newcommand{\beq}{\begin{eqnarray}}
\newcommand{\eeq}{\end{eqnarray}}
\newcommand{\bea}{\begin{eqnarray}}
\newcommand{\eea}{\end{eqnarray}}
\newcommand{\be}{\begin{equation}}
\newcommand{\ee}{\end{equation}}
\newcommand{\vs}{\vc\sigma}
\newcommand{\vt}{\vc\tau}
\newcommand{\vcs}[1]{\bf{#1}^{'}}
\newcommand{\vc}[1]{\text{\boldmath $#1$}}
\newcommand{\g}{\gamma}
\newcommand{\G}{\Gamma}
\newcommand{\om}{\omega}
\newcommand{\sv}{\vc{\sigma}}
\newcommand{\tv}{\vc{\tau}}
\newcommand{\s}{\sigma}
\newcommand{\da}{\delta}
\newcommand{\pb}{\overline{\p}}
\newcommand{\Li}{\Ld_{int}}
\newcommand{\Gh}{\hat\G}
\newcommand{\var}[2]{\frac{\delta #1}{\delta #2}}
\newcommand{\bmatr}{\begin{pmatrix}}
\newcommand{\ematr}{\end{pmatrix}}
\newcommand{\sums}[3]{\sum^{\textit{$#3$}}_{\textit{$#1$}=\textit{$#2$}}}
\newcommand{\suminfs}[1]{\sum^{\infty}_{\textit{#1}=0}}
\newcommand{\suminfss}[1]{\sum^{\infty}_{\textit{#1}=-\infty}}
\newcommand{\Psib}{\overline{\Psi}}
\newcommand{\psib}{\overline{\psi}}
\newcommand{\Lcal}{{\cal L}}
\newcommand{\p}{\partial}
\newcommand{\se}{\Sigma}
\newcommand{\intg}[3]{\int_{#1}^{#2} \text{d}#3}
\newcommand{\intu}[2]{\int #1 \text{d}#2}
\newcommand{\intn}[2]{\int \text{d}^#1 #2}
\newcommand{\td}{\text{d}}
\newcommand{\Rear}{Rearrangement }
\newcommand{\rear}{rearrangement }
\newcommand{\dfp}[1]{\frac{\td^4#1}{(2\pi)^4} }
\newcommand{\dtp}[1]{\frac{\td^3#1}{(2\pi)^3} }
\renewcommand{\d}{\textrm{d}}

\newcommand{\diff}[2]{{\frac{\partial #1}{\partial #2}}}
\newcommand{\diffq}[2]{{\partial^2 #1\over\partial #2^2}}
\newcommand{\difftot}[2]{{d#1\over d#2}}
\newcommand{\difftotq}[2]{{d^2 #1\over d #2^2}}
\newcommand{\diffo}[1]{\partial^{#1}}
\newcommand{\diffu}[1]{\partial_{#1}}
\renewcommand{\do}[1]{\partial^{#1}}
\newcommand{\du}[1]{\partial_{#1}}

\newcommand{\mc}[1]{\mathcal{#1}}
\newcommand{\Ld}{\mathcal{L}}

\newcommand{\tauiso}{\vec{\tau}}
\newcommand{\ud}{\text{d}}
\newcommand{\ex}{\text{e}}
\newcommand{\im}{\text{i}}
\newcommand{\id}{\mathbbm{1}}

\newcommand{\nf}[1]{\text{\font\nicefont=cmsy12\nicefont #1}}
\renewcommand{\Im}{\text{Im}}
\renewcommand{\Re}{\text{Re}}

\renewcommand{\sl}[1]{\slashed{#1}}
\newcommand{\slp}{p\makebox[0.11 em][r]{\slash}}
\newcommand{\dirac}{\left(i\sl{\partial}-m\right)}
\newcommand{\deltafkt}[2]{\delta^{#1}({#2}-{#2}')}
\newcommand{\oofpi}[1]{\frac{1}{(4\pi)^{#1}}}
\newcommand{\ootpi}[1]{\frac{1}{(2\pi)^{#1}}}
\newcommand{\oover}[1]{\frac{1}{#1}}
\newcommand{\eps}{\varepsilon}
\newcommand{\rot}{\text{rot}}
\renewcommand{\div}{\text{div }}
\newcommand{\bra}{\left<}
\newcommand{\ket}{\right>}
\newcommand{\ab}{_{\alpha\beta}}

\newcommand{\NNW}{NN-Wechselwirkung}
\newcommand{\NNI}{NN interaction}
\newcommand{\ia}{interaction}
\newsavebox{\ZitName}
\newenvironment{Zitat}[1]
{\em\small\sbox{\ZitName}{\textit{#1}}} {\hspace*{\fill}---
\usebox{\ZitName}\\\\}

\newcommand{\nts}{\rho^s_{(3)}}
\newcommand{\nt}{\rho_{(3)}}
\newcommand{\ns}{\rho^{s}}
\newcommand{\sn}[1]{\sum_{#1}n_{#1}}
\newcommand{\jb}{\vc{j}_B}
\newcommand{\Ms}{{M^*}}
\newcommand{\Es}{{E^*}}
\newcommand{\ks}{{\vc{k}^*}}
\newcommand{\Cs}{\frac{\G^2_\sigma}{m_\sigma^2}}
\newcommand{\Cd}{\frac{\G^2_\delta}{m_\delta^2}}
\newcommand{\Cr}{\frac{\G^2_\rho}{m_\rho^2}}
\newcommand{\Co}{\frac{\G^2_\om}{m_\om^2}}
\newcommand{\jt}{\vc{j}_{(3)}}

\newcommand{\panda}{$\overline{\mbox P}$ANDA~}

\newcommand{\w}{\omega}

%% file: introduction.tex
\section{Introduction}\label{sec:Intro}
\subsection{Emergence of Subnuclear Particle Physics}
\begin{figure}
\begin{center}
\includegraphics[width = 10cm]{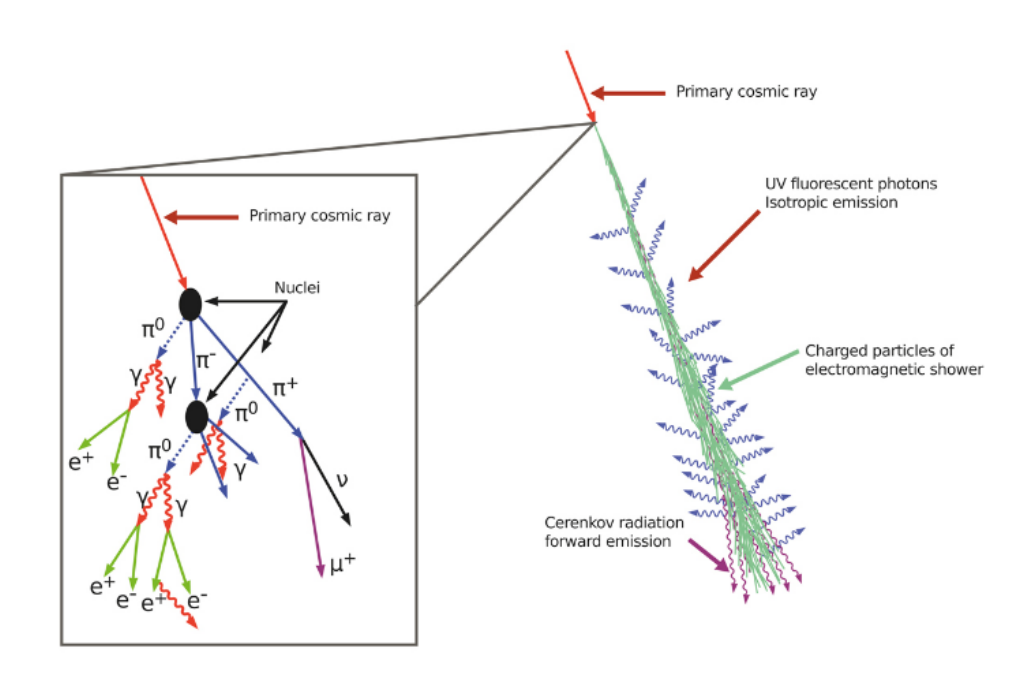}
\caption{Illustration of an air shower resulting from the interaction of an incoming high-energy cosmic particle with atmospheric nuclei. The figure is
        adapted from Ref.~\cite{Bietenholz:2014dlr}. }
\label{fig:CosmicShower}
\end{center}
\end{figure}

The first recorded observation of particles different from the material forming the everyday terrestrial environment
was made as early as 1912 by Victor Franz Hess (1883-1964) during high-altitude balloon campaigns~\cite{Hess:1912srp}. First considered merely a curiosity, the wide-reaching importance of Hess's observations soon became evident.  For his groundbreaking research, Hess was awarded the Nobel Prize in Physics in 1936, along with Carl David Anderson (1905-1991), who discovered the positron in cosmic rays~\cite{Anderson:1933pos}, thereby marking the beginning of the era of antimatter research. The modern understanding of air showers produced by cosmic rays is illustrated in Fig.~\ref{fig:CosmicShower}.

The epochal scientific revolution of quantum physics, accelerating in the decade after 1920, served as a primer for 
Cosmic-ray research, which has made it a thriving field of study. A significant step forward was made by Patrick Maynard Stuart Blackett (1897-1974) in collaboration with Giuseppe Paolo Stanislao ``Beppo'' Occhialini (1907-1993) in constructing a coincidence Wilson cloud chamber.  In 1933, he confirmed the positron and, in the years that followed, charged particle showers, pair production, and pair annihilation processes, which he explained based on the newly formulated Dirac theory\footnote{Paul Adrien Maurice Dirac (1902-1984) shared the 1933 Nobel Prize in Physics with Erwin Schr\"odinger (1887-1961) for \textit{the discovery of new productive forms of atomic theory.}}. Cloud chambers, operated by supersaturated water vapor, were invented much earlier by the Scottish physicist Charles Thomson Rees Wilson (1869-1959), who was awarded the Nobel Prize in Physics in 1927 for \textit{his method of making the paths of electrically charged particles visible through the condensation of vapor}.

\begin{figure}
\begin{center}
\includegraphics[width = 10cm]{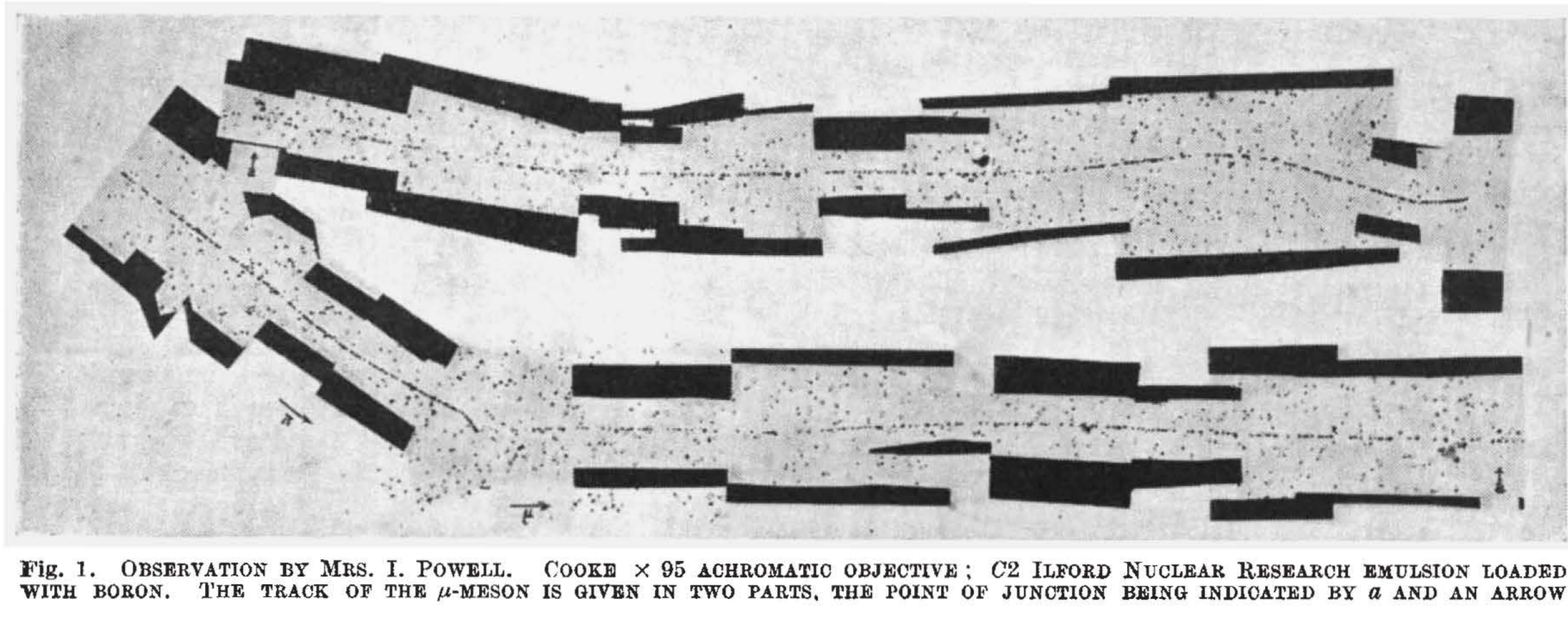}
\caption{The photographic emulsion plate showing the first observation of - as known today - the weak decay of a pion,
$\pi^-\to \mu^- \bar{\nu}_\mu$. The muon-antineutrino $\bar{\nu}_\mu$ is not observed. Photo, including the original caption, taken from Ref.~\cite{Lattes:1947mx}.
}
\label{fig:PiDecay}
\end{center}
\end{figure}

In the late 1930s, Cecil Frank Powell (1903-1969) joined cosmic-ray research. He and Occhialini optimized the technology and the use of photographic plates, which became essential for cosmic-ray research—and for many years thereafter. Exposing the plates at high altitude to cosmic rays, Lattes, Muirhead, Occhialini, and Powell recorded, for the first time, the decay of a pi-meson,
$\pi^- \to \mu^- \bar{\nu}_\mu$, shown in Fig.~\ref{fig:PiDecay} and only later understood to be caused by weak interaction~\cite{Lattes:1947mx, Lattes:1947my}. This measurement can be considered the start of research on hadron production, first evidenced by the track recording the existence of the decaying particle, and second by identifying that particle and its properties through the decay products. Basically, the same scheme is applied in modern production experiments at accelerator facilities, albeit at a much more involved technological level. In 1950, Powell and Occhialini won the Nobel Prize for \textit{their seminal work}.

In their first paper on the pion observation of 1947~\cite{Lattes:1947mx}, Lattes, Occhialini, and Powell write: \textit{We represent the primary mesons by the symbol $\pi$, and the secondary by $\mu$}. In the second paper of 1947~\cite{Lattes:1947my}, they casually refer to the particles as \emph{pions} and \emph{muons}. There is a good reason to state that this is the only documented birth and baptism certificate of the pion. The actual inventor of - at that time - such a hypothetical particle, however, was Hideki
Yukawa (1907-1981)~\cite{Yukawa:1935int1, Yukawa:1937int2} who postulated the existence of a \emph{meson} (lent from the ancient Greek language for \emph{medium}, \textit{i.e.}, between electron and nucleon masses). By that hypothesis, Yukawa could explain the short-range forces between nucleons in accordance with the tiny size of the nuclei. After the discovery of a medium-mass particle of about 200~$m_e$ in cosmic-ray showers, the new particle was initially considered to be the Yukawa meson. By the work of Lattes \textit{et al.}, it became clear that the earlier found particle was not the Yukawa meson, but the pion was the proper candidate, while the muon was of a different nature. For a short while, the muon was, in fact, considered as \textit{the light meson} but soon the distinct nature of the muon as a lepton and fermion\footnote{Half-integer spin-particles are named after Enrico Fermi (1901-1954) who received the Nobel Prize in 1938 \textit{for his demonstrations of the existence of new radioactive elements produced by neutron irradiation, and for his related discovery of nuclear reactions brought about by slow neutrons.}} was understood. 

Interestingly, Yukawa himself apparently never claimed the name pion or pi-meson openly as his origin. In his original papers of 1935 and 1937, Yukawa denotes the \emph{new quantum} as \emph{U-field}~\cite{Yukawa:1935int1, Yukawa:1937int2}.  In his Nobel lecture of 1949~\cite{Yukawa:1949vwn} and in a paper from that year~\cite{Yukawa:1949rmp}, Yukawa seems to have accepted the naming of his meson as pion or pi-meson by using those names repeatedly. 

In his own words, Yukawa summarizes in~\cite{Yukawa:1949rmp} the situation of 1949 by stating:
\emph{This rather puzzling situation of the meson theory changed in 1947 due to the experiment by the Rome group on the decay of negative mesons on the one hand, and the discovery by the Bristol group of two kinds of mesons in cosmic rays on the other.},
where the \emph{Bristol group} refers to Lattes \textit{et al.} and the \emph{Rome group} refers to an earlier paper from 1943 by Nereson and Rossi~\cite{Nereson:1943if}. By denoting the muon as \emph{negative meson}, Yukawa continues to use the notation that had been established after about 1935, when it was thought that there would be only a single meson.
\vspace{0.5cm}

\section{Hadron Production on Accelerators}
As early as 1948, in parallel with cosmic ray research, new opportunities for hadron physics emerged when Gardner and Lattes published the first results on the
\emph{Production of Mesons by the 184-Inch Berkeley Cyclotron}~\cite{Gardner:1948pkt}, demonstrating that hadron research could be performed under laboratory conditions. The circulating beam of $380~\mathrm{MeV}$ alpha particles inside the cyclotron passed through the thin carbon target, producing mesons and a bunch of other particles. The negatively charged mesons were sorted out by the magnetic field and roughly focused at the edge of a stack of photographic plates. Besides using carbon targets, a few experiments were conducted on beryllium, copper, and uranium targets.

The modern era of particle physics — lasting until today — started with the advent of particle accelerators that were able to deliver intense beams of highly energetic particles. In 1953, the Cosmotron synchrotron at BNL delivered proton beams with a design energy of $3.3~\mathrm{GeV}$. A year later, the Bevatron at Berkeley Laboratory became operational with proton beams of even $6.2~\mathrm{GeV}$. With these machines, cosmic ray mesons and V-particles could be produced and studied in the laboratory. The most eminent Bevatron result was — and is — the discovery of hadronic antimatter by first observing the antiproton in 1955~\cite{Chamberlain:1955ns} and shortly after also the antineutron in 1956~\cite{Cork:1957nu}, for which Owen Chamberlain (1920-1959) and Emilio Gino Segr\'{e} (1905-1989) were honored with the Nobel Prize in Physics in 1959 for \emph{the discovery of the antiproton, a sub-atomic antiparticle}.

\subsection{Entering Strange Territory and Symmetry Violation}\label{sec:Strange}
\begin{figure}
\begin{center}
\includegraphics[width = 10cm]{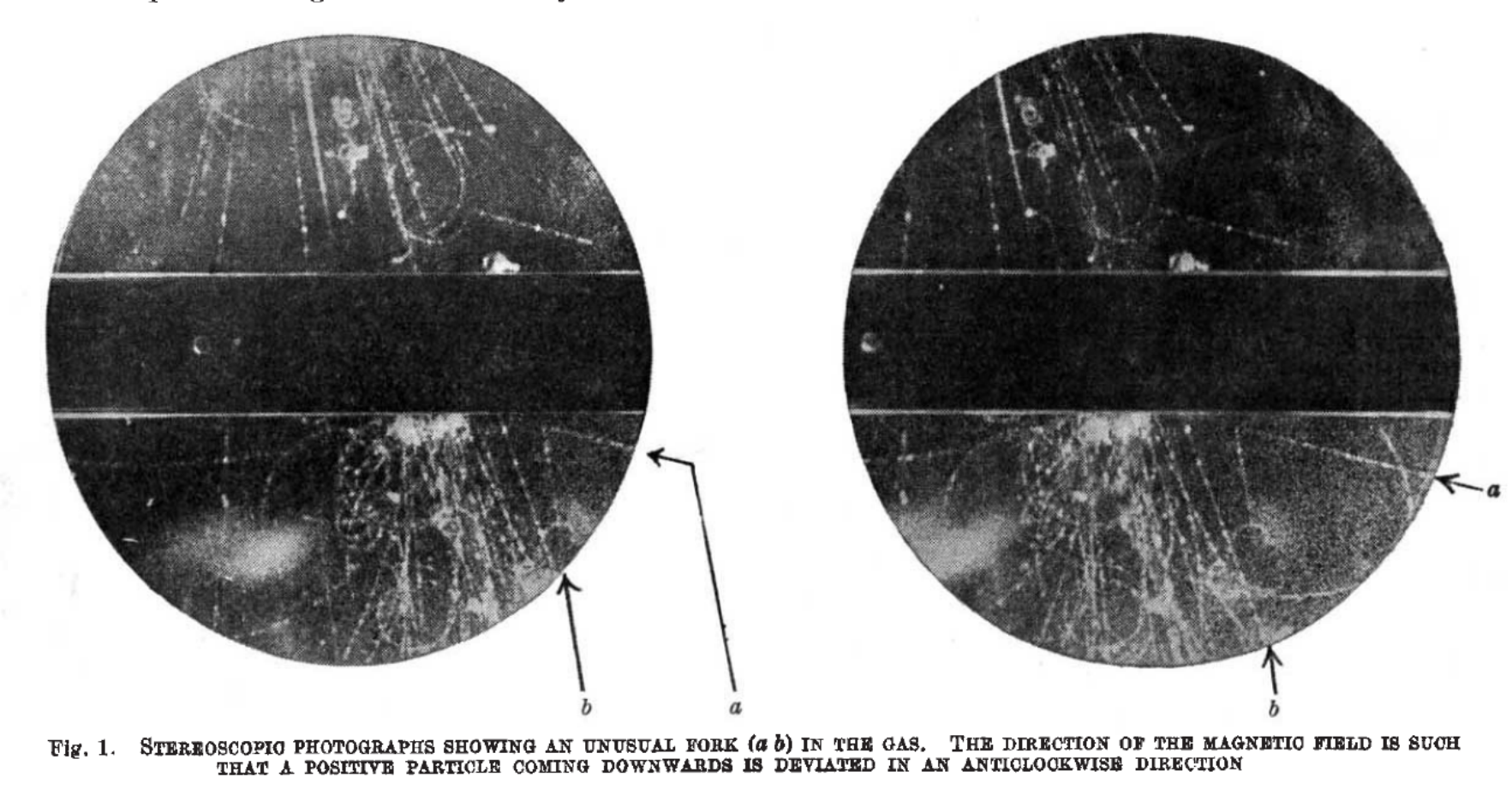}
\caption{Photograph of the cloud chamber event recording the decay of a heavy neutral particle of two-pronged fork \emph{V-particle} pattern. The cloud chamber of size
$(240\times 400\times 460)~\mathrm{mm^3}$ is on exhibition in the Science Museum of the Imperial College of London, Great Britain. Photograph taken from Ref.~\cite{Rochester:1947mi}. }
\label{fig:KaonV}
\end{center}
\end{figure}

The years around 1947 actually played a key role in hadron discoveries. Rochester and Butler reported on Wilson cloud chamber observations indicating \emph{Evidence for the Existence of New Unstable Elementary Particles}~\cite{Rochester:1947mi}, showing an unusual decay pattern resembling a two-pronged fork, later denoted as
V-particles~\cite{Armenteros:1951vd}, with a mass of $980~m_e\leq m < m_p$ ($m_e \simeq 
511~\mathrm{keV}$ and  $m_p\simeq 938.3~\mathrm{MeV}$ are the electron and proton rest masses, respectively) - fitting well with what was shortly thereafter named the $K$-meson.  In 1950, Hopper and Biswas of Melbourne observed another neutral particle decaying by the same V-like pattern observed by Rochester and Butler, but with a much heavier mass of about 2370~$m_e$, which surprisingly matches the rest mass of the $\Lambda$ hyperon. The characteristic decay pattern of these strange particles, shown in Fig.~\ref{fig:KaonV}, led to the name
\emph{V-particles}~\cite{Rochester:1993yn}. Hence, the first hadrons - a meson and a baryon - with strangeness content were observed, although it still took another 15~years before a concise theoretical scheme was formulated. 

In the years that followed, the strangeness sector could be explored systematically, resulting in a whole spectrum of higher-lying hyperon states. The observation of the non-conservation of parity (P) in hyperon decays was the first experimental datum of a broken symmetry at the level of elementary particles. In many cases, however, detailed spectroscopic studies were beyond the technological capabilities of that time. A typical case is the $\Lambda(1520)$ excitation. That 
$J^\pi=\frac{3}{2}^-$ excited state of the $\Lambda$ hyperon 
was originally observed as an elastic resonance in kaon scattering on a proton 
target in 1962~\cite{Ferro-Luzzi:1962jcn}. However, it lasted about 50~years and 
required an electroproduction experiment and the advanced detector setup at JLab before the pole position for the first hyperon of that rather sharp resonance ($\Gamma \simeq 18~\mathrm{MeV}$) could be determined precisely for 
$\Lambda(1520)$ ~\cite{Qiang:2010ve}.

A far-reaching result of the Cosmotron and the Bevatron is related to parity 
non-conservation, \textit{i.e.}, left-right or mirror symmetry violation. Breaking mirror symmetry means that the image would be different from the original, \textit{e.g.}, placing pears in front of a symmetry-broken mirror might result in an image showing apples. The story started in 1949, Rosemary Brown of the Powell 
group~\cite{Brown:1949mj, Sheehy:2024nat}, identified on an emulsion plate, exposed to cosmic rays, an unexpected decay event. A hitherto unknown object, given the name \textit{$k$ particle}, with a mass of about $900~\mathrm{m_e}$ decayed by two distinct \textit{theta} ($\theta^+$) and \textit{tau} ($\tau^+$) modes. For a long time, the validity of those data and a few rare follow-up observations by other groups were considered questionable. By 1953, a total of 11~$k$ particle events were known. When the Cosmotron and the Bevatron delivered their respective first beams, campaigns were initiated on that issue, rapidly increasing the number of $k$ particle observations and the related theta- and tau-decays, which are now obtained under controllable laboratory conditions. The two separate decay modes were identified as given by two and three pions, respectively~\cite{Orear:1956tau}. When Chinowsky  and 
Steinberger\footnote{Jack Steinberger (1921-2020) was a recipient of the 1988 Nobel Prize in Physics, along with Leon M. Lederman (1922-2018) and Melvin Schwartz (1932-2006), for \textit{the discovery of the muon neutrino.}} in 1954 could prove experimentally the intrinsic negative parity of pions~\cite{Chinowsky:1954zz}, it was clear that the $k$ particle suffers a parity-violating decay.  Shortly after, the famous
\textit{Wu experiment}, performed by Chien-Shiung Wu (1912-1997), cleared the skies by showing convincingly that weak interaction indeed violates parity conservation~\cite{Wu:1957my}\footnote{To the surprise of many, Chien-Shiung Wu did NOT receive the Nobel Prize for her \textit{pioneering 
Experimentum Crucis}, thereby sharing the fate of many other women in science}. At that time, the nomenclature was changed to $K$ meson or kaon. A posteriori, it's evident that the puzzling $k$ particle decay was indeed the first observation of the weak decay of the kaon. It is noteworthy to mention that, in 1954, Chinowsky and Steinberger also showed that charged and neutral pions have slightly different rest masses~\cite{Chinowsky:1954mju}.
 
Describing the striking discovery of parity non-invariance as a fundamental interaction, which is a game-changer for elementary particle physics, does not overemphasize the importance of that observation. In 1956, Tsung-Dao (T.D.) Lee (1926-2024) and Chen Ning (C.N.) Yang (1922-2025) formulated a concise theory explaining phenomena that, at that time, were totally unexpected and counterintuitive. In their seminal paper titled \textit{Question of Parity Conservation in Weak Interactions}~\cite{Lee:1956qn}, honored by the 1957 Nobel 
Prize in physics, they wrote: \textit{The way out of the difficulty is to assume that
parity is not strictly conserved, so that $\theta^+$ and $\tau^+$ are
two different decay modes of the same particle, which
necessarily has a single mass value and a single lifetime}.
The Nobel Prize was awarded in equal parts to C.N.~Yang and T.D~Lee \textit{for their penetrating investigation of the so-called parity laws, which have led to important discoveries regarding elementary particles}.

Since their discovery, neutral kaons $K^0$ together with their anti-particle counterparts $\bar{K}^0$ have been a laboratory for research on fundamental aspects of particle physics. 
They were the first hadrons to show oscillation phenomena through the formation of short-lived ($K_S$, $t_{1/2} \approx 8\cdot 10^{-11}~\mathrm{s}$) and long-lived ($K_L$, $t_{1/2} \approx 5\cdot 10^{-8}~\mathrm{s}$) superpositions of the flavor (strangeness) eigenstates $K^0$, $S=-1$ and $\bar{K}^0$, $S=+1$, respectively. The differences in lifetime are due to the differences in the phase space available for the two-pion and three-pion decay channels: the larger two-pion phase space causes a larger decay width and a shorter lifetime than that of the three-pion decay. 

Once a $P$ violation in weak interactions of hadrons had to be accepted, it was assumed that at least the combined charge conjugation and parity (CP) symmetry would be conserved. However, 
in an experiment at the AGS at BNL, James Watson Cronin (1931-2016) and 
Val Logsdon Fitch (1923-2015), together with James H. Christenson and Rene Turlay, observed in 1964 a CP-violating 2~$\permil$ difference in the decay of $K_L$ compared to 
$K_S$~\cite{Christenson:1964fg}. Cronin and Fitch were awarded the Nobel Prize in Physics in 1980 for \textit{the discovery of violations of fundamental symmetry principles in the decay of neutral K-mesons}. Since then, CP violation has  been detected in the decay of 
charm~\cite{LHCb:2019hro} and beauty mesons~\cite{BaBar:2001pki, Belle:2001zzw}. 
Surprisingly, CP violation was not observed for a long time in baryon decays until a very recent experiment at LHCb found CP violation in the decay of $\Lambda^0_b$ vs. 
$\bar{\Lambda}^0_b$~\cite{LHCb:2025ray}. 

As early as 1967, Andrey Dmitrievich Sakharov (1921-1989)\footnote{In later years, Sakharov engaged in human rights activities, turning into a Soviet dissident, and received in 1975 the Nobel Prize for Peace \textit{for emphasizing human rights around the world}.} published in~\cite{Sakharov:1967dj} the famous three Sakharov conditions on the cosmological matter-antimatter asymmetry problem. The second paradigm demands a violation of $C$ and $CP$ symmetry. The modern understanding of CP violation at the microscopic level in hadrons is related to a \textit{CP-violating phase factor} in the Cabibbo–Kobayashi–Maskawa (CKM) 
matrix~\cite{Kobayashi:1973fv} (named after Nicola Cabibbo (1935-2010), Makoto Kobayashi (born 1944), and Toshihide  Maskawa (1940-2021)). Kobayashi and Maskawa were awarded half of the 2008 Nobel Prize in Physics for \textit{the discovery of the origin of broken symmetry, which predicts the existence of at least three families of quarks in nature}. The CKM matrix describes the mixing of quarks under weak interactions, which is also relevant in another open nuclear physics problem, namely, neutrinoless double beta decay. In combination, these research areas connect sub-nuclear and cosmological phenomena.  The present status is that time reversal, combined with $CP$ symmetry, is conserved, forming together the \textit{triple-symmetry $CPT$}. 

As a side remark, it might be of interest that Richard Feynman (1918-1988) is said to have placed a 50~US Dollar bet against parity 
violation~\cite{APSNews:2001} at a conference in 1955, underlining how unreal such a conjecture appeared at the time of the first rumors of an experimental signature. Feynman is eminently known for \textit{his contributions to the development of quantum electrodynamics}, for which he received the Nobel Prize in Physics in 1965, jointly with Julian Schwinger (1918-1994) and Shinichiro Tomonaga (1906-1979). Moreover, Feynman's name is engraved forever in \textit{Feynman diagrams}, which are indispensable tools for modern QFT research in any theoretical context. The diagram shown in Fig.~\ref{fig:titlepage} summarizes the topic of this article in a well-defined, albeit rather condensed, graphical language of general rules~\cite{Itzykson:1980rh, Cheng:1985bj}. 

\subsection{Charm and Beyond}\label{sec:Charm}
The focus of this article is on the physics of hadrons made of up ($u$), down ($d$), and strange ($s$) valence quarks and anti-quarks, respectively. Charm physics is discussed in due detail elsewhere in this Encyclopedia~\cite{Friday:2025gpj}. However, since the charm quark is the partner of the strange quark in forming together the \textit{second generation} of the quark family, we briefly give an overview of the exciting history of charm discovery.

As early as 1964, the idea arose that there might be more flavors than $\{u,d,s\}$. James Bjorken (1934-2024) and Sheldon Glashow (born 1932) hypothesized the existence of another quark, which they named charm ($c$). As 
Glashow declared in 1976 in a newspaper article in \textit{New York Times} the charm quark received its name because of \textit{the symmetry it brought to the Subnuclear world}, as cited in~\cite{Riordan:1987}. Experimentally, a signature of charm was first observed in 1974 in two independent experiments at BNL by Samuel~C.C.~Ting (born 1936) and 
collaborators~\cite{E598:1974sol} and at SLAC by the group of Burton Richter (1931-2018)~\cite{SLAC-SP-017:1974ind}. Both teams found a narrow resonance at an energy of $3.1~\mathrm{GeV}$ with a width of only $5~\mathrm{MeV}$. A particularly strong confirmation of the existence of charm was that the BNL group used the purely hadronic nuclear reaction $p \mathrm{Be} \to e^+e^- X$, while the SLAC group observed the new resonance in purely leptonic $e^+e^-$ annihilation reactions. Thus, this proved the ubiquity of the phenomenon.  
The BNL group gave that state the name $J$, the competitors from SLAC chose $\psi$, and finally, as a compromise and in honor of the work of both teams, the two labels merged into the official name $J/\psi$. 

Ting and Richter met on November 15, 1974; they issued a press release about their discoveries. The charm story then took off: On November 21, 1974, the SPEAR@SLAC collaboration already reported the observation of $\psi^\prime(3700)$, an excited state of the $J/\psi(3100)$. Both particles are charmonium states, composed of $c\bar{c}$ valence quarks. The modern spectroscopic notation is $J/\psi(1S)$  and $\psi^\prime(2S)$, following the constituent quark model and indicating that $\psi^\prime$ is understood as a radial (compressional) excitation of $J/\psi$. The latest mass values are $M_{J/\psi}=3096.900\pm 0.006~\mathrm{MeV}$ and $M_{\psi^\prime}=3686.097\pm 0.010~\mathrm{MeV}$~\cite{PDG:2024cfk}.

The two group leaders, Ting and Richter, were awarded the 1976 Nobel Prize, honoring the great achievement of the two teams for \textit{discovering the subatomic $J/\psi$ particle}. Their epochal discoveries became known as the \textit{November Revolution} in particle physics. 

In 1986, Matsui and Satz predicted  \textit{$\ldots$ that $J/\psi$ suppression in nuclear collisions should provide an unambiguous signature of quark-gluon plasma 
formation}~\cite{Matsui:1986dk}. 
Since then, $J/\psi$ has played a special role in searching for QGP in heavy-ion collisions at ultra-relativistic energies at RHIC and the LHC. 

A multitude of \textit{hidden charm} charmonium states and \textit{open charm} mesons and baryons have been observed, containing a charmed quark or anti-quark to which $\{u,d,s\}$ partners and the respective anti-quarks are attached. There is strong evidence that charmed hadrons may develop exotic configurations, \textit{e.g.}, tetra-quark states or molecular-like states formed by D-mesons. These so-called $X, Y, Z$ states fall out of the systematics of the otherwise very successful quark models for quarkonia. Their spectroscopy is discussed elsewhere in this Encyclopedia~\cite{Hanhart:2025bun}. 

As an explanation for the still unsatisfactorily solved CP violation puzzle, Makoto Kobayashi and Toshihide Maskawa predicted in 1973 the existence of a third family of quarks~\cite{Kobayashi:1973fv}. That conjecture was partly confirmed in 1977 by the discovery of 
the bottom ($b$) or beauty quark~\cite{E288:1977xhf} and completed by the observation of the top ($t$) quark in 
1995~\cite{CDF:1995wbb, D0:1995jca}.  Both the bottom and top quarks were first detected at FNAL.  

The Nobel Prize of 2008 was awarded one half to Yoichiro Nambu (1921-2015) \textit{for the discovery of the mechanism of spontaneous broken symmetry in subatomic physics}, the other half was shared by Makoto Kobayashi (born 1944) and Toshihide Maskawa (1940-2021) \textit{for the discovery of the origin of the broken symmetry which predicts the existence of at least three families of quarks in nature}. Bottom and top quarks together represent the third generation of the quark family.

Charmonium-like hadron configurations have been found for bottom quarks ($\Upsilon$ family) and were also expected to exist in the top ($t$) quark sector. The latter is indeed confirmed by the first observation of a $t\bar t$ vector meson by the CMS experiment at the LHC~\cite{CMS:2025dzq}. In Fig.~\ref{tab:quarks}, a table summarizing the three quark generations and their properties is shown.

\section{Theoretical Approaches and Interpretation of Data}
\subsection{Aspects of QCD}
QCD is the fundamental theory of strong interactions and thus of hadrons\footnote{Lev Borisovich Okun (1929-2015) introduced the term \textit{hadron} (ancient Greek for \textit{heavy} or \textit{strong})  in a plenary talk at the 1962 International Conference on High Energy Physics.}. The beauty and peculiarities of Quantum Chromodynamics (QCD) as a non-Abelian quantum gauge field theory of intrinsically non-perturbative character will be discussed elsewhere in this Encyclopedia. Here, an overview is provided of the aspects that are of special interest and relevance for hadron physics and the production of hadrons.

In the early days of nuclear and elementary particle physics, the plethora of particles observed in cosmic-ray and high-energy experiments made it improbable that they were elementary particles. Obviously, a new approach was necessary, allowing one to bring order to the data. In 1961, Yuval Ne'eman (1925-2006) introduced the classification of hadrons by means of the SU(3) flavor symmetry group,  exploiting the representation of the SU(3) group in terms of eight linearly independent 3-by-3 matrices and three elementary degrees of freedom. At the same time, Murray Gell-Mann (1929-2019) also independently pursued similar group-theoretical research and named the approach the Eightfold Way in a book from 1964~\cite{Gell-Mann:1961omu, Gell-Mann:1964ook}.

\begin{figure}
\begin{center}
\includegraphics[width = 8cm]{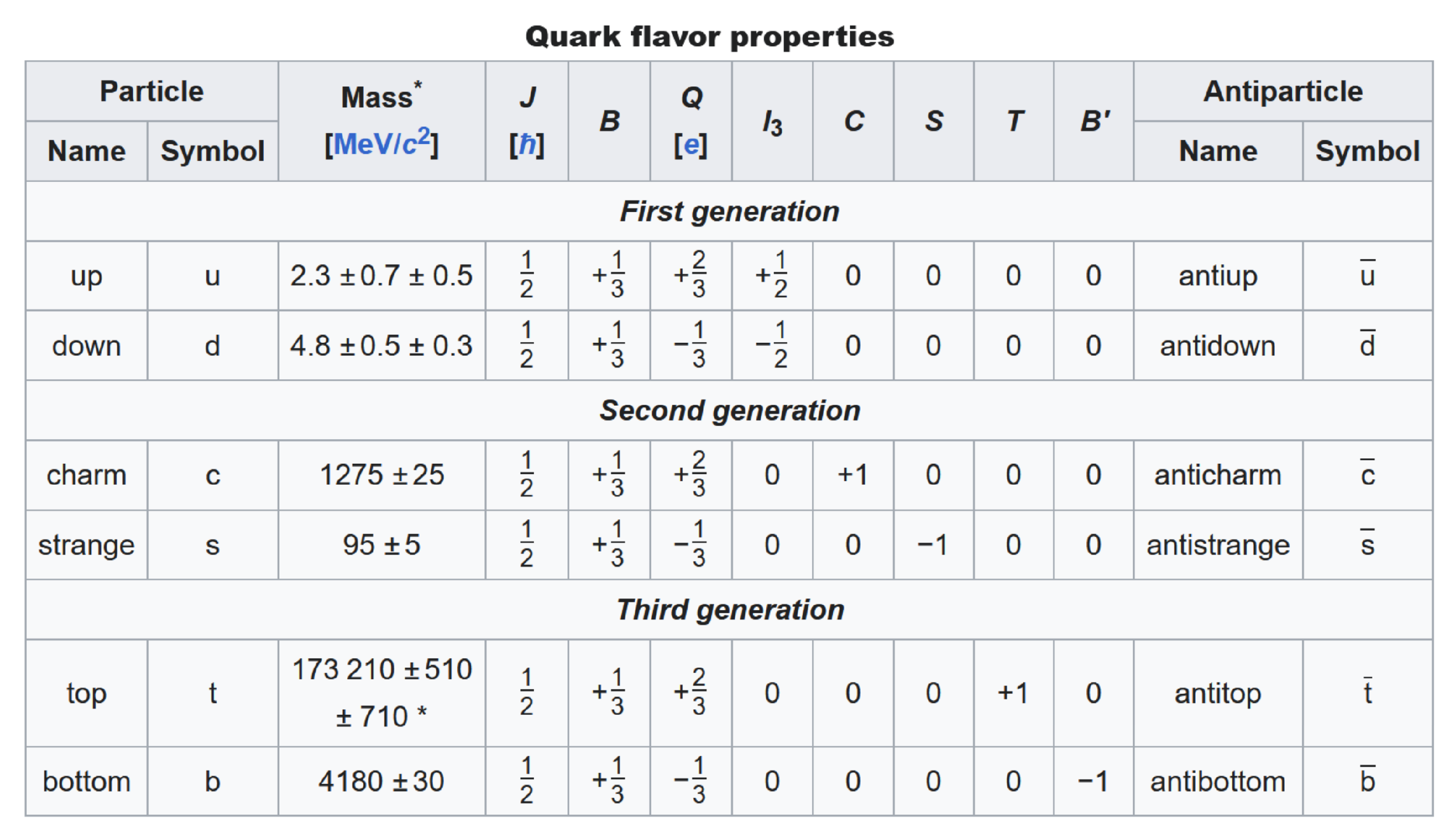}
\caption{Properties of the current quarks as appearing in the QCD Lagrangian of the first $(u,d)$, second $(c,s)$, and third $(t,b)$ generation~\cite{PDG:2024cfk}. As a generalization of isospin, originally introduced for $\{u,d\}$ systems and used above to distinguish proton and neutrons by the eigenvalues of the third component of the isospin operator $I_3=\pm \frac{1}{2}$, heavier quarks are characterized by the flavors $c, s, t, b$ with values $0,1$. In this chapter, only production processes involving hadrons built of the first and second quark generations are considered. The figure is
        adapted from Ref.~\cite{AnaSofiaCarvalho-SoaresQuarks:2025wik}. }
\label{tab:quarks}
\end{center}
\end{figure}

In parallel to  Ne'eman and Gell-Mann, George Zweig (born 1937) developed his own version of a SU(3) scheme of strongly interacting particles~\cite{Zweig:1964ruk, Zweig:1964jf}, mostly during his stay as a post-doctoral researcher at CERN. His contributions to QCD are engraved in the OZI rule, accounting, \textit{e.g.}, for the hindered decay of the $\varphi(1020)$ vector meson.  Thus, by 1964, two competing group theoretical approaches were available, relating to the abstract mathematical structure of a non-commuting Lie algebra (named after the mathematician Sophus Lie (1842-1899) to elementary particle physics. Although at that time, the authors insisted on essential differences between their theories, there were obvious overlapping similarities. Both approaches needed three elementary entities in order to place mesons and baryons on the same footing. Zweig chose the name \textit{aces}, inspired by the joker cards of card games.
However, in a paper of 1964~\cite{Gell-Mann:1964ewy}, Gell-Mann introduced  \textit{A Schematic Model of Baryons and Mesons}. That paper can be considered foundational to the theory of elementary particle physics: Gell-Mann points out that \textit{A simpler and more elegant scheme can be
constructed if we allow non-integral values for the
charges}. He also denotes the constituents of his model as \textit{quarks} by explicitly referring to the novel \textit{Finnegans Wake} by James Joyce, and that naming was finally adopted by the community. 
\footnote{``\textit{Three quarks for Muster Mark}'' is a line from James Joyce's novel ``\textit{Finnegans Wake}.'' (It took James Augustine Aloysius Joyce (1982-1941) 16~years to complete this experimental novel.) The phrase is notable because it provided the name ``\textit{quark}'' for a fundamental subatomic particle, chosen by Murray Gell-Mann. Gell-Mann initially considered the word ``\textit{kwork}'' before encountering the Joyce passage and finding it more fitting. While the original line seems to rhyme ``\textit{quark}'' with ``\textit{Mark,}'' Gell-Mann proposed a justification for pronouncing it ``kwork'' by associating it with the pub owner's call of ``\textit{Three quarts for Mister Mark.}''}  In Figure (Table)~\ref{tab:quarks}, the quarks and their properties are listed. 
Murray
Gell-Mann received the Nobel Prize in Physics in 1969 for his \textit{contributions and discoveries concerning the classification of elementary particles and their interactions}.

\begin{figure}
\begin{center}
\includegraphics[width = 11cm]{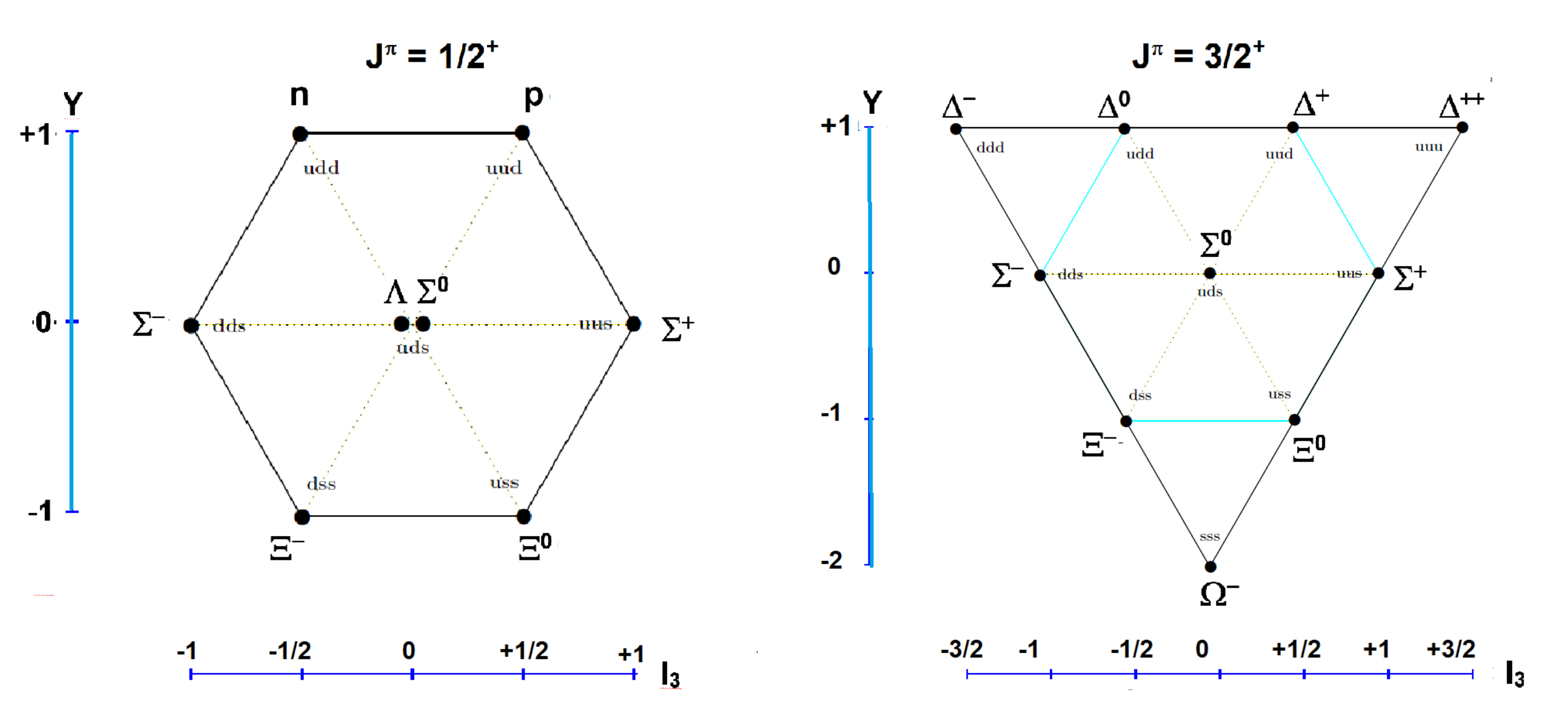}
\caption{The first two baryon SU(3) flavor multiplets are given by an octet (left) and a decuplet (right). The valence quark content of the baryons is indicated explicitly. The vertical axes represent the hypercharge $Y=S+B$, given by the strangeness $S$ and baryon number $B$; the horizontal axes indicate the third component $I_3=Q-\frac{1}{2}Y$ of the isospin $I$, which also includes the charge number $Q$. The group theoretical background and construction of these kinds of diagrams are discussed in depth in textbooks, see
\textit{e.g.},~\cite{Cheng:1985bj}. }
\label{fig:Multiplets}
\end{center}
\end{figure}

\begin{figure}
\begin{center}
\includegraphics[width = 12cm]{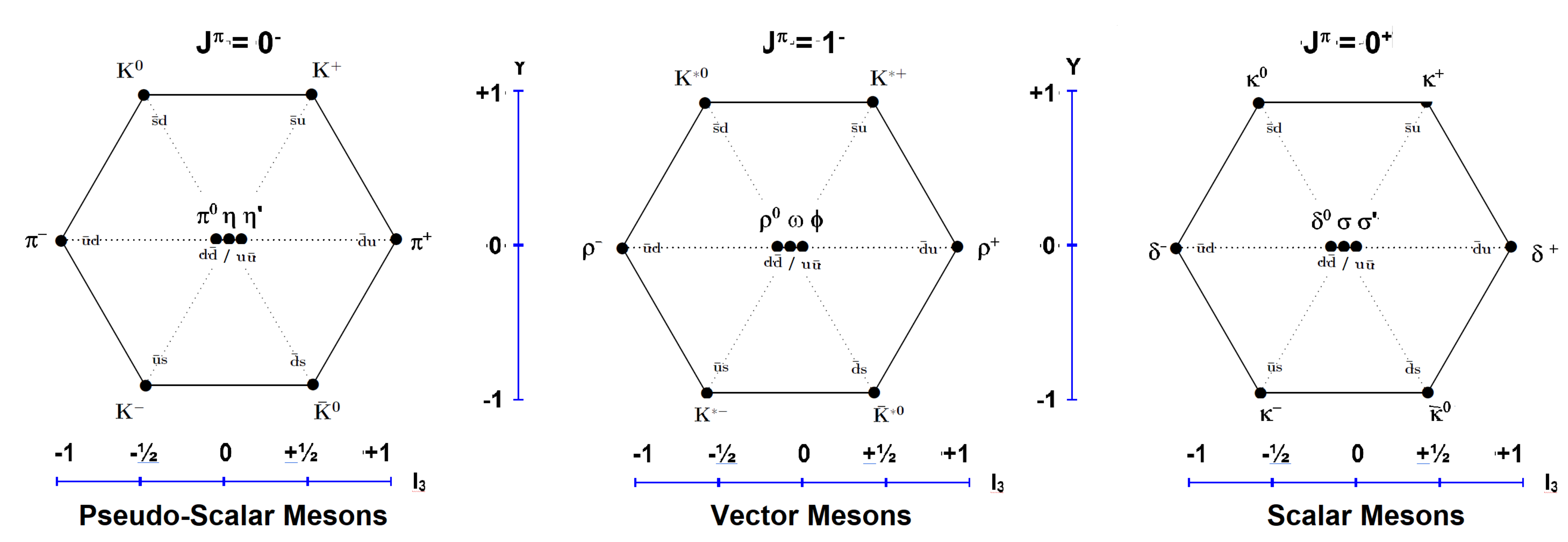}
\caption{The lowest meson SU(3) flavor nonets of pseudo-scalar (left), vector (center), and scalar mesons (right) as functions of hypercharge Y and isospin I$_3$. Different to the baryons, mesonic two-quark systems of $\{u, d, s\}\otimes\{\bar{u},\bar{d},\bar{s}\}$ structure arrange into SU(3) octets and accompanying singlets.
The isoscalar (charge-neutral) octet mesons obtain an isoscalar singlet partner.  The singlet mesons are included into the octet diagrams, giving rise to the octet-singlet pairs $\{\eta,\eta^\prime\}$, $\{\omega,\phi\}$, and $\{\sigma\simeq f_0(500),\sigma'\simeq f_0(980)\}$ in the pseudo-scalar, vector, and scalar meson sectors, respectively. Singlets and octets are mixed by, in principle, arbitrary mixing angles. A widely used choice is \emph{Ideal Mixing}, chosen such that the isoscalar octet mesons $\eta,\omega,\sigma$ do not contain a $s\bar{s}$ component. Electromagnetic effects lead to $\omega -\rho$ mixing, and correspondingly, $\sigma - \delta^0$ mixing has to be expected. Scalar mesons may contain multi-pion, \textit{i.e.}, multi-$q\bar q$ components. The complexities of scalar mesons are especially visible in the isovector scalar meson $\delta \simeq a_0(980)$. $a_0(980)$ is decaying preferentially into $\eta\pi$, finally ending in three pions, but also $K\bar K$ and $\eta^\prime\pi$ decays are reported~\cite{PDG:2024cfk}. The physical counterpart of the $\kappa$ meson is the $K^\ast_0(700)$ meson of a notoriously uncertain pole 
structure~\cite{PDG:2024cfk}. Thus, the physically observed mesons are not identical with the bare SU(3) mesons.}
\label{fig:Nonets}
\end{center}
\end{figure}

Group theory predicts highly valuable ordering schemes by which particles of equal symmetry class can be arranged in multiplets.  Graphical representations of the lowest baryon octet and decuplet are shown in Fig.\ref{fig:Multiplets}. The pseudo-scalar, scalar, and vector meson nonets are displayed in Fig.~\ref{fig:Nonets}. 

Although group theory does provide neither absolute mass scales nor absolute (numerical) values of coupling constants, relations among coupling constants and sum rules for the masses within a given multiplet can be derived, typically reducing the problem to the determination of a few free parameters.  
In~\cite{Gell-Mann:1962yej}, Murray Gell-Mann exploited those relations to identify the already known two-pion rho-meson ($\rho$) resonance as part of a vector meson octet. He also predicted that there must be an isoscalar partner of the rho-meson decaying into three pions, which he named the omega-meson ($\omega$). Moreover, at the International Conference on High-Energy Nuclear Physics, Geneva, 1962, Gell-Mann used the mass relations to predict that the still missing last decuplet member of strangeness $S=-3$, \textit{i.e.}, the $\Omega^-$ hyperon, would have the mass $M^{(theo)}_{\Omega}=1676~\mathrm{MeV}$. Two years later, the particle was indeed found at Brookhaven by Barnes \textit{et al.}~\cite{Barnes:1964pd, Barnes:1964ga} at a mass of
$M^{(exp)}_{\Omega}=1674 \pm 3~\mathrm{MeV}$\footnote{The BNL 80-in. hydrogen bubble chamber was exposed to a mass-separated beam of $5.0~\mathrm{BeV/c}$ K~mesons at the Brookhaven AGS. About 100,000 pictures were taken, containing a total K~track. The analysis led to a \textit{single} event associated with the $\Omega^- \to n K^0 \pi^-$ decay.}. That extraordinarily good agreement provided strong support and confidence for the quark model of hadrons. 

Independent of Gell-Mann,  Susumo Okubo (1930-2015)  was pursuing his own studies of unitary symmetry in strong interactions at Rochester, leading to mass relations of the same type~\cite{Gell-Mann:1961omu, Okubo:1961jc, Okubo:1962zzc}. Since then, the SU(3) mass relations are known as Gell-Mann-Okubo (GMO) mass formulas\footnote{Deviations from the  GMO mass formula predictions often provide important hints for mass shifts caused by mixing effects from dynamical self-energy.}.

The interpretation, prediction, and description of hadron spectra in terms of constituent quarks paved the way for QCD. It was clear that Gell-Mann's quark model could not be the final answer to strong interaction physics. A missing link was the lack of a theory incorporating the forces between quarks.  In 1973, Harald Fritzsch, Murray Gell-Mann, and Heinrich Leutwyler published a letter~\cite{Fritzsch:1973pi} in which they presented the principles of a quantum field theory of strong interactions, pointing out the \textit{Advantages of the Color Octet Gluon Picture} already in the title of their article.    

A defining feature of QCD is the SU(3) color group, physically represented by three color charges.  Color confinement had to be imposed in order to account for the empirical observation that hadrons exist only in color singlet states, \textit{i.e.}, they are colorless. 

A central role is played by the spontaneously broken chiral symmetry, which leads to a complex, non-perturbative QCD vacuum composed of gluon and quark condensates. The pseudo-scalar mesons shown below in Fig.~\ref{fig:Nonets} represent, for example, the Goldstone bosons of broken chiral symmetry in the $u, d, s$ sector.

The works of Yuval Ne'eman, Murray Gell-Mann, and George Zweig were the long-awaited breakthroughs towards a new understanding of hadrons in terms of a few elementary degrees of freedom. The theory of Fritzsch, Gell-Mann, and Leutwyler provided the missing links for a full-scale theory of strong interaction physics. The modern understanding of hadrons and their interactions, emerging from those ground-breaking works, is centered around elementary massive quark fermion fields, interacting by gluon gauge vector fields as the force carriers of strong interactions. Different from the photons of QED and the weak interaction gauge bosons $Z^0, W^\pm$, gluons carry color charges and interact with each other. QCD is a non-Abelian quantum gauge field theory of highly non-linear character. The complexities manifest in a rich spectrum of particles, but also in peculiarities such as running coupling constants, color confinement, and quarks with fractional baryon and charge numbers. 

Long before the advent of the SM, Peter Higgs (1929-2024) proposed in 1964 a general mechanism for mass generation, including a multiplet of 
bosons\footnote{Particles with integer spin are named bosons in honor of Satyendra Nath Bose (1894-1974).}. At the same time, Fran\c{c}ois Englert (born 1932) and Robert Brout (1928-2011) independently discovered the same mechanism. After an initial period of ignorance, in the late 1960s, the community became aware of the deep relevance of that mechanism and adopted it informally as \textit{Higgs mechanism}. 
In 2012, indeed, a boson of about the right mass was found at 
CERN~\cite{ATLAS:2012klq, CMS:2012wnd}. Further studies confirmed that this was, in fact, the long-sought (neutral) Higgs boson $H^0$ (colloquially called the ``God Particle'') of mass $m_{H} = (125.20\pm 
0.11)~\mathrm{GeV}$~\cite{PDG:2024cfk}.  Higgs and Englert received the Nobel Prize in 2013 \textit{for the theoretical discovery of a mechanism that contributes to our understanding of the origin of the mass of subatomic particles, which was recently confirmed through the discovery of the predicted fundamental particle by the ATLAS and CMS experiments at CERN's Large Hadron Collider}. 

With the discovery of the Higgs boson, the Standard Model (SM) of elementary particle physics has achieved the status of an intrinsically closed theory. The combined predictive power of QCD and electroweak theory is confirmed in all experimental 
tests. Moreover, until now, the SM has survived all attempts to search for physics beyond the SM.  

Electroweak theory was developed in the 1960s by Sheldon Glashow (born 1932), Mohammad Abdus Salam (1926-1996), and Steven Weinberg (1933-2021). They received equal parts of the 1979 Nobel Prize in Physics \textit{for their contributions to the theory of the unified weak and electromagnetic interactions between elementary particles, including, inter alia, the prediction of the weak neutral current}. 

\subsection{From QCD to Hadrons}
Hadron physics investigates condensed asymptotic mass eigenstates of QCD, held together by the strong forces among quarks and gluons, and additionally stabilized by the surrounding polarization cloud. Hadrons are a generic new form of bound states. Attempts to separate the constituents fail, except under the extreme conditions of a highly compressed and heated quark-gluon plasma. Spectroscopic studies are demanding due to their reliance on powerful accelerators and detector systems. 

However, attempts to explore the interior of hadrons have been made since the beginning of modern elementary particle physics. 
First access to the intrinsic properties of the nucleon was gained as early as 1955 by Robert Hofstadter (1915-1990). Hofstadter was a pioneer in electron-nucleus scattering and used the same methods to measure the electromagnetic form factors of the 
proton~\cite{Hofstadter:1955ae, Hofstadter:1961zz}. Those measurements and the neutron data of Sobottka \textit{et al.}~\cite{Sobottka:1960neutron} finally led to two revolutionary conclusions: Firstly, the nucleon is an extended object. Secondly, protons and neutrons both have a charged core, which, in a proton, is surrounded by a cloud of neutral pions, while in neutrons, the internal charge is compensated by a cloud of oppositely charged pions. Hofstadter shared with M\"{o}ssbauer the Nobel Prize in 1961  
\textit{for his pioneering studies of electron scattering in atomic nuclei and for his subsequent discoveries concerning the structure of nucleons}.   

On the energy scale of modern hadron physics, the Hofstadter experiments, using electron beams of about $100~\mathrm{MeV}$, were conducted at the lowest limit of quasi-elastic scattering with limited resolving power, where the collective features of the target prevail.  About a decade later, when beams of much higher energy were available, deep-inelastic scattering revealed many more details of nucleon structure, as was exploited by Friedman \textit{et al.} around 1970~\cite{Friedman:1972}. Friedman started his research in Hofstadter's group at Stanford. He then began a collaboration with Kendall on probing nuclear targets with the $22~\mathrm{GeV}$ electron beam at SLAC in the late 1960s. The much increased resolving power is compatible with an internal three-particle structure of the proton, in perfect agreement with the constituent quark model. Jerome Isaac Friedman (born 1930) received the 1990 Nobel Prize in Physics, together with Henry Kendall (1926-1999) and Richard Taylor (1929-2018) \textit{for their pioneering investigations concerning the deep inelastic scattering of electrons on protons and bound neutrons, which have been of essential importance for the development of the quark model in particle physics}.

These works were key contributions to the still ongoing efforts to unravel the secrets of hadron structure. Hadrons are understood as confined, color-neutral systems of quarks, fermions of spin $s=\frac{1}{2}$ with fractional electric charge, flavor, and color, interacting via gluons, colored spin-$1$ and self-interacting vector gauge fields.  Renormalization group techniques and regularization methods are indispensable components of the QCD mathematical toolbox. Spontaneous chiral symmetry breaking is a defining mechanism of QCD. The complexities of the non-Abelian structures of QCD lead to higher-order non-linear self-interactions, which inhibit solving the deeply nested field equations in closed form. Only in the ultra-high energy regime of asymptotic freedom beyond about $250~\mathrm{GeV}$, where perturbative treatments become possible and leading-order results can be derived analytically, was the work of David J. Gross (born 1941), H. David Politzer (born 1949) and Frank Wilczek (born 1951) \textit{honored with the Nobel Prize (in 2004) for their discovery of asymptotic freedom in the theory of the strong interaction}.

Mesons and baryons are the asymptotic mass eigenstates of QCD, but they are fundamentally different in character.
Considered as systems of valence quarks and gluons, mesons are formed by an even number of valence quarks, while baryons are defined as particles containing an odd number of valence quarks. In both cases, an arbitrary number of gluons is allowed in addition. Since quarks are Fermions with spin $s=\frac{1}{2}$  and positive parity, baryons carry half-integer spin quantum numbers $J_B=\frac{1}{2},\frac{3}{2},\frac{5}{2},\ldots$, while mesons have integer spin quantum numbers $J_M=0,1,2\ldots$. Moreover, baryons play a special role in cosmology because they account for protons and neutrons, along with a minor contribution from electrons, in the matter content of the Universe.

\begin{figure}
\begin{center}
\includegraphics[width = 5cm]{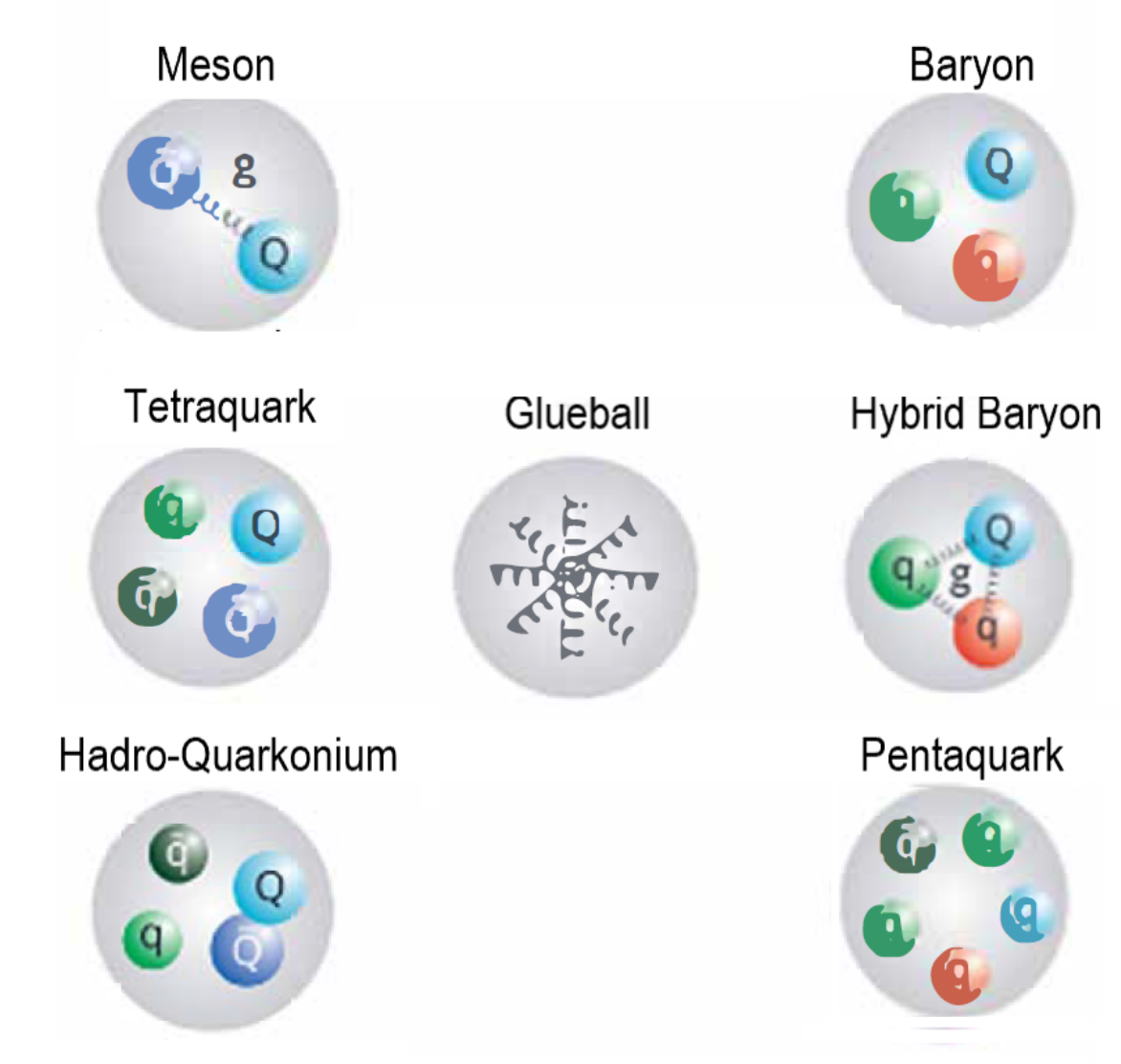}
\caption{Illustration of elementary quark-gluon configurations of mesons (left) and baryons (right). The object in the center is a glueball, \textit{i.e.}, an extremely exotic hadron consisting solely of self-interacting gluons. }
\label{fig:QGConfig}
\end{center}
\end{figure}

The, at first sight, seemingly simple picture of hadrons emerging from QCD is, in fact, deceptive. The complex field-theoretical dynamics lead to a large variety of spectroscopic configurations. The basic constituents, quarks and gluons, interact in a manifold manner, which will lead to a variety of intrinsic configurations forming the inner confined core of a hadron. A selection of expected elementary configurations is displayed in Fig.~\ref{fig:QGConfig}. The core region of a physical hadron will contain a mixture of those configurations, \textit{i.e.}, hadron wave functions are typically quantum mechanical superpositions of multi-quark, quark-gluon, gluon-gluon, \textit{etc.} components - all embedded on top of a highly dynamical vacuum. The core region, albeit confined to an incredibly small volume of about $0.5\times 
10^{-45}~\mathrm{m^3}$, is only part of the complete hadron. The quark-gluon core polarizes the vacuum and, as a consequence, is surrounded by a cloud of virtual particles, mainly consisting of pairs of pions and other mesons.

Asymptotic freedom on one side and quark confinement on the other side are defining aspects of strong interaction physics. Gluon self-interactions and effects from the spontaneously broken chiral symmetry are important ingredients for constituent quark and hadron masses. The highly non-linear nature of QCD inhibits analytical or perturbative solutions for most parts of the accessible energy region, except for the highest energies, such as those reachable at the LHC at CERN. The mathematical complexities of QCD at energies relevant to hadron structure, spectroscopy, and interactions of hadrons can only be treated numerically. LQCD is the method of choice, successfully used to explore mesons and baryons through QCD principles.


%% file: MethodsCC.tex
\section{Hadron Spectroscopy}\label{sec:HadroSP}
\subsection{QCD and Phenomenology}
\begin{figure}
\begin{center}
\includegraphics[width = 10cm]{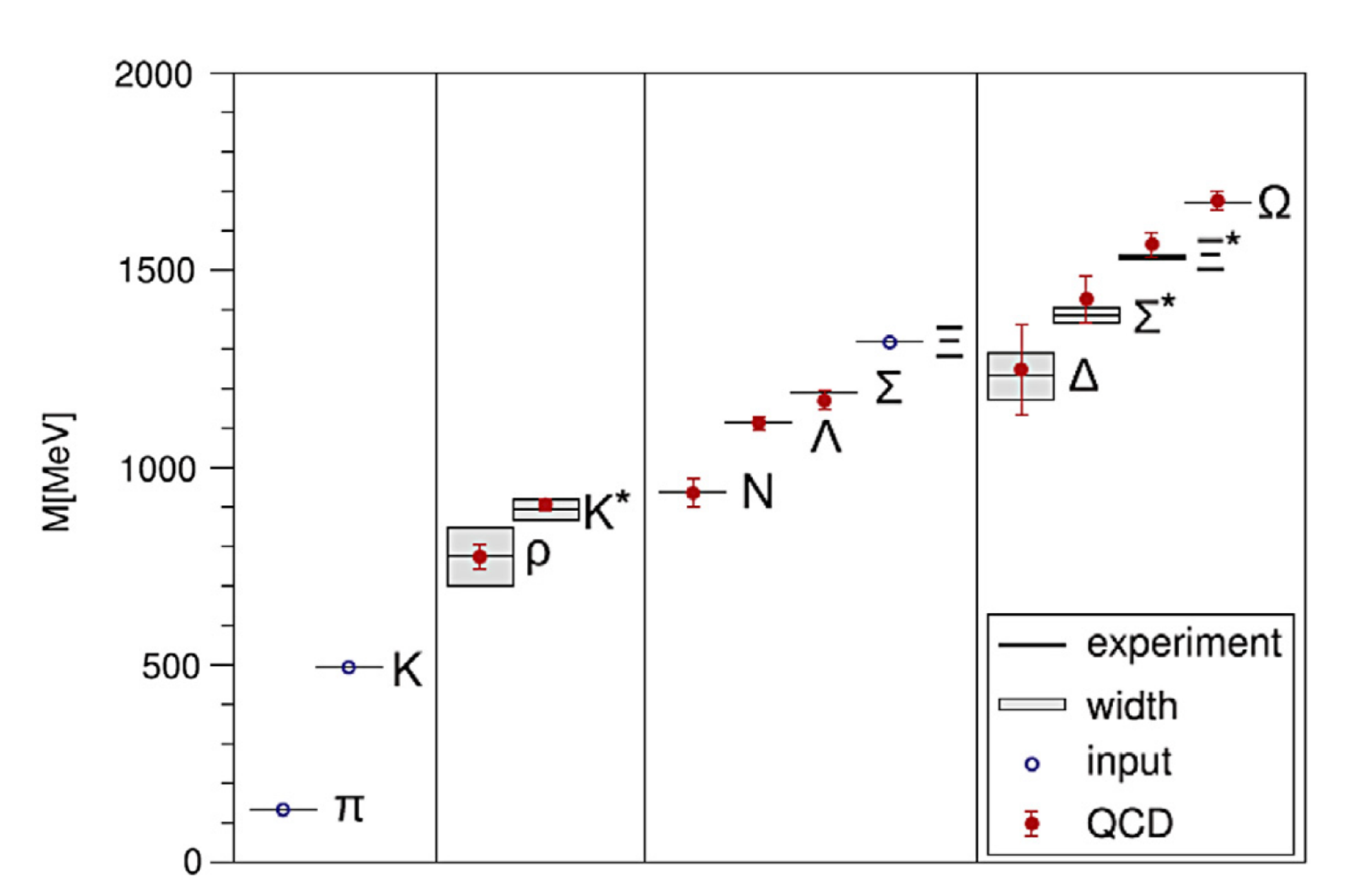}
\caption{The light hadron spectrum of QCD as obtained by QCD. Horizontal lines and bands are the experimental values with their decay widths. QCD results are shown by solid circles. Vertical error bars represent combined statistical systematic error estimates. The pion ($\pi$), $S=-1$ kaon ($K$), and the  $S=-2$ cascade baryon $\Xi$ have no error bars, because they were used to set the $u, d, s$ quark masses and the overall scale, respectively. The figure is adapted from Ref.~\cite{Lenske:2018bgr} using the LQCD results of Durr \textit{et al.}~\cite{Durr:2008zz}.}
\label{fig:Hadrons_LQCD}
\end{center}
\end{figure}

While the high-energy sector of QCD is well understood based on perturbative QCD, the low-energy limit remains a field with many open questions, reflecting the highly nontrivial task of handling a non-perturbative quantum field theory. Formally, the QCD path integral is the proper starting point. As discussed
elsewhere in this encyclopedia, two alternative numerical schemes are currently pursued: In LQCD, a set of coupled correlation functions is derived from the path integral and propagated on a Euclidean (named after Euclid (about 325 — 265 BC), an ancient Greek mathematician) space-time grid, where the numerically produced 
energy levels~\cite{LQCD:2025}. One of the first successful LQCD descriptions of the mass spectrum of the octet mesons and the octet and decuplet baryons, respectively, was achieved by the Budapest-Wuppertal Collaboration~\cite{Durr:2008zz, Fodor:2012gf}. The mass spectrum is displayed in Fig.~\ref{fig:Hadrons_LQCD}. Functional methods follow a different strategy by deriving sets of coupled Dyson-Schwinger equations (DSE) (named after Freeman John Dyson (1923-2020) and Julian Schwinger) by performing variational functional derivatives of the path integral~\cite{Eichmann:2025wgs}. 

While LQCD and DSE approaches address hadron spectroscopy from the inner sector of the quark core of hadrons, effective field theories and phenomenological reaction models focus on the properties that reflect the nature of hadrons as dynamically generated composite states. Compositeness is another aspect of the dualism hidden in baryons and mesons, as observed in the resonances in scattering and production experiments. Until today, phenomenological approaches based on an underlying covariant field theory have been an important source of spectroscopic information on the quantum numbers of resonances, their formation, and their decay. The models are well-suited for large-scale numerical coupled-channel calculations using the partial-wave formalism. Meson production on the nucleon is described by the excitation of baryon resonances in photo- and meson-induced reactions. Baryon resonances are identified through the traces they leave as complex energy poles in the partial wave scattering matrix elements, which, in many cases, are strongly affected by coupled channel effects.

At masses above the pion-nucleon threshold, baryon resonances must be regarded as superpositions of molecular-like, loosely bound, or unbound meson-nucleon configurations. They compete with the genuine QCD-type confined quark core configurations, as illustrated in Fig.~\ref{fig:QGConfig}. 
A given resonance will be composed of an arbitrary mixture of two building blocks. In some cases, the quark-type configurations will dominate; in others, the molecular configurations may prevail. An intensively and controversially debated candidate for the latter type of states is the $\Lambda^\ast(1405)$ ($S=-1$) resonance, which falls outside the systematics of quark models but is characterized by a pronounced pole structure in the complex plane. The chiral unitary model~\cite{Oller:2000fj} predicted a two-pole structure from a superposition of $\pi\Sigma$ and $\bar{K}N$ components. Years later, improved experimental and theoretical studies concluded that $\Lambda^\ast(1405)$ is dominantly a $\bar{K}N$ composite; however, they did not fully exclude $\pi\Sigma$ admixtures~\cite{Kamiya:2016jqc}.
 
\section{The Constituent Quark Model and Hadron Spectroscopy}
Once QCD  had established quarks and gluons as the constituents of hadrons, interest in nucleon spectroscopy was inevitable. However, in the early years of the QCD epoch, long before lattice QCD appeared on the horizon, the complexity of QCD dynamics did not allow for the explicit study of the non-perturbative regime of hadrons. However, in order to account for at least the most essential aspects, phenomenological models were developed. Bag models emphasize the confinement aspect; Skyrmion-inspired approaches concentrate on topological aspects, chirality, and soliton-like properties of hadrons. covariant energy density functionals, such as the still popular Nambu-Jona-Lasinio model, were derived, and the so-called sigma model, existing in linear and non-linear versions, has been extremely helpful in understanding spontaneous chiral symmetry breaking, the appearance of chiral condensates, and many other fundamental mechanisms inherent to hadrons. A detailed discussion, which goes beyond the purpose of this article, can be found in excellent topical monographs that address these topics in breadth.  

\begin{table}[htb!]

\centering \caption{
The mass spectrum of constituent quarks is used in the Constituent Quark Model. Masses are derived from the data of meson and baryon masses, such that the experimental masses are reproduced by the sum of constituent quark masses. The 
quark generations are indicated by separating lines. For further details, see 
Ref.~\cite{Entem:2025bqt}.
}

\vspace{2mm}
{%
\begin{tabular}{|c|c|}
\hline
Constituent Quark & Mass [MeV] \tabularnewline
\hline
Up (u)      & 336 \tabularnewline
Down (d)    & 340 \tabularnewline
\hline
Strange (s) & 486 \tabularnewline
Charm (c)   & 1,550 \tabularnewline
\hline
Bottom (b)  & 4,730 \tabularnewline
Top (t)     & 177,000 \tabularnewline
\hline
\end{tabular}} \label{tab:CQMMasses}
\end{table}

For spectroscopic purposes of hadrons, a model was required that could especially treat the quark degrees of freedom, because they were expected, for good reason, to prevail in the lowest baryon and meson resonances. Hence, oriented toward Gell-Mann's former quark model, attempts started to add dynamics to the group-theoretical counting schemes. This was successful once gluons and other effects, such as polarization self-energies from the Dirac sea, were eliminated by ascribing their contributions to the effective masses of quarks—a standard method of quantum many-body theory, \textit{e.g.}, applied successfully in solid state and nuclear physics,  and, as will be seen below, also in hadron spectroscopy. A working and solvable model was derived by replacing the \textit{bare} QCD  current quarks with effective \textit{constituent} quarks. Constituent quarks incorporate the contributions to hadron spectra from the unresolved sectors of QCD. They can be understood as being surrounded by a polarization cloud of virtual excitations of the neglected degrees of freedom.  A comparison of the current and constituent quark masses, Fig.~\ref{tab:quarks} and Table~\ref{tab:CQMMasses}, respectively, leads to the interesting conclusion that light quarks are affected the most by polarization effects, which may be interpreted as the polarization of the Dirac sea being important.   

A still widely used CQM is the model of
Isgur and Karl~\cite{Isgur:1978xj, Isgur:1978wd, Isgur:1978xb, Isgur:1979be, Copley:1979wj, Isgur:1980hh, Chao:1980em, Isgur:1987ht}. It was an important step forward in understanding and predicting hadron spectra on phenomenological grounds. Mesons are described by quark-antiquark ($q\bar{q}$) configurations and baryons as states of three valence quarks ($qqq$). The gluon component in hadron masses - actually accounting for a major part of the masses - is taken into account as sketched above, namely by assigning to the quarks effective constituent masses that are derived empirically from data. In Table~\ref{tab:CQMMasses}, the CQM masses up to the top quark are listed. 

The constituent quarks interact through a Coulomb-like static 
potential\footnote{Charles-Augustin de Coulomb (1736-1806) derived first the $1/r^2$ behavior of forces between electric charges, corresponding to a potential $\sim 1/r$.}, modeling gluon exchange, and are enclosed by a confinement potential, modeling color confinement. The confining potential, originally introduced in the early 1970s by Eichten \textit{et 
al.}~\cite{Eichten:1974af}, known as the Cornell potential, is typically chosen as a power law in the radial coordinate. Spin- and flavor-dependent residual interactions are introduced in order to describe the fine structure of hadron spectra. 
Originally, the Cornell model was developed for heavy quarkonium spectroscopy. However,  Godfrey and Isgur showed in the mid-1980s \cite{Godfrey:1985xj}, that the CQM was indeed able to reproduce meson spectra surprisingly well, from the low-mass \textit{u,d,s} multiplets to the region of heavy quarks.  
Modern versions of the Cornell potential and its off-springs are a standard tool for the modeling of hadrons, firmly established in spectroscopic studies of heavy mesons in the charm and bottom sectors. The CQM is discussed in breadth elsewhere in this 
Encyclopedia~\cite{Entem:2025bqt}, including also a concise derivation on QCD grounds.

The CQM plays an important role in counting, ordering, and classifying hadron states. For example, the CQM wave functions of baryons composed of three fermions are given by
\be
    |\Phi_{3q}\ran = |\psi_{color}\ran \otimes |\psi_{flavor}\ran \otimes |\psi_{spin}\ran \otimes |\psi_{space}\ran \>,
\ee
which, in total, must be antisymmetric. Color neutrality demands that the color component be in a singlet state, which is antisymmetric by definition and construction. Thus, the remaining flavor-spin-space part must be symmetric, which implies that the spin-flavor and space parts belong to the same symmetry class, \textit{i.e.}, both are symmetric, antisymmetric, or of mixed symmetry. As explained in detail by 
Giannini~\cite{Giannini:1990np} and in many textbooks, the symmetry condition imposes strict constraints on the allowed baryon configuration. In the $\{u,d,s\}$ sector, the spin and flavor degrees of freedom can be combined conveniently to form an SU(6) group. Combinatorics leads immediately to a total of 216 allowed spin-flavor configurations. The respective irreducible representations are arranged into four SU(6) multiplets, given by a 20-plet of antisymmetric states, two 70-plets of mixed symmetry, and a 56-plet of symmetric states. The 20-plet and the 70-plets include flavor singlet states, which are typically neglected due to a lack of empirical evidence. The remaining irreducible components are flavor octets and decuplets, both with total spin $S=\frac{1}{2}$ and $S=\frac{3}{2}$, respectively. Evidently, a plethora of baryon states will be obtained by combining the SU(6) spin-flavor group structures and space parts provided by the O(3) group.

\subsection{Counting Resonances}
Soon, it was realized that there was an obvious discrepancy between the number of resonances predicted by the rather successful constituent quark model and those identified experimentally. Nathan Isgur (1947-2001) was probably the first to introduce the term \textit{missing resonances}~\cite{Koniuk:1979vw}, thus honoring the mismatch between the more than 400 baryon states predicted group-theoretically by $SU(6)\times O(3)$ multiplet rulings (three 70-plets and four 56-plets)  - about the same number is obtained by LQCD - and the number of experimentally verified resonances plus proton and neutron, which amounts to a little more than 100 safely confirmed states PDG2024~\cite{PDG:2024cfk}. 
The problem of \emph{missing resonances} is a major issue in baryon spectroscopy. Final answers regarding the number of excited states of the nucleon and their spectral properties are still pending. Solutions are sought both experimentally and theoretically. 

The current understanding of the internal quark-gluon composition of a hadron is illustrated in Fig.~\ref{fig:QGConfig}. The observed states, however, are the result of additional interactions with the set of meson-baryon scattering states that possess the same total quantum numbers. Hence, the complete picture of a hadron will necessarily lead to a multi-configuration problem, superimposing the internal, confined quark-gluon components and the external meson-baryon configurations, either as a virtual polarization cloud or as on-shell continuum channels. Obviously, quark models by themselves will account for such additional dynamical effects only in a weak coupling limit, where mass shifts and other effects can be subsumed into effective, phenomenologically determined parameters.

In general, polarization effects, discussed below, induce self-energies that shift the bare QCD configurations in mass and redistribute the spectroscopic strength over the eigenstates of the coupled system. Formally, the eigenenergies of the interacting systems will be found to have moved into the complex plane, where the imaginary parts define the decay and the lifetime of the polarized states. 
Thus, baryons heavier than the nucleon-pion system acquire rather short lifetimes because they may decay either by strong (and, to a minor degree, electromagnetic) interactions or, as in the case of hyperons, by weak interactions. 

The spin-isospin multiplet of $\Delta_{33}(1232)$ states, belonging to the 
$J^P=\frac{3}{2}^+$-baryon decuplet, is a typical case: As a QCD state, the Delta-resonance is understood as a spin-isospin vector ($\Delta S=1=\Delta I$) excitation of the nucleon by reorienting the spin and isospin of at least one quark, resulting in the $(J, I)=(\frac{3}{2}\frac{3}{2})$ configuration. That particular configuration is coupled, however, strongly to the continuum of pion-nucleon $P$-wave scattering states, which induces a rapid decay within a time span of $t_{1/2}\sim 10^{-23}~\mathrm{sec}$, corresponding to a spectral distribution of full width at half maximum (FWHM) $\Gamma_\Delta\simeq 120~\mathrm{MeV}$. In contrast to other baryon resonances, the Delta-resonance is prominently excited in practically all types of reactions, from virtual and real photo-excitation by electrons and photons to neutral and charged current neutrino-nucleon and neutrino-nucleus reactions, as well as in hadron scattering on both elementary proton targets and nuclear targets. The description of those multi-channel and multi-configuration phenomena is the domain of the coupled channels methods discussed in the following sections.

\subsection{Notations}
The CQM  spectroscopic notation is $|B{} ^{2S+1}L_J\ran_{\sigma} $, where $B=N,~\Delta$ denotes an octet or a decuplet, $L=S,~P,~D\ldots$ denotes the total orbital momentum, and $|L-S|\leq J \leq L+S$ is the total angular momentum. The flavor symmetry character is specified by $\sigma = A, M, S$ for \textit{Asymmetric} (A),\textit{ Mixed} (M), and \textit{Symmetric} (S).

Traditionally, a different notation is used in meson-nucleon spectroscopy, where states are classified by $L_{2I,2J}$ according to their pion-nucleon partial wave quantum number $L=S,~P,~D\ldots$, twice the total pion-nucleon isospin $I=\frac{1}{2},\frac{3}{2}$, and twice the total spin $J$. In that notation $I=\frac{1}{2}$, nucleon-like states are denoted by $P_{11}$, $P_{13}\ldots P_{12J}$, \textit{e.g.}, the nucleon is $P_{11}(940)$  and $P_{11}(1440)$, which is the Roper resonance (discovered in 1964 by L.~David Roper (born 1935)). Thus, the resonance (pole) mass is added to distinguish the resonances within the same partial wave. 

The non-strange $I=\frac{3}{2}$ decuplet states are labeled by $\Delta_{3/2, J}$, where the best-known case is the lowest Delta-isobar resonance, $\Delta_{33}(1232)$, which was discovered by Enrico Fermi and his group in 1952.  For pion-nucleon S-wave configurations and the higher partial waves $L\geq 2$, the standard spectroscopic notation $L_{2I,2J}$ is used, combined with the resonance mass. There are ambiguities in the choice of resonance masses. From a theoretical point of view, unstable, decaying states are characterized by the spectroscopic pole in the complex plane, which includes full information on self-energies from the interactions with the surrounding continuum. However, frequently, the so-called \textit{Breit-Wigner mass} (named after Gregory Breit (1899-1981) and Eugene Paul Wigner (1902-1995)\footnote{Wigner received the Nobel Prize in Physics in 1963 for \textit{his contributions to the theory of the atomic nucleus and elementary particles, particularly through the discovery and application of fundamental symmetry principles.}} ) is used instead~\cite{willenbrock:2025briefhistorymass}.

The international Particle Data Group (PDG), founded in 1957 by Murray Gell-Mann (1929-2019) and Arthur 
Hinton Rosenfeld (1926-2017) (the latter named ``Godfather of Energy 
Efficiency'')~\cite{Gell-Mann:1957uuj}) uses a slightly different notation, as encountered in the 
latest PDG edition~\cite{PDG:2024cfk}: The proton and the neutron are distinguished by $p$, 
$I_3=+\frac{1}{2}$ and $n$, 
$I_3=-\frac{1}{2}$, thus accounting for the slight breaking of isospin symmetry in the nucleon iso-doublet, as reflected in the p-n mass difference of about $1~\mathrm{MeV}$. Higher mass members of 
the $I=\frac{1}{2}$ nucleon family are denoted by $N(M_n)J^\pi$, where $\pi=(-1)^{L+1}=\pm 1$ is 
the parity of the state, composed of the orbital angular momentum $L$ of the partial wave and the 
intrinsic (negative) parity of the pion. $M_n$ is the resonance mass, also specifying the position 
of the state in the spectral sequence.

The strangeness sector, $S=-1$ and $S=-3$  iso-singlet ($I=0$) baryons are denoted by $\Lambda(M_k)J^\pi$ and  $\Omega(M_k)J^\pi$, respectively.
Correspondingly, the iso-triplet ($I=1$) baryons of strangeness $S=-1$ and the iso-doublet ($I=\frac{1}{2}$) of strangeness $S=-2$ are classified by $\Sigma(M_k)J^\pi$ and $\Xi(M_k)J^\pi$, respectively. By historical reasons, the generic names Lambda ($\Lambda$), Sigma ($\Sigma^{0,\pm}$), Cascade ($\Xi^{0,-}$), and Omega ($\Omega^-$) are used for the lowest octet and decuplet hyperons.

For obvious reasons, most of the measurements are performed on proton (hydrogen) targets. 
Neutron data are obtained mainly from deuterium targets, which are used as a surrogate for non-existent elementary neutron targets, taking advantage of the loosely bound deuteron 
with a large distance between its constituents. The proton, as a free particle, was 
identified first by Ernest Rutherford (1871-1937) in 1920. However, already in 1908, he was 
awarded the Nobel Prize in Chemistry for \textit{his investigations into the 
disintegration of the elements and the chemistry of radioactive substances}. The 
neutron was discovered by James Chadwick (1891-1974) in 1932, who received the Nobel Prize 
in Physics in 1935 for \textit{his discovery of the neutron}. 

\input{Polarization.tex}
\section{Coupled Channels Models for Baryon Spectroscopy}\label{sec:CCModels}
\subsection{Overview}\label{sec:CCOverview}
Modeling hadron production, and especially the search for resonances on a quantitative level, is a demanding task. Several groups have developed coupled channel theories, derived sophisticated models suitable for practical work, and converted the theoretical results into numerical codes, ready for large-scale analysis of the data measured at the experimental sites and, not least, also serving to prepare for experimental campaigns. A detailed review of these ground-breaking developments is given elsewhere in this 
encyclopedia~\cite{Oller:2025leg}.
\begin{figure}
\begin{center}
\includegraphics[width = 10cm]{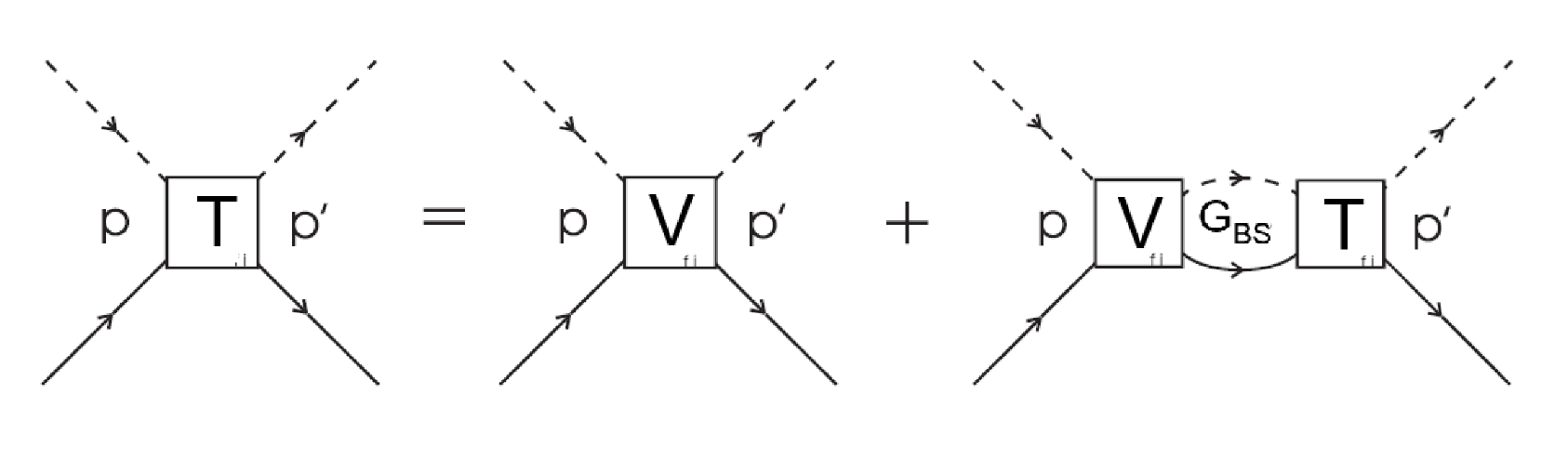}
\caption{Diagram illustrating the Bethe-Salpeter equation of the T-matrix for scattering of a hadron or photon (dashed line) on a nucleon or any other baryon (full line). The bare interactions, \textit{i.e.} the Born terms, are denoted by $V$; the scattering amplitude $T$ includes the full scattering series, summed to all orders.  The evolution of the intermediate system is described by the Bethe-Salpeter propagator $\mathcal{G}_{BS}$ (see text). }
\label{fig:BSE}
\end{center}
\end{figure}

The existing approaches utilize a covariant description of the scattering process. The major step is to construct, from an effective Lagrangian, the interaction matrix elements\footnote{Joseph-Louis Lagrange (1736-1813) developed the formalism of analytical classical mechanics and the variational methods, which, in quantized form, became the essential tools of QFT.} These serve as input to a system of coupled Bethe-Salpeter equations (named after Hans Albrecht Eduard Bethe (1906-2005) and Edwin Salpeter (1924-2008))\footnote{Hans Bethe received the Nobel Prize in Physics in 1967 for \textit{his work on the theory of stellar nucleosynthesis}}, which is illustrated in Fig.~\ref{fig:BSE}. Couplings between channels, 
\textit{e.g.}, the aforementioned case of $|KN\ran \leftrightarrow |\pi\Sigma\ran$ coupling and interference in $\Lambda(1405)$ spectroscopy, must be taken into account. In Fig.~\ref{fig:sut}, the elemental ingredients of coupled channel approaches for hadron production on the nucleon are displayed in diagrammatic form.

In short-hand notation, the BSE for the transition matrix  (T-matrix) reads:
\be\label{eq:BSE}
    T(p^\prime,p;w) = V(p^\prime,p;w)+
    \int \frac{\mathrm d^4 q}{(2\pi)^4} V(p^\prime,q;w) 
    \mathcal{G}_{BS}(q;\sqrt s) T(q,p;w)  \>,
\ee
where $V$ is the matrix containing the full set of elemental channel interactions. $w=\sqrt{s}$ is the available center-of-mass energy.
$p$ and $p'$ are the incoming (outgoing) hadron (on-shell) four-momenta.
Intermediate propagation is described by the BS propagator $\mathcal{G}_{BS}$. The integration over the energy variable $q^0$ is performed in closed form by contour integration, leading to the reduced propagator 
$G_{BS}(\mathbf{q};\sqrt{s})$. That step, in fact, requires special attention because it involves a projection to positive energy states, hence effectively eliminating the vacuum contributions. The price to pay is that interactions must implicitly account for the eliminated degrees of freedom by effective, phenomenological coupling constants. 
A reduction scheme  widely used, \textit{e.g.}, in NN scattering, is the Blankenbecler-Sugar (BbS) approach (Richard Blankenbecler (born 1933) and 
Robert Sugar~\cite{Blankenbecler:1965gx}). The BbS projection reduces the full BSE to a three-dimensional problem by preserving covariance and two- and three-body unitarity. 

The solution of a large system of coupled integral equations defined in complex algebra is a formidable numerical task, even for modern computing facilities. In order to optimize the numerical workload, the full CC problem is solved piecewise by first constructing the K-matrix and then retrieving the full T-matrix in a second step.

For that goal, the reduced propagator $G_{BS}$ is decomposed into its real and imaginary parts. Accordingly, the (reduced) BSE separable equations can be rewritten in terms of two (nested) equations.
The scattering series generated by Re$(G_{BS})$ can be summed separately, unaffected by Im$(G_{BS})$ leading to the \emph{K-matrix equation} 
\be
    K(\mathbf{p},\mathbf{p}';w) = 
    V(\mathbf{p},\mathbf{p}') + 
    \int \frac{d^3q}{(2\pi)^3} V(\mathbf{p},\mathbf{q}) \mathrm{Re} \left[G_{BS}(q;w)\right]K(\mathbf{q},\mathbf{p}';w) \>,
\ee
which accounts for all off-shell contributions. The momentum integrals must be regularized, which is accomplished by attaching to the bare interaction $V$ vertex form factors. A widely used, but not unique choice is
\be
    F(q^2) = \frac{\Lambda_q^4}{\Lambda_q^4 + (q^2 - m^2)^2} \>,
\ee
where $q^2$ denotes the four-momentum squared of the involved particle of mass $m$, which mediates the interactions.

The second step consists of reconstructing the complete scattering amplitude by solving the BSE, now given in the form
\be
    T(\mathbf{p},\mathbf{p}';w) = 
    K(\mathbf{p},\mathbf{p}';w) + i \int\frac{d^3q}{(2\pi)^3} T(\mathbf{p},\mathbf{q};w) \mathrm{Im}\left[G_{BS}(q;w)\right]K(\mathbf{q},\mathbf{p}';w) \>.
\ee
This is still a system of coupled integral equations, but it has a much simpler structure than before because the imaginary part of $G_{BS}$ is given by Dirac $\delta$-distributions in energies and/or powers of momenta, projecting the intermediate channels to their respective on-shell kinematics. 
The reduced BSE is finally solved by expanding the matrix elements into partial wave components, through which the remaining integration over the angles of the momenta in the chosen reference frame is performed analytically. 

Denoting the partial wave components by their total spin $J$, orbital angular momentum $L$, and parity $P$, a further reduced system of coupled integral equations is obtained:
\be
    T^{JLP}_{f\,i}(w)=K^{JLP}_{f\,i}(w)  
    + {\rm i}\sum_{j}\int^{\infty}_{\mu_{j0}} d\mu_j K^{JLP}_{j\, i}(\mu_j,w)A_j (\mu_j;w)T^{JLP}_{f\,j}(\mu_j;w)  \>.
\label{eq:BSE_PW}
\ee
Final, initial, and intermediate $meson-baryon$ channels are denoted by $f,i,j$, respectively. Spectral distributions of the intermediate configurations are taken into account by integrating over the respective mass distribution $A_j(\mu_j)$, thus allowing a detailed description of unstable states like $f_0$ or $\sigma(500)$, $\rho(770)\ldots$ attached to a stable baryon, or the propagation and interaction of decaying baryons such as $\Delta_{33}(1232)$. 

For stable particles of mass $m_j$, the mass distributions reduce to $A_j=\delta(\mu_j-m_j)$. 
In the other cases, the integrals are evaluated using numerical integration formulas, thus replacing the integrals with a sum over a discrete set of mesh points. Then, the integral equations reduce to a system of coupled linear equations that are solved numerically. The essence of such a treatment is that unstable states are represented by a discrete distribution of states with fractional spectroscopic strengths, as defined by the value of the spectral distribution at the mesh points and the integration weights, see 
\textit{e.g.},~\cite{Shklyar:2014kra}. Unitarity is fulfilled as long as $V$ is Hermitian. 

\subsection{Perturbative Treatments: Photon-Hadron Channels and $K$-matrix Born Approximation}\label{sec:Born}
The much weaker electromagnetic coupling constant, 
$e^2 \sim \alpha_f\simeq 1/137 \ll g^2_{\pi N}\sim 14$, allows the photo-production reaction channels to be treated perturbatively in the leading order of the $\gamma N$, $\gamma N^\ast$, and photon-hadron vertices in general. Thus, the T-matrix elements involving a photon are treated in the lowest order Born approximation, $T_{\gamma h}\to V_{\gamma h}$,  where $h$ denotes a hadron. 

A significant reduction of the numerical effort is achieved by the so-called $K$-matrix Born approximation by using $K\equiv V$~\cite{Pearce:1990uj}. This approximation corresponds to neglecting the real part of $G_{BS}$. Hence, it is, in fact, a pole approximation since the intermediate propagation is frozen to the on-shell contributions produced by the remaining imaginary part of $G_{BS}$. 

The systematic term-by-term summation of scattering series and similar perturbative series goes back to Max Born (1882-1970). Max Born shared the 1954 Nobel Prize in Physics with Walther Bothe (1891-1957) for his \textit{fundamental research in quantum mechanics, especially in the statistical interpretation of the wave function}. Moreover, Max Born's book of 1933 on \textit{Optik}~\cite{Born:1933} played a central role in the theory and experiments on gamma spectroscopy. In 1959, the book was updated, extended by contributions from other authors, and republished in English, experiencing several reprints~\cite{Born:1999ory}. 

\subsection{Interaction Potential and Scattering Matrix}
\begin{figure}
	\centering
	\includegraphics[width=10cm,clip]{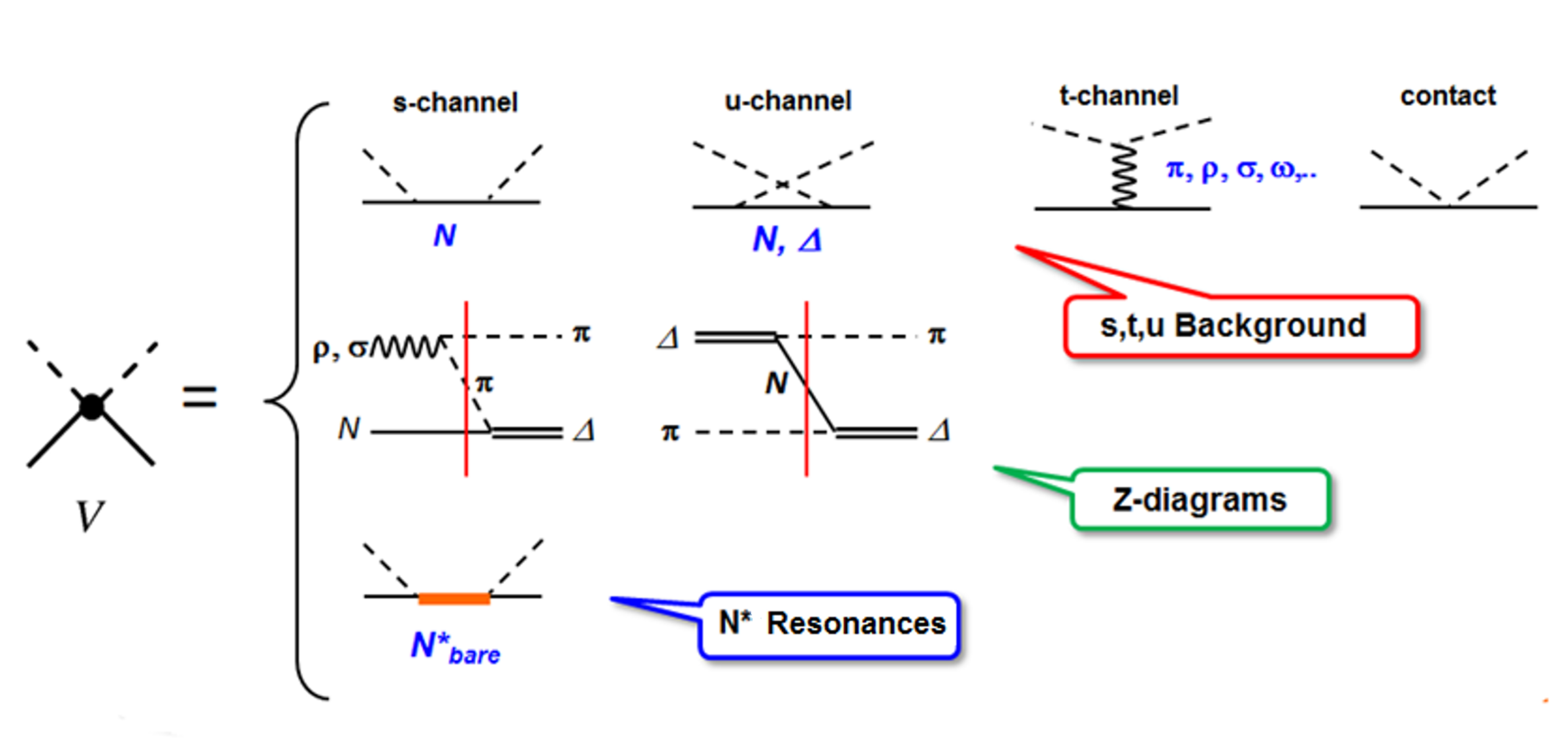}
      \caption{Structure of the tree-level interaction potential $V$. $s-$, $u-$, and $t-$~channel interactions defining the non-resonant background contributions are shown in the first line, including contact terms which are chosen such that gauge invariance is assured. The so-called \emph{z-diagrams}, displayed in the second line, are generic for the double-pion channels. $s-$~channel resonance interactions are depicted in the last line. Time is running from left to right.}
	\label{fig:sut}
\end{figure}

The hadron-hadron interaction matrix $V$ is built from a sum of $s$-, $u$-, and $t$-channel
Feynman diagrams (originally introduced by Richard Feynman for visualizing QFT/QED amplitudes) of the type shown in Fig.~\ref{fig:sut}. They are derived from the underlying effective model Lagrangians, \textit{e.g.}, see~\cite{Feuster:1997pq} for the one used in the Giessen Model (GiM).
In order to reduce the number of model parameters, the non-resonant background terms should be derived consistently from the $u$- and $t$-channel diagrams resulting from the model Lagrangian. 

In symbolic short-hand notation, the bare Born-term interaction for a reaction $a\to b$ at center-of-mass energy $w=\sqrt s$ is written as
 \be 
    V_{ab}=V^{(sut)}_{ab}+V^{(Z)}_{ab}+V^{(R)}_{ab} \>,
 \ee
where the contributions represent the first, second, and third lines of Fig.~\ref{fig:sut}, respectively, where
\be 
    V^{(R)}_{ab}=\sum_Q\frac{V_{aQ}V_{Qb}}{w-m_Q} \>.
 \ee
In each partial wave channel $[JLP]$, a  hadronic interaction matrix is found 
\be
\mathcal{V}^{[JLP]}=
\left(
  \begin{array}{ccccc}
    V_{\pi \pi} & V_{\pi\rho} & V_{\pi \sigma} & V_{\pi K} & \ldots \\
    V_{\rho \pi} & V_{\rho \rho} & V_{\rho \sigma} & V_{\rho K} & \ldots  \\
    V_{\sigma \pi} & V_{\sigma \rho} & V_{\sigma \sigma} & V_{\sigma K} & \ldots \\
    V_{K \pi} & V_{K \rho} & V_{K \sigma} & V_{K K} & \ldots \\
    \ldots & \ldots & \ldots & \ldots & \ldots \\ 
  \end{array}
\right)^{[JLP]}
\ee
where $V_{\pi \pi}\equiv  V_{\pi B \pi B^\ast}$ for baryons $B,B^\ast=N,N^\ast,\Lambda,\Sigma \ldots$, and correspondingly for the other meson channels.. 

After deriving the K-matrix, the hadronic partial wave T-matrices are defined by a system of  coupled linear equations with the formal solution
\be 
    \mathcal{T}^{[JLP]}=\left(1-i\rho_C\mathcal{K}^{[JLP]} \right)^{-1}\mathcal{K}^{[JLP]}\approx \left(1-i\rho_C\mathcal{V}^{[JLP]} \right)^{-1}\mathcal{V}^{[JLP]} \>,
\ee
where the K-matrix Born approximation is also indicated. The phase space factor $\rho_C$ is produced by evaluating the integral over the delta-distributions of $\mathrm{Im}[G_{BS}]$.

Adding the perturbatively determined photo-production T-matrix elements, the complete T-matrix is finally obtained
\be
\mathcal{T}^{[JLP]}=
\left(
  \begin{array}{cccccc}
    T_{\gamma \gamma} & T_{\gamma\pi} & T_{\gamma \rho } & T_{\gamma \sigma} & T_{\gamma K} & \ldots \\
    T_{\pi \gamma} & T_{\pi\pi}& T_{\pi\rho} &  T_{\pi \sigma} & T_{\pi K} & \ldots \\
    V_{\rho \gamma } & T_{\rho \pi} & T_{\rho \rho} & T_{\rho \sigma} & T_{\rho K} & \ldots  \\
    V_{\sigma \gamma} &T_{\sigma \pi} & T_{\sigma \rho} & T_{\sigma \sigma} & T_{\sigma K} & \ldots \\
    V_{K \gamma} &T_{K \pi} & T_{K \rho} & T_{K \sigma} & T_{K K} & \ldots \\
    \ldots & \ldots & \ldots & \ldots & \ldots & \ldots \\
  \end{array}
\right)^{[JLP]}.
\ee

\subsection{Coupled Channel Projects for Partial-Wave Analyses}
Historically, Adrien-Marie Legendre (1752-1833) was the first to find the multipole expansion of 
the gravitational two-body potential~\cite{Legendre:1782} for which he had to invent his famous system of orthogonal polynomials, since then indispensable for any partial wave expansion. Rayleigh's book on the \textit{Theory of Sound}~\cite{Rayleigh:1877} established PWA as a tool in classical wave mechanics. In quantum scattering theory, Partial Wave Analysis (PWA) is the mathematical tool used to calculate reaction amplitudes for given interactions. On the practitioner's side, PWA is the key technique for determining reaction amplitudes by fitting 
scattering data and concluding on the underlying interaction. That task is, in fact, a non-trivial mathematical problem corresponding to the solution of, in general, an ill-posed inversion.
problem, as was pointed out some time ago on quantum-mechanical grounds, \textit{e.g.}, by Andrei Nikolaevich Tikhonov~\cite{Tikhonov:1977} (1906-1993).

Over the years, in hadron physics, several CC approaches have been developed, out of which a few long-term projects have emerged. The investigations of H\"ohler's~\cite{Hoehler:1971zz} and Cutkosky's group~\cite{Cutkosky:1979fy} have the merit of establishing computational methods in hadron spectroscopy as tools for systematically enlarging and interpreting the database.  Coupled-channel (CC) approaches have proven to be an efficient tool for extracting hadron properties from experiments. They are firmly established as indispensable workhorses for research on hadron spectroscopy, playing a dominant role, especially in baryon spectroscopy.

In that tradition, the SAID project is probably the one with the largest long-term impact on the community. SAID, initiated by Richard A. Arndt, L. David Roper and their group at Virginia Polytechnic Institute and State University has since continuously fostered research at The George Washington University since 1999~\cite{SAID}, providing compilations of experimental data, constant updates of CC methods, and results. The easy online accessibility is an important service to the community and strongly supports research on hadron spectroscopy.  Unlike the other CC approaches referred to below, SAID does not assume ad-hoc resonance contributions but derives partial wave amplitudes directly from data, followed by an analysis of pole positions in the complex energy plane and Breit-Wigner parameters. The latter approach has natural limits because of: (i) revealing wide resonances up to a width $\Gamma < 500~\mathrm{MeV}$, (ii) missing transitions of small Branching Ratios $BR < 4\%$, and (iii) tending (by construction) to miss narrow resonances if $\Gamma < 20~\mathrm{MeV}$. However, in SAID, narrow resonances are accounted for by a modified PWA method~\cite{Arndt:2003ga, Arndt:2003xz}. The computational SAID approach accounts for energy dependencies of self-energies based on the Chew-Mandelstam method.

The MAID project, connected to the MAMI facility at Johannes Gutenberg University of Mainz, has been following similar routes. 
The Bonn-Gatchina (BnGa) PWA project of Bonn University and the Saint Petersburg Nuclear Physics Institute (PNPI) at Gatchina\footnote{Gatchina is a town in $40~\mathrm{km}$ south of 
Saint Petersburg and by about $700~\mathrm{km}$ far from Moscow.} Collaboration and centered at the ELSA laboratory at Bonn University is another influential CC activity~\cite{BnGa}. BnGa utilizes the so-called \textit{N/D} approach (see~\cite{Anisovich:2016vzt}) and otherwise incorporates loop corrections into the K-matrix approach in a spirit similar to SAID. The BnGa, MAID, and SAID solutions are accessible online.

The J\"ulich-Bonn-Washington Collaboration (J\"u-Bo), including J\"ulich 
Research Center, Bonn University, and The George Washington University have been formulating a coupled channels model~\cite{Huang:2011as, Ronchen:2024phm}. That approach, in fact, is based largely on the work of Haberzettl \textit{et al.}~\cite{Haberzettl:2011zr}.

Over a long period, the Giessen group at Justus Liebig-University of Giessen has been developing its own covariant CC model, GiM. Initiated in the early 1990s by Ulrich Mosel, the project has grown over a few decades into a full-fledged and versatile tool for baryon spectroscopy through meson 
photoproduction on the nucleon and subsequent spectroscopy of the decay channels. The GiM utilizes a K-matrix approach strictly oriented toward Lagrangian methods, with special attention to the consistency of interactions~\cite{Feuster:1998cj}. The theoretical background of GiM is closely oriented to the discussions in the previous paragraphs. Applications range from single pion, eta, and kaon production and investigations of vector meson production to the population of two-pion channels and Compton scattering~\cite{Cao:2017njq}\footnote{Named after Arthur Holly Compton (1892-1962), who won the 1927 Nobel Prize in Physics for \textit{his discovery of the Compton effect, which demonstrated the particle nature of electromagnetic radiation}.}.  A recent overview of the GiM is found in~\cite{Lenske:2018bgr}. 

\subsection{Quantum Interference in Hadron Spectra}
At the time of writing this article, the science and technology world is celebrating 100~years of quantum mechanics. Among others, coherence of wave functions and the interference of matrix elements are defining properties of quantum systems. While in atomic, molecular, and solid state systems, quantum interference is actively used to manipulate the systems (see, \textit{e.g.}, 
Ref.~\cite{Ott:2013Fano}), the vastly different conditions encountered in nuclear and sub-nuclear systems usually inhibit direct actions at a comparable level. Important signatures for interference phenomena are irregularities in line shapes, observed as significant deviations from Lorentz (named after Hendrik Antoon Lorentz (1853-1928)\footnote{Lorentz shared the 1902 Nobel Prize in Physics with Pieter Zeeman (1865-1943) for \textit{their discovery and theoretical explanation of the Zeeman effect} (named after Pieter Zeeman).}) (or Gaussian (named after Carl Friedrich Gauss (1777-1855)) shapes, as expected for isolated, non-interacting resonances. In nuclear spectra, exceptional line shapes have been seen in a few. 
cases~\cite{Baur:1977ayb, Orrigo:2006rd, Cavallaro:2017wju} and interpreted in the Fano formalism\footnote{Ugo Fano (1912-2001), originally a member of Fermi's team,  \textit{was a master at understanding how radiation interacts with matter. His work set the agenda for much of modern atomic physics, and had a broad sweep across the field, as stated in an obituary published in~\cite{Clark:2001Fano}.}~\cite{Fano:1961zz}, being used also in atomic and laser physics in~\cite{Ott:2013Fano}. In~\cite{Cao:2014qna, Cao:2014vca}, that formalism was applied to interpret irregular line shapes of the $D\bar D$ decay modes of those as mentioned earlier $\psi(3770)$ resonance.} 

Though the phenomenon of quantum-mechanical interference has been known for many years, there are still open questions, not only with respect to hadron physics but also in electronic systems, see, \textit{e.g.}, Ref.~\cite{Gong:2024csi}. In the review~\cite{Azimov:2009ta}, Azimov discusses how the interference of resonances may and does work. A rich source of data is on rare decay modes of well-known resonances, which demonstrate a wide variety of possible different manifestations of interference. Some special kinds of resonance interference, which have not yet been sufficiently studied and understood, are also briefly considered. The interference may provide useful experimental procedures to search for new resonances with arbitrary quantum numbers, including exotic ones, and to investigate their properties.

\begin{figure}[htb!]
\centering
{
    \includegraphics[width=0.5\textwidth,keepaspectratio]{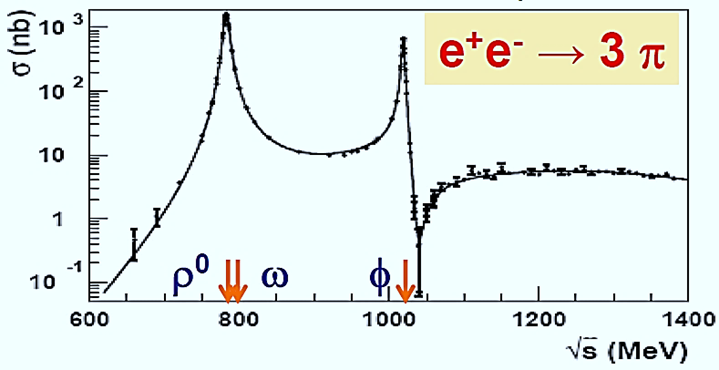} 
}

\centerline{\parbox{\textwidth}{
\caption[] {\protect\small
The $e^+e^- \to \pi^+\pi^-\pi^0$ cross section measured by the SND Collaboration and collected in~\cite{Achasov:2006xb}.
The curve is the fit with the $\rho^0$, $\omega$, $\phi$, $\omega^\prime$, and $\omega^{\prime\prime}$ resonances.
Red arrows shown $\rho^0$, $\omega$, and $\phi$ thresholds.}
\label{fig:int} } }
\end{figure}

In Figure~\ref{fig:int} resonance interference is illustrated for $\rho^0$, $\omega$-, and $\phi$-meson production in the reaction $e^+e^- \to \pi^+\pi^-\pi^0$. Total cross sections collected by the SND Collaboration are shown. For the $3\pi$ decay case, $\Gamma(\rho^0\to 3\pi) = 0.015~\mathrm{MeV}$ (Isospin symmetry violation), $\Gamma(\omega\to 3\pi)  = 7.58~\mathrm{MeV}$, and  $\Gamma(\phi \to 3\pi)  = 0.65~\mathrm{MeV}$ (Zweig rule violation)~\cite{PDG:2024cfk}. 
It has a clear BW-like peak in the ($\rho^0$, $\omega$) region and a bump–dip structure in the 
$\phi$ region. The background near $\phi$ changes slowly compared to the nearly standard interference curve. 
Instead of the $\phi$–peak, there are both bumps and dips, each having a form different from BW; 
max/min shifts from $\phi$-mass. There is a similar shift for the $\rho$-mass. Then, the $\rho$
contribution here deforms the $\omega$-tail.

\begin{figure}[htb!]
\centering
{
    \includegraphics[width=0.45\textwidth,keepaspectratio]{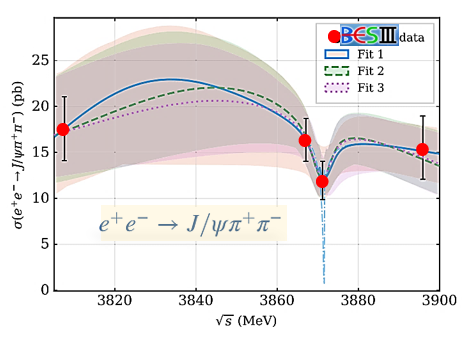} 
}

\centerline{\parbox{\textwidth}{
\caption[] {\protect\small
The line shapes for the three best fits~\cite{Baru:2024ptl} to the
BESIII data~\cite{BESIII:2022ner} for the reaction $e^+e^- \to J/\psi \pi^+\pi^-$ after convolution with the energy spread function. As an example, the blue dashed line shows the line shape for fit~1 without the effect of the energy spread. The $1\sigma$ error bands
correspond to the uncertainty propagated from the data.}
\label{fig:int1} } }
\end{figure}

The recent BESIII  cross section data for the reaction $e^+e^- \to J/\psi \pi^+\pi^-$~\cite{BESIII:2022ner} have been interpreted as an observation of the $X(3872)$ resonance~\cite{Baru:2024ptl} via a destructive interference of a sharp resonance and large background contributions (this is a natural way for the inelastic reaction, see, for instance,~\cite{Strakovsky:2023kqu}) (Fig.~\ref{fig:int1}). The large non-resonance contribution magnifies a small resonance effect, as described in different contexts by the Fano mechanism \cite{Cavallaro:2017wju}. The observed dip is small, but visible, and more statistics would be highly desirable.

The data displayed in Figs.~\ref{fig:int} and \ref{fig:int1} demonstrate the direct interference between two resonances decaying into the same final states\footnote{A while ago, Richard Feynman said: \textit{When looking at Maxwell equations, it is hard to imagine how beautiful the rainbow is.}. Then Yakov Azimov 
(1938-2016) extended it: \textit{Everybody knows that the interference does exist. But it is not always easy to imagine how it will work in a particular case.}}.


%% file: Polarization.tex
\section{Coupled Channels Methods to Meson Production and Hadron Spectroscopy}\label{sec:CCSpec}
\subsection{Interacting Hadrons}\label{ssec:OpenClose}
The quark-gluon configurations displayed in Fig.~\ref{fig:QGConfig} are the building blocks of the QCD core and could be taken as a basis for extended studies. The core eigenstates are obtained by diagonalization within the quark-gluon configuration space, which is demanding but not prohibitive.  The result is a spectrum of energy-sharp, discrete eigenstates due to confinement. CQM and bag models of hadron structure are coming closest to that kind of picture. However, for describing physical hadrons realistically, it must be taken into account that these model states are embedded in the continuum of hadronic scattering states\footnote{In molecular, atomic, and nuclear physics, such states are known as \emph{bound states in the continuum} (BIC).}.

As theoretical constructs, the bare quark-gluon states correspond to closed channels with inseparable constituents under normal environmental conditions.
The (multi-)hadron-hadron scattering states are open channels with asymptotically free particles. Depending on the case, they are composed either of pure mesons and baryon-antibaryon pairs for $B=0$ systems or of mixed meson-baryon scattering states in $B\neq 0$ objects. If there would be 6-quark objects, \textit{e.g.}, the notorious and experimentally still searched for $S=-2$ H-dibaryon hypothesized by 
Jaffe~\cite{Jaffe:1976yi}, baryon-baryon channels would be involved as well. The 
observation of the non-strange $d^\ast(2380)$ dibaryon, probably a double-$\Delta^0$ 
configuration, was reported by the WASA-COSY collaboration~\cite{WASA-at-COSY:2011bjg}. Theoretical studies find rather compact d-baryon wave 
functions~\cite{Farrar:2023wvm}.  In either case, the set of discrete channel quantum 
numbers, defined by the coupled irreducible representations of the hadronic 
constituents, must coincide with those of the core\footnote{It's worth 
noting that the best known and stable dibaryon is the deuteron, although it is not of the 
compact stature expected for exotic states.}.
Functional methods and LQCD will, in principle, account for such effects, as the good reproduction of the experimental mass spectrum in Fig.~\ref{fig:Hadrons_LQCD} shows. 

The basic mechanisms governing the interplay of quark-gluon and hadron-hadron components of hadron wave functions are easily understood in the Hamiltonian formulation (named after William Rowan Hamilton (1805-1865)). The description of physical hadrons demands solving a coupled-channel problem. 

\subsection{Interactions of QCD Core and Hadron Scattering Configurations}
Consider a QCD core state $|Q\ran$ of bare mass $M_Q$ which is coupled by the interactions $V_{QC}=V^\dag_{CQ}$ to a set of open hadron scattering channels $C$ of total channel mass $M_C=\sum_{h\in C}m_h$, given by the sum over the masses $m_h$ of all asymptotically free hadrons $h=p,n,\pi\ldots$. Standard projection methods allow us to recast the set of coupled equations by a fully equivalent dispersive, non-Hermitian polarization self-energy operator\footnote{An operator $A$ is Hermitian if $A=A^\dag=(A^T)^\ast=(A^\ast)^T$ does not change under transposition ($T$) and complex conjugation ($\ast$). A real number $x$ is Hermitian, a complex number $z=x+iy$ is non-Hermitian. Any non-Hermitian operator can be decomposed into a Hermitian and an anti-Hermitian operator (named after Charles Hermite (1822-1901)).}  acting on the quark-gluon core:
\be
    \Sigma_Q(w_\alpha)=
    \sum_C\int \frac{d^3q}{(2\pi)^3}\frac{|V_{QC}(\mathbf{q})|^2}{w_\alpha-w_C(q)+i\eta} \>,
\ee
where $\eta\to 0+$ and $w_\alpha=\sqrt{s_\alpha}$ denote the invariant energy brought in by an external probe $\alpha$. 

By means of the Cauchy formula (named after Augustin-Louis Cauchy (1789-1857)), the propagator is decomposed into a principal value real part and an imaginary part given by a Dirac $\delta$-distribution. Following a widely used notation, the self-energy operator is written as
\be
    \Sigma_Q(w_\alpha) = \Delta_Q(w_\alpha)-i\frac{1}{2}\Gamma_Q(w_\alpha) \>.
\ee
The real part is
\be\label{eq:Shift}
    \Delta_Q(w_\alpha)=
    \sum_C\int \mathcal{P}\frac{d^3q}{(2\pi)^3}\frac{|V_{QC}(\mathbf{q})|^2}{w_\alpha-w_C(q)} \>
\ee
to be evaluated as a Cauchy Principal Value $\mathcal{P}$ (named after Augustin-Louis Cauchy). The imaginary part, expressed by the width
\be\label{eq:Width}
    \Gamma_Q(w_\alpha)=2 \sum_C\Theta(w_\alpha-M_C)\rho_C(q_{\alpha C})\lan|V_{CQ}(q_{\alpha C})|^2\ran \>
\ee
is determined by the sum of the residues at the poles of the propagator appearing at the equivalent on-shell energy of the intermediate channels. For two-body channels, $M_C=M+m$, the invariant on-shell three-momenta are defined by the positive root of
\be
    q^2_{\alpha C}=\frac{1}{4w^2_\alpha}(w^2_\alpha-(M+m)^2)(w^2_\alpha-(M-m)^2)\geq 0 \>,
\ee  
where, different from the real part, only channel states with $M+m < w_\alpha$ contribute. The momentum integration has led to the  kinematical (phase space) factor
\be
    \rho_C(q_{\alpha C})=\frac{q_{\alpha C}E_{M}(q_{\alpha C})E_{m}(q_{\alpha C})}{\pi w_\alpha} \>,
\ee
where $E_m(q)=\sqrt{q^2+m^2}$.
The integration over the orientations of the intermediate momentum coupling results in the angle-averaged matrix element
\be
    \lan|V_{CQ}(q_{\alpha C})|^2\ran=\frac{1}{4\pi}\int d\hat{\mathbf{q}}|V_{CQ}(\mathbf{q})|^2_{|q=q_{\alpha C}} \>.
\ee
Including the self-energy, \textit{i.e.}, the coupling to the open hadron channels, the bare (model) states $|Q\ran$ are transformed into the polarized - in this model the physical - resonances $|R\ran$. They are determined by a generalized, dispersion eigenvalue problem. The spectrum of resonances is defined by the (complex) roots $\mathcal{M}_R$ of the equation.  
\be\label{eq:EigenV}
    M_Q + \Delta_Q(\mathcal{M}_R)-i\frac{1}{2}\Gamma_Q(\mathcal{M}_R) - \mathcal{M}_R = 0 \>.
\ee
The solutions are complex eigenvalues $\mathcal{M}_R=M_R-\frac{i}{2}\Gamma_R$, implying a finite lifetime $t_{1/2}=\hbar/\Gamma_R$ (using $c=1$) and a mass shift $\Delta_R=M_R-M_Q$. 

A closer inspection of the dispersion equations reveals that if the QCD configuration $|Q\ran$ interacts with the $N^{(Q)}_C$ channel states $|C\ran$,  Eq.~(\ref{eq:EigenV}) has a total of $N_R=N^{(Q)}_C+1$ solutions of discrete complex eigenvalues $\mathcal{M}_{R_n}$. Hence, for each fundamental QCD configuration, the channel coupling will produce a spectrum of $N^{(Q)}_C$ satellite states $|R_n\ran$, $n\leq N^{(Q)}_C$ plus a solution corresponding to the unperturbed state $|Q\ran$ of mass eigenvalue $\mathcal{M}_{R_n}=M_{R_n}-\frac{i}{2}\Gamma_{R_n}$. The spectroscopic strength of $|Q\ran$ is distributed over the spectrum of core-like eigenstates. Typically, but in detail depending on the channel spectrum and the interactions $V_{CQ}$, the state of energy closest to $M_Q$ will acquire the largest spectroscopic factor.

To maintain analyticity, theoretical approaches should account for the energy dependence of all 
parts of the self-energies. However, for obvious reasons, little to nothing is known about the 
intrinsic structures of self-energies because their full knowledge is equivalent to the exact 
solution of the complete CC problem. In practice, typically the width, \textit{i.e.}, the imaginary 
part of the channel self-energy is approximated by functional forms of the proper behavior at the 
threshold. Close to the threshold of a partial wave with orbital angular momentum $\ell$ and channel 
momentum $\mathbf{q}$, the coupling matrix elements behave as $V_{CQ}(|\mathbf{q}|)\sim q^\ell$, 
implying a width of $\Gamma_{CQ}\sim q^{2\ell+1}$. Widely used parameterizations are the so-called 
Blatt-Weisskopf form factors (named after John Markus Blatt (1921-1990) and Victor Frederick 
Weisskopf (1908-2002) $F_\ell(q/q_0)$~\cite{Blatt:1952ije}. These form factors account for 
the centrifugal barrier effects, which determine the behavior of an outgoing scattering wave at the 
interaction radius. They are given by spherical Hankel functions (named after Hermann Hankel (1839-
1873). In~\cite{VonHippel:1972fg}, they were generalized to hadron resonance scattering. Explicit 
formulas and practical applications are discussed, \textit{e.g.}, in Ref.~\cite{PDG:2024cfk}.

Once the imaginary part of a self-energy is fixed, the real part is consistently obtained through dispersion theory. The Chew-Mandelstam method (named after Geoffrey Foucar Chew (1924-2019) and Stanley Mandelstam (1928-2016)) accomplishes that goal by evaluating a once-subtraction dispersion integral, where the widths are parameterized by Blatt-Weisskopf form factors. In the SAID approach, these effects are incorporated~\cite{Workman:2012hx}. 

\subsection{Polarization Self-energies in Hadron Scattering Amplitudes}
For understanding the core-channel interplay in hadron production processes,  the channel self-energies induced by the core states need to be derived:
\be \label{eq:SigmaC}
    \Sigma_C(\w)=\sum_Q V_{CQ}\frac{1}{\w - M_Q}V_{QC}=\sum_Q x_{CQ}\frac{\Gamma_Q}{\w - M_Q} \>,
\ee
where $\Gamma_Q=\sum_C|\rho_C\lan V_{CQ}\ran|^2$ and $x_{CQ}=\rho_C|\lan V_{CQ} \ran|^2/\Gamma_{Q}$ are the branching ratios, to be evaluated at the appropriate channel momenta. As a consequence, the scattering amplitude of channel $C$ becomes the sum of the scattering amplitudes from the generic channel interactions and the amplitude produced by $\Sigma_C$. The latter gives rise to resonances
that interfere with the weakly energy-dependent in-channel amplitudes. As a function of energy, the respective partial wave cross sections will show peak structures on top of a smooth background. Occasionally, the resonances are accompanied by an interference pattern, which may lead to irregular line shapes. 

The scattering channels will also interact among themselves. In combination with kinematical and structural conditions, \textit{e.g.}, openings of particle emission thresholds or energy-dependent accidental cancellations or enhancements among resonance and background components, the channel-channel interaction amplitudes may produce resonance-like structures in the partial wave cross section which, however, being of purely dynamical origin, will fall outside the systematics expected by symmetries and the related selection rules. Care must be taken to identify, isolate, and separate model-dependent effects, which are achieved by comparing independently derived results from different approaches. 
Careful studies of the evolution of the amplitudes in the complex plane by Argand diagrams (named after Jean-Robert Argand (1768-1822))~\cite{Argand:1806} will reveal the nature of the structure. Dalitz plots, introduced in 1953 by Richard Henry Dalitz (1925-2006) while studying kaon decay~\cite{Dalitz:1953cp}, are indispensable tools for investigations of three-body reaction channels by revealing correlations between decay products. 

\subsection{Channel Coupling in Meson-Meson Scattering}
The above formalism is a general scheme, applicable to any interacting quantum system at any scale. The same types of basic mechanisms are acting in baryon-baryon, meson-baryon, and meson-meson systems. Prominent cases include the formation of pion-nucleon resonances like the first excited state of the nucleon represented by the well-known Delta-resonance $P_{33}(1232)$ of width $\Gamma_{\Delta}=120~\mathrm{MeV}$, corresponding to a lifetime of about $t_{1/2}\sim 10^{-23}$~sec. The widths and finite lifetimes of the members of the $J^P=1^-$ vector meson octet indicate the same trend. The isovector ($I=1$) $\rho(770)$ meson, for example, is embedded in the $\pi\pi$ $P$-wave continuum, resulting in a decay width of $\Gamma_{\rho}\sim 150~\mathrm{MeV}$ with a two-pion branching ratio of 99.9\%. Isospin selection rules affect the rho-meson by excluding the $\rho^0 \to \pi^0\pi^0$ decay channel.

Occasionally, the $\rho$ meson is considered to be the gauge boson of a broken \emph{hidden local chiral symmetry} which, however, is distinct from the broken \emph{global chiral symmetry} of which the pion is the respective Goldstone (named after Jeffrey Goldstone (born 1933)) boson (named after Satyendra Nath Bose)~\cite{Georgi:1990chi}. Moreover, the non-strange vector mesons $\rho,\omega,\phi$ play a central role in Sakurai's Vector Meson Dominance (VMD) model (named after Jun John Sakurai (1933-1982))~\cite{Sakurai:1960ju} where they account for the (virtually admixed) hadronic part of the photon, through which the photon couples to hadronic matter.

The $\rho$- and $\omega$-mesons were first detected in 1961 at LBL~\cite{Maglich:1961rtx}. 
Remarkably, the $\omega(782)$ meson, being the isoscalar partner of the $\rho$-meson, is a rather long-lived object with a small width of about $\Gamma_{\omega}\sim 8~\mathrm{MeV}$. The increase in lifetime/decrease in decay width by about a factor of 20 is an effect of isospin conservation.
Inhibiting the decay of an isoscalar vector meson into two pions. An even more extreme case is the $\phi(1020)$ meson, which is the singlet partner of the $\w$-meson. Although located in a mass region with many open channels, a width as small as $\Gamma_\phi=4.249\pm 0.013~\mathrm{MeV}$~\cite{PDG:2024cfk} is observed. The long life-time of the $\phi$-meson is caused by the dominant $s\bar s$ structure which requires the creation of $u\bar u$ or $d\bar d$ pairs out of the vacuum before a decay into lighter mesons, \textit{e.g.}, the prevailing $\phi \to K\bar K$ process, can occur, as expressed in the OZI rule which is also responsible for extraordinarily small widths of $c\bar{c}$ states like $J/\psi$. 

Since currently dedicated meson production facilities are not available, $N \pi \to N^\prime 2\pi$ and $N \gamma \to N^\prime 2\pi$ reactions are used to investigate the otherwise inaccessible $\pi\Delta(1232)$ and $\rho(770) N$ final states, also aiming to separate these final states kinematically (see, for instance, Ref.~\cite{Sarantsev:2025lik}). Such indirect methods are also applied to extract hadron-hadron cross sections from hadron production data in high-energy nucleon-nucleus and nucleus-nucleus collisions at the operating hadron facilities.  


%% file: PionPhotoproduction.tex
\section{Single Meson Photoproduction}\label{sec:PiPhoto}
\subsection{Prelude}
The determination of the resonance properties for all accessible baryon states is a central objective of hadron and nuclear physics. The extracted resonance parameters provide a crucial body of data for understanding the dynamics of nucleon excitations and the 
spectral distributions, essential for testing phenomenological models of the nucleon, calibrating lattice QCD calculations as a way of connecting non-perturbative hadron physics to QCD. The spectrum of $N^\ast$ and $\Delta^\ast$ non-strange baryon resonances with masses up to about $2~\mathrm{GeV}$ is probably the best-studied sector of hadron physics. Meson-nucleon
scattering and photoproduction of mesons on nucleon and nuclear targets have led to a wealth of data, which has also been supplemented lately by combined systematic studies of kaon and associated hyperon production experiments. Properties of the known resonances continue to become better determined as experiments involving polarized beams, targets, and recoil measurements are expanded and refined. With increasing energy, multi-meson channels will open. They provide the opportunity to identify states that are weakly coupled to two-body channels, such as $\pi +N$. Since production reactions involving the octet-neglecting the $\eta'$ singlet of pseudo-scalar mesons and the related baryon resonances are studied the best, they will be the focus of the following discussions. To a lesser degree, production reactions leading to the nonet of vector mesons have also been investigated and will be addressed selectively. Thus, in the following sections, reactions involving the baryon octet and the decuplet, Fig.~\ref{fig:Multiplets} and two of the three meson nonets of Fig.~\ref{fig:Nonets} will be discussed. 

Experimentally, light scalar mesons typically suffer from short lifetimes and ill-defined spectral distributions; see,\textit{e.g.}, the experimental status of the $f_0(500)/\sigma$  meson~\cite{PDG:2024cfk}. However, after decades of research, the properties of the lowest scalar meson are encapsulated as discussed comprehensively in~\cite{Pelaez:2015qba} and reviewed in detail in a separate article of the encyclopedia~\cite{Pelaez:2025wma}. Diffuse objects such as $f_0(500)$ remain a challenge for coupled channel approaches because, in principle, they require solving at least a numerically very involved coupled three-body problem in order to properly account for the cross-talk between the various interacting many-particle channels.
A brief account of sigma meson production on the nucleon is given in one of the following sections in the context of double pion production.

\subsection{Single Pion Photoproduction}
\subsubsection{Pion Photoproduction on the Proton}
Single pseudoscalar meson photoproduction involves the interaction of a photon with a free proton, a bound neutron, or an entire nucleus. The growth of the database on photoproduction of $\pi^{0,+}$ meson production on the proton is depicted in Fig.~\ref{fig:ProtonD}. For studies of the baryon spectrum, one is normally interested in the first two of these. Thus, a spin-1 particle (the photon, two helicity states) and a spin-$\frac{1}{2}$ particle (the nucleon) react to produce a spin-0 particle (the pseudoscalar meson) and a spin-$\frac{1}{2}$ particle (the recoiling baryon). This yields eight spin combinations, of which four are possible within the parity-conserving strong interaction that has occurred. EM interaction does not conserve isospin, so multipole amplitudes contain isoscalar and isovector contributions of the EM current.

The four combinations are represented as amplitudes (Eqs.~(\ref{eq:eq1}) and (\ref{eq:eq2})), the exact form of which is a matter of choice. 
A theory of pion photoproduction was constructed in the 1950s. Kroll and Ruderman~\cite{Kroll:1953vq} were the first to derive model-independent predictions in the threshold region, a so-called
low energy theorem (LET), by applying gauge and Lorentz invariance to the reaction $\gamma N\to\pi N$. The general formalism for this process was developed by Chew, Goldberger, Low, and Nambu (Geoffrey Foucar Chew (1924-2018), Marvin Leonard ``Murph'' Goldberger (1922-2014), Francis Eugene Low (1921-2007), and Yoichiro Nambu (1921-2015)\footnote{Nambu was awarded half of the Nobel Prize in Physics in 2008 for \textit{the discovery in 1960 of the mechanism of spontaneous broken symmetry in subatomic physics, related at first to the strong interaction's chiral symmetry and later to the electroweak interaction and Higgs mechanism}.}). The results are known as \textit{CGLN amplitudes}~\cite{Chew:1957tf} and a few years later helicity amplitudes~\cite{Jacob:1959at} were derived. About a decade later, Berends, Donnachie, and Weaver analyzed the existing data in terms of a multipole decomposition and extracted the various multipole amplitudes contributing in a region up to an excitation energy of $500~\mathrm{MeV}$~\cite{Berends:1967vi}. These amplitudes are vital inputs to low-energy descriptions of hadron physics based on chiral perturbation theory (ChPT)~\cite{Hilt:2013fda}.  Within any of these bases, 16 possible bilinear combinations are referred to as the ``observables.''

For the proton
\begin{equation}
    A(\gamma p\to\pi^0p) = A^{(0)} + \frac{1}{3}A^{(1/2)} + \frac{2}{3}A^{(3/2)}~~~\mathrm{and}~~~
    A(\gamma p\to\pi^+n) = \sqrt{2}\biggl(A^{(0)} + \frac{1}{3}A^{(1/2)} - \frac{1}{3}A^{(3/2)}\biggl)
    \>.
\label{eq:eq1}
\end{equation}
For the neutron
\begin{equation}
    A(\gamma n\to\pi^0n) = \sqrt{2}\biggl(-A^{(0)} + \frac{1}{3}A^{(1/2)} + \frac{2}{3}A^{(3/2)}\biggl)~~~\mathrm{and}~~~
    A(\gamma n\to\pi^0n) = -A^{(0)} + \frac{1}{3}A^{(1/2)} + \frac{2}{3}A^{(3/2)}
    \>.
\label{eq:eq2}
\end{equation}
Proton data alone do not allow for the separation of isoscalar and isovector components~\cite{Drechsel:1992pn}.
Accurate evaluation of EM couplings $N^\ast \to \gamma N$ and $\Delta \to \gamma N$ from meson photoproduction data remains a paramount task in hadron physics. Only with good data on both proton and neutron targets can one hope to disentangle isoscalar and isovector EM couplings of various  $N^\ast$ and $\Delta^\ast$ resonances~\cite{Watson:1954uc, Walker:1968xu}, as well as the isospin properties of non-resonant background amplitudes.

The lack of $\gamma n\to\pi^-p$ and $\gamma n\to\pi^0n$ data on free, isolated neutrons\footnote{Apart from lower-energy ($< 700~\mathrm{MeV}$), there is data for the inverse $\pi^-$ photoproduction reaction, 
$\pi^-p\to \gamma n$~\cite{SAID}. This process is free from complications associated with a deuteron target. 
However, there is a major disadvantage to using $\pi^-p\to\gamma n$: there is a large background 
from $\pi^- p \to \pi^0 n \to \gamma\gamma n$, whose cross section is 5 to 500 times larger 
than $\pi^- p \to \gamma n$, } which does not allow us to be as confident about the determination of neutron couplings relative to those of protons.

Since the neutron targets do not exist, it remains necessary to use nuclear ones. In this case, when extracting information on the elementary reaction involving the bound neutron from nuclear data, one should take into account the nuclear-medium effects, \textit{i.e.}, the final-state interaction (FSI) and Fermi-motion effects~\cite{Migdal:1955ab, Watson:1952ji}\footnote{It is impossible to measure FSI experimentally~\cite{Tarasov:2015sta}. Obviously, in the case of polarized measurements, the FSI corrections are small or consistent with experimental uncertainties.} 

\begin{figure}[htb!]
\centering
{
    \includegraphics[width=0.45\textwidth,keepaspectratio]{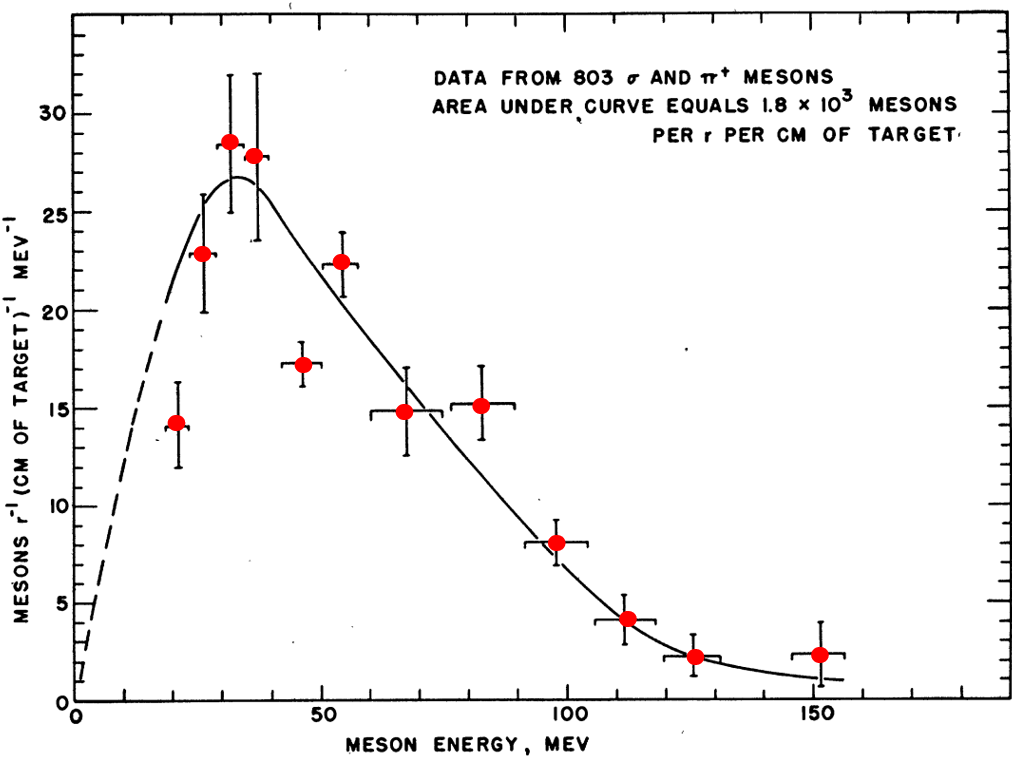} 
}

\centerline{\parbox{\textwidth}{
\caption[] {\protect\small
Distribution of positive pion energies from photon energy of $335~\mathrm{MeV}$~\cite{McMillan:1949em}. The apparent lower limit on the energy is caused by the fact that the energies are computed as if the mesons originated in the center of the carbon block. The dashed line is simply a guess as to the trend of the distribution at low energies, which was used in the integration leading to the total cross section.
}
\label{fig:pion} } }
\end{figure}

\begin{figure}[htb!]
\centering
{
    \includegraphics[width=0.65\textwidth,keepaspectratio]{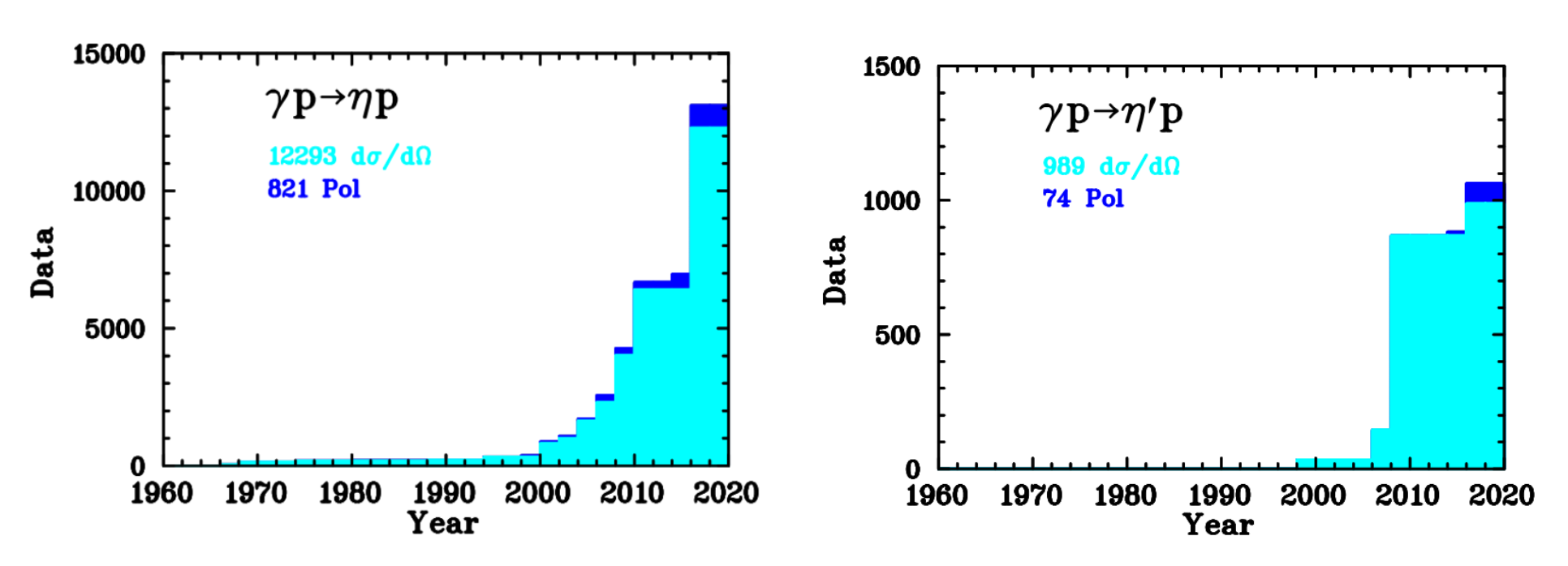}
}

\centerline{\parbox{\textwidth}{
\caption[] {\protect\small
Database for $\gamma p\to\pi^0 p$ (left) and $\gamma p\to\pi^+ n$ (right). Experimental data from the SAID database~\cite{SAID} selected for 1996 through 2018. Right: Amount of data as a
function of time. Full SAID database. The data is shown as a stacked histogram. Light shaded – cross sections, dark shaded – polarization data. The figure is
        adapted from Ref.~\cite{Ireland:2019uwn}.
}
\label{fig:ProtonD} } }
\end{figure}

Measurements of pion photoproduction on both proton and quasi-free neutron targets have a very long 
history, starting about 70~years ago with the discovery of the pion by the University of Bristol 
group~\cite{Lattes:1947mx}. Two years later, at the 1949 Spring Meeting of the US National Academy 
of Sciences, a preliminary account was given of some observations of mesons produced by the 
$335~\mathrm{MeV}$ photon beam from the Berkeley 
synchrotron~\cite{McMillan:1949em}\footnote{Edwin Mattison McMillan (1907-1991) 
shared the 1951 Nobel Prize in Chemistry with Glenn Theodore Seaborg (1912-1999), \textit{credited 
with being the first to produce a transuranium element, neptunium}. In addition, McMillan co-
invented the synchrotron with Vladimir Iosifovich Veksler (1907-1966).}  
Starting with the use of bremsstrahlung facilities, pioneering results for $\gamma p\to\pi^+ n$ 
(Fig.~\ref{fig:pion})~\cite{McMillan:1949em} and for 
$\gamma p\to\pi^0 p$~\cite{Steinberger:1950equ} were obtained. 
One can possibly understand that the pion family is a triplet. Finding a neutral pion is much more 
difficult since it does not leave marks in photo-emulsion or Wilson chambers. 

In addition to the $\gamma p\to n \pi^+$ reaction on the free proton, measurements were performed at Berkeley on the loosely bound, quasi-free neutron in deuterium. Using a deuterium target, $\gamma n\to p \pi^-$ reactions were studied at the photon energy $318\pm 10~\mathrm{MeV}$~\cite{White:1952zz}. Hence, within three years, positive and negative pion photoproduction experiments were accomplished.

Despite all the shortcomings of the first measurements (such as large normalization uncertainties, 
wide energy and angular binning, limited angular coverage, and so on), these data were crucial 
for the discovery of the first excited $\Delta$ and nucleon states using PWA for $\pi N$ elastic 
scattering data. The $\Delta(1232)3/2^+$ was determined by Enrico Fermi's 
group~\cite{Anderson:1952nw}. While the second one $N(1440)1/2^+$, called Roper resonance, or colloquially simply ``Roper'', in honor of 
L. David Roper (born 1935), who discovered this baryonic state, came several 
years after the $\Delta$-isobar~\cite{Roper:1964zza}\footnote{As an anecdote, it is worth 
mentioning that Lev Davidovich Landau, when becoming aware of Fermi's discovery of the Delta 
resonance, did not believe in the existence of such a state, arguing \textit{A width of 
$120~\mathrm{MeV}$ - what is that? The pion will make a quarter of the circle around the nucleon, 
and that is supposed to be a pion-nucleon bound state?} The concept of a meson-nucleon resonance 
was waiting to be established $\ldots$.} 

\begin{figure}[htb!]
\centering
{
    \includegraphics[width=0.65\textwidth,keepaspectratio]{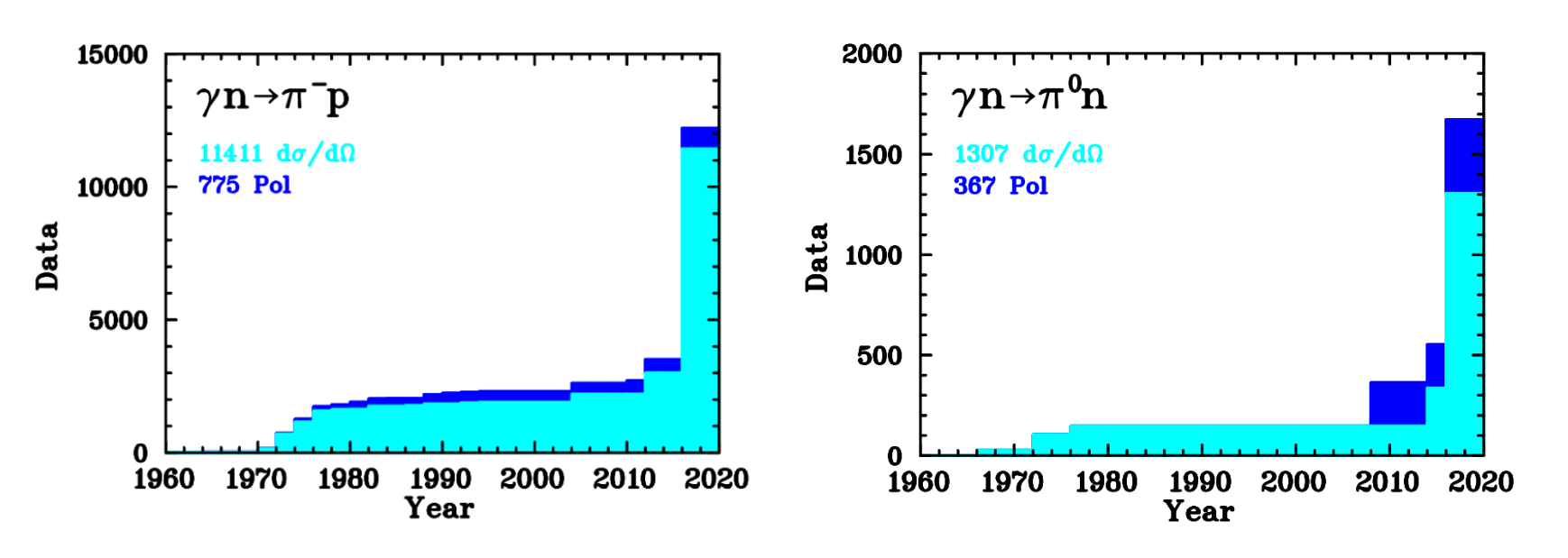}
}

\centerline{\parbox{\textwidth}{
\caption[] {\protect\small
Database for $\gamma n\to\pi^- p$ (left) and $\gamma n\to\pi^0 n$ (right). Experimental data from the SAID database~\cite{SAID} selected for 1996 through 2018. Right: Amount of data as a
function of time. Full SAID database. The data is shown as a stacked histogram. Light shaded – cross sections, dark shaded – polarization data. The figure is
        adapted from Ref.~\cite{Ireland:2019uwn}.
}
\label{fig:neutron} } }
\end{figure}

\subsubsection{Pion Photoproduction on the Neutron}
The ``neutron'' database is significantly smaller than the ``proton'' one.  The majority of single pseudoscalar meson photoproduction from the ``neutron'' target came from EM facilities at BNL and JLab, USA; MAMI, Germany; GRAAL, France (Fig.~\ref{fig:neutron})~\cite{Ireland:2019uwn}. 

\begin{figure}[htb!]
\centering
{
    \includegraphics[width=0.45\textwidth,keepaspectratio]{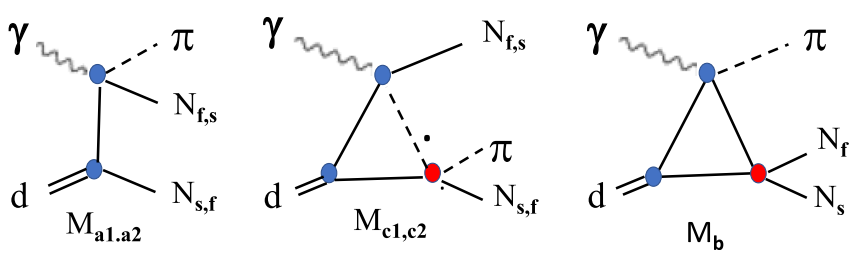} 
}

\centerline{\parbox{\textwidth}{
\caption[] {\protect\small
Feynman diagrams for the leading components of the $\gamma D\to \pi NN$ amplitude. a: Impulse approximation (IA), b: $NN$-FSI, and c: $\pi N$-FSI. Filled red circles show FSI vertices. Wavy, dashed, solid, and double lines correspond to the photons, pions, nucleons, and deuterons, respectively.
}
\label{fig:IA} } }
\end{figure}

Studies of the $\gamma n\to \pi^- p$ and $\gamma n\to \pi^0n$ reactions can be carried out in quasi-free kinematics with deuteron targets.  The reactions $\gamma d\to \pi^- p(p)$ and $\gamma d \to \pi^0 n(p)$ in these kinematics involve a fast, knocked-out nucleon and a slow proton
spectator; the slow proton is assumed not to be involved in the pion production process. In this quasi-free region, the reaction mechanism corresponds to the ``dominant'' impulse  
approximation (IA) diagram in Fig.~\ref{fig:IA}(a) with the slow proton emerging from the 
deuteron vertex. Here, the differential cross section on the deuteron can be related to that on
the neutron target in a well-understood way, see Ref.~\cite{Briscoe:2021siu} and references therein. Fig.~\ref{fig:IA} illustrates this dominant IA diagram, as well as the leading terms of FSI corrections. 

An energy- and angle-dependent FSI correction factor, R(E, $\theta$), can be defined as the ratio between the sum of three dominant diagrams in Fig.~\ref{fig:IA} and IA (the first of the diagrams). This can then be applied to the experimental $\gamma d$ data to obtain a two-body cross section for $\gamma n\to \pi^-p$ and $\gamma n\to \pi^0n$.

The GWU-ITEP FSI calculations (see Ref.~\cite{Briscoe:2021siu} and references therein) are available over a broad energy range (threshold to $E = 2.7~\mathrm{GeV}$), and for the full CM angular range ($\theta = 0^\circ$ to $180^\circ$). Overall, the FSI correction factor $R < 1.0$, while its value varies from 0.70 to
0.90 depending on the kinematics. The behavior of R is very smooth with respect to the pion production angle. There is a sizable FSI effect from the $S$-wave part of $pp$-FSI at small angles. 

R(E, $\theta$) is used as the FSI correction factor for the CLAS quasi-free $\gamma d\to  \pi pN$ cross section averaged over the laboratory photon energy bin width. Note that the FSI correction grows rapidly in the forward direction ($\theta < 30^\circ$).
There are currently few measurements in this regime, so the uncertainty due to FSI for this reaction at forward angles does not cause much concern. The contribution of uncertainty in FSI calculations to the overall systematic normalization uncertainty is estimated to be about 2\%–3\% (the sensitivity to the deuteron wave function is 1\%, and to the number of
steps in the integration of the five-fold integrals is 2\%. For the CLAS measurements, no sensitivity was found regarding the value of proton momentum used to determine whether or not it is a spectator.

The $\gamma n\to \pi^0n$ measurement is much more complicated than the case of $\gamma n\to \pi^-p$ because the $\pi^0$ can come from both neutron and proton initial states. The GW-ITEP studies have shown that photoproduction cross sections from protons and neutrons are generally not equal~\cite{Tarasov:2015sta}. For $\pi^0$ photoproduction on proton and neutron targets, one has
\begin{equation}
    A(\gamma p\to\pi^0p) = A_v + A_s~~~\mathrm{and}~~~
    A(\gamma n\to\pi^0n) = A_v - A_s \>,
\label{eq:eq3}
\end{equation}
where $A_v$ and $A_s$ are the isovector and isoscalar amplitudes, respectively. Therefore, if $A_s  \neq 0$ then the $\gamma p$ and $\gamma n$ amplitudes are not equal.

\begin{figure}
	\centering
        \includegraphics[width=10cm,clip]{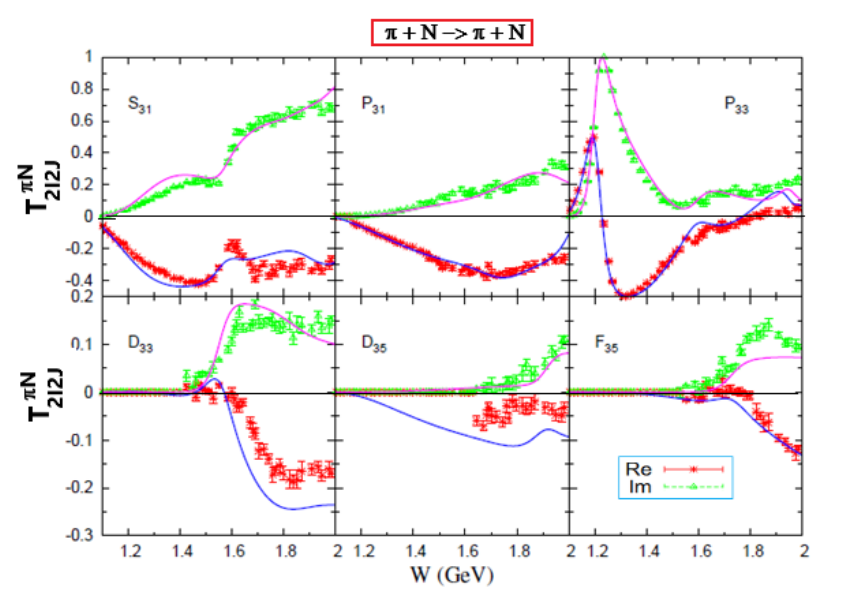}
	\caption{The elastic $\pi N$ partial wave scattering amplitudes in the isospin $I = 3/2$ channel. The solid magenta lines and the green
triangle points are the imaginary parts of the GiM and GWU/SAID
 amplitudes, respectively. The blue solid line and the red stars are the respective real parts of the GiM and GWU/SAID amplitudes.}
	\label{fig:piN_GimGWU}       
\end{figure}

\subsubsection{Elastic Scattering of Pions on a Nucleon}
Elastic scattering of hadrons is the process in which various kinds of interactions are reflected in a summarized manner. As emphasized in 
section~\ref{sec:CCSpec}, CC dynamics are much more involved than would be concluded from the elemental s-, t-, and u-channel meson-nucleon diagrams shown in Fig.~\ref{fig:sut}. Induced interactions and the resulting polarization self-energies may be of paramount importance. In practice, however, their contributions will depend on the configuration space and, to some extent, on the general methodology of the approach. In order to identify and understand such dependencies, comparisons of model calculations are indispensable. In 
Fig.~\ref{fig:piN_GimGWU}, the isospin 
$I=\frac{3}{2}$ partial wave scattering amplitudes obtained by the GiM approach (lines) and from the GWU/SAID project (symbols) are displayed for the entire spectrum of S-, P-, D-, and F-waves. Most prominently seen is the $P_{33}(1232)$ Delta resonance in the upper right panel, for which the two CC solutions indeed agree the best. Overall, the calculations are in fair agreement, albeit with detailed differences that are clearly visible, especially in the real parts of the amplitudes. They mainly reflect differences in the CC configuration space and different strategies for treating (or including at all) experimentally less confirmed resonances in the fitting procedures. The importance of such problems grows with energy and orbital angular momentum $L$. Overlapping and interfering structures complicate the clear identification of resonance structures. 

\subsection{$eta$ and $eta'$ Photoproduction on the Nucleon}\label{sec:etaProd}
\begin{figure}[htb!]
\centering
{
    \includegraphics[width=0.35\textwidth,keepaspectratio]{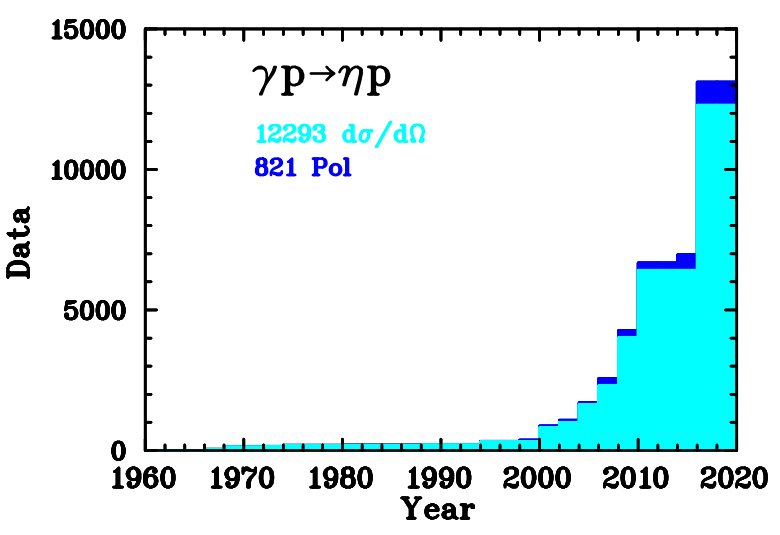}~~~ 
    \includegraphics[width=0.35\textwidth,keepaspectratio]{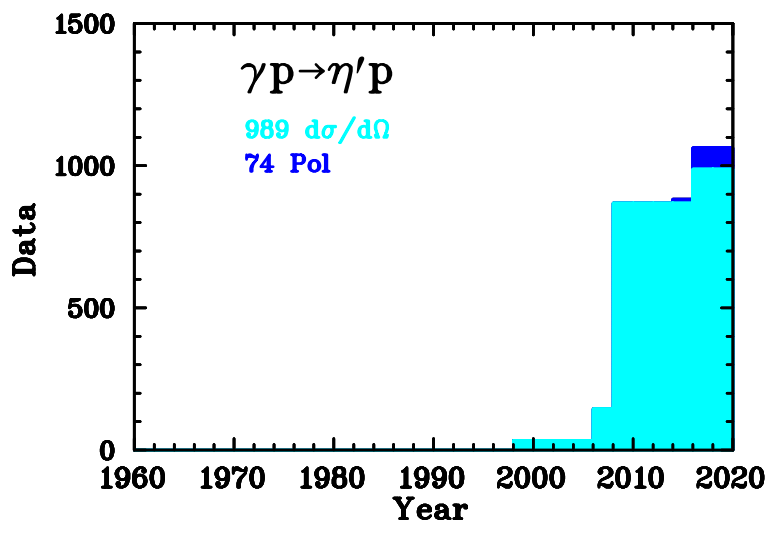}
}

\centerline{\parbox{\textwidth}{
\caption[] {\protect\small
Database for $\gamma p\to\eta p$ (left) and $\gamma p\to\eta' p$ (right). Experimental data from the SAID database~\cite{SAID} selected for 1996 through 2018. Right: Amount of data as a function of time. Full SAID database. The data is shown as a stacked histogram. Light shaded – cross sections, dark shaded – polarization data. The figure is
        adapted from Ref.~\cite{Ireland:2019uwn}.
}
\label{fig:eta} } }
\end{figure}

Overall, the experimental activities on the pseudoscalar-isoscalar $\eta(548)$ and 
$\eta^\prime(958)$ mesons have a short history, mainly because they are charge-neutral particles and, as such, are notoriously hard to detect. Compared to the pion case, the database for the isoscalar partner mesons is much smaller, as is evident from (Fig.~\ref{fig:eta}). 

In a pure quark-picture, one would assign $\eta\sim [q\bar{q}]$ as given by light $u,d$-quarks and $\eta^\prime\sim [s\bar{s}]$ as given by the heavier $s\bar{s}$-quarks. However, that assumption is in vain due to the remarkably large mass splitting of $eta$ and $eta^\prime$ mesons by about $500~\mathrm{MeV}$. Moreover, the SU(3)-singlet $\eta'$ meson is much heavier than the octet subset of the pseudoscalar nonet, which plays a central role in the spontaneous breaking of the $U(1)_A$ axial 
symmetry~\cite{Leutgeb:2019lqu}. The mass splitting is caused by large mixing effects involving a \textit{bare} gluonic flavor singlet $\eta_0$ state, as expressed by the famous Witten-Veneziano mass formula~\cite{WITTEN:1979u1a, VENEZIANO:1979u1a}
\be  
m^2_\eta+m^2_{\eta'}=m^2_{K}+m^2_{\eta_0}
\ee
which is satisfied for $m^2_{\eta_0}\simeq 0.71\mathrm{GeV}^2$. At Jefferson laboratory, the JEF (JLab Eta Factory) program aims to perform precision measurements of various $\eta$ and $\eta'$ decay channels. The JEF collaboration expects especially clean data sets on the rare neutral decays of the two
isoscalar mesons \cite{Somov:2024jiy} .

Eta-production on the nucleon has a slightly different focus.  Understanding the dynamics of $\eta$-meson production and, \emph{vice versa}, the decay of nucleon resonances into the nucleon-eta exit channel is of ongoing interest in hadron spectroscopy. The $\eta$-meson photoproduction on the proton has been measured with high precision by the Crystal Ball Collaboration at MAMI~\cite{McNicoll:2010qk}. These high-resolution data  provide a new step forward in understanding the reaction dynamics and in the search for a signal from the
``weak'' resonance states. The main result reported in~\cite{McNicoll:2010qk} is a very clean signal for a dip structure around $W=1.68~\mathrm{GeV}$, seemingly confirming older 
data~\cite{Dugger:2002ft, Crede:2009zzb, Bartholomy:2007zz, Bartalini:2007fg}. This raised the question of the origin of that structure, eventually indicating the appearance of a new, narrow, possibly exotic resonance state.

The study aimed to extend the previous coupled-channels analysis of the $\gamma p\to \eta p$ reaction by including the data from the new high-precision  
measurements~\cite{McNicoll:2010qk}.
The main question is whether the $\eta p$ reaction dynamics can be understood in terms of the established resonance states or whether a new state has to be introduced, thus confirming previous conjectures. A major issue for the analysis is unitarity and a consistent treatment of self-energy effects, as evidenced by the total decay width of resonances. Since the latter are driven by hadronic interactions, the analysis of photo-production data requires knowledge of the hadronic transition amplitudes as well. Hence, a coupled-channels description, such as the GiM, is an indispensable necessity.

As discussed in detail in Ref.~\cite{Shklyar:2012js}, various relevant meson-baryon coupling constants were newly determined in the context of this work through large-scale coupled-channels calculations. This gave rise to improved constraints on the interaction parameters and the derived resonance parameters, \textit{i.e.}, masses and widths.
Representative examples are the mass and width of the $D_{13}(1520)$ resonance, $M=1516\pm10~\mathrm{MeV}$, and $\Gamma=106\pm4~\mathrm{MeV}$, which agree with and confirm the values obtained earlier by Arndt \emph{et al.}~\cite{Arndt:2006bf}. It is interesting to note that the mass of this resonance, deduced from pion photoproduction, tends to be $10~\mathrm{MeV}$ lower than the values derived from the pion-induced reactions~\cite{PDG:2016}. The second $D_{13}(1900)$ state has a very large decay width. That state is likely to be related to the $D_{13}(2080)$ two-star resonance, proposed by the Particle Data Group in the 2024 issue of the Particle Physics Review~\cite{PDG:2024cfk}.

\begin{figure}
	\centering
        \includegraphics[width=9cm,clip]{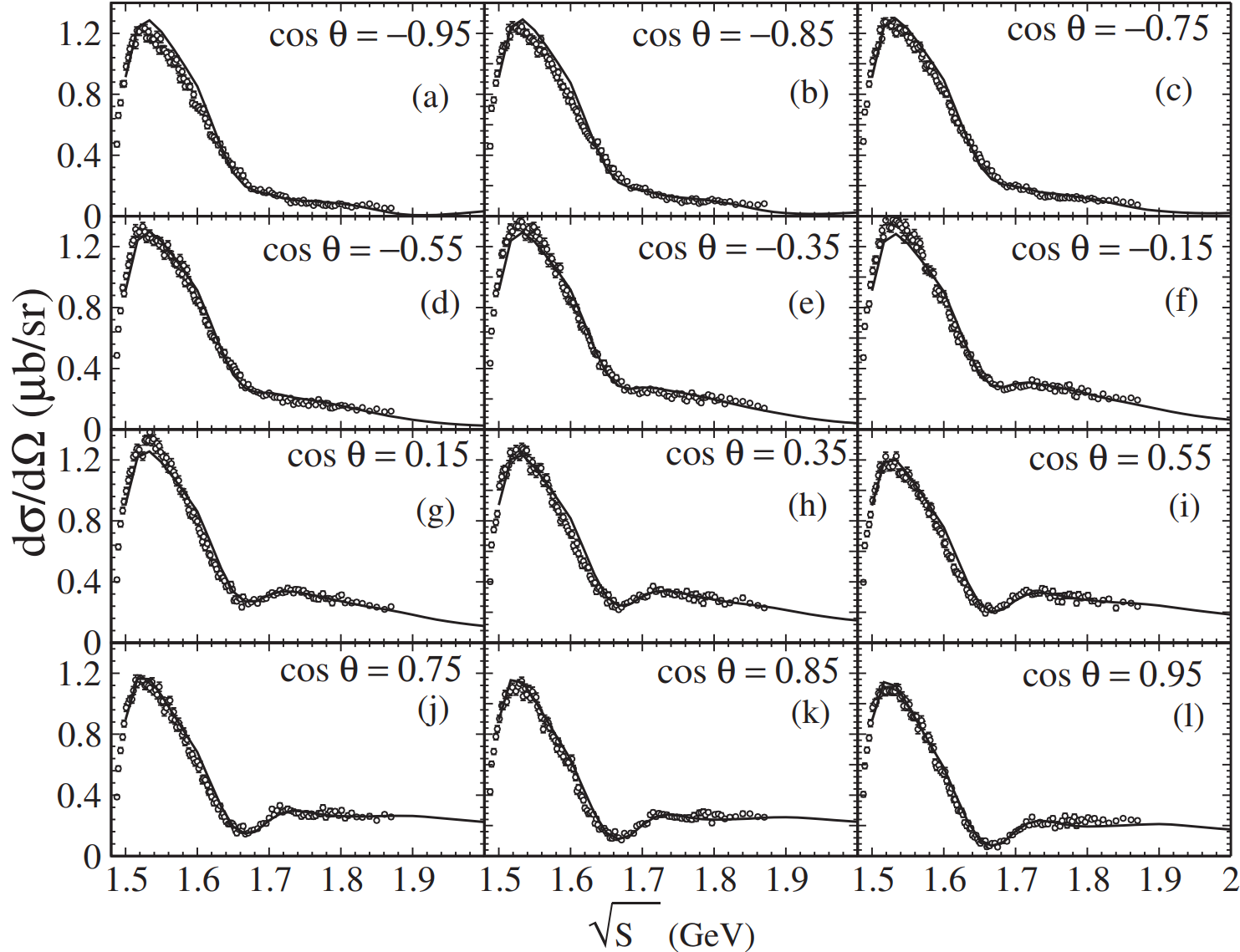}
	\caption{GiM results (full lines) of differential $\eta p$ cross section are compared to data from MAMI (symbols)
        data~\cite{McNicoll:2010qk}. The figure is adapted from Ref.~\cite{Shklyar:2012js}.}
	\label{fig:fig-2}       
\end{figure}

The results of the calculation of the $\eta$-photo production channel are shown and compared with the experimental data in Fig.~\ref{fig:fig-2}.

The calculations demonstrate very satisfactory agreement with
the experimental data in the entire kinematical region. The first peak is related to the $S_{11}(1535)$
resonance contribution. Similar to the $\pi^- p \to \eta n$ reaction, the $S_{11}(1650)$ and $S_{11}(1650)$
states interfere destructively, producing a dip around $W=1.68~\mathrm{GeV}$.
The coherent sum of all partial waves leads to a more pronounced effect from the dip
at forward angles. 
\begin{figure}
	\centering
        \includegraphics[width=10cm,clip]{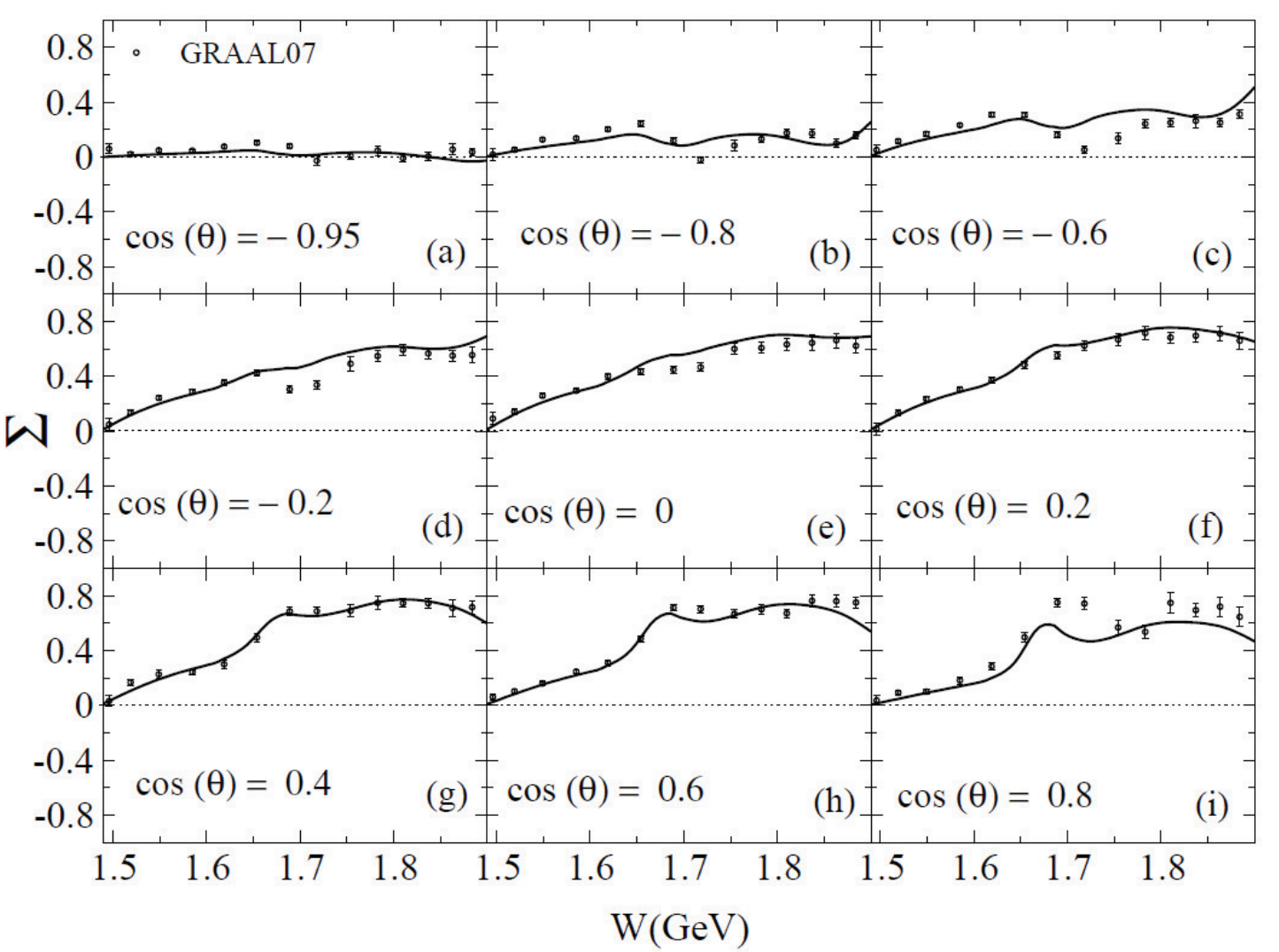}
	\caption{GiM results for the photon-beam asymmetry $\Sigma$ in $\eta$ photoproduction on the proton compared to GRAAL data~\cite{Bartalini:2007fg}. The figure is
        adapted from Ref.~\cite{Shklyar:2012js}.} 
	\label{fig-3}       
\end{figure}

In Fig.~\ref{fig-3},  results for the photon-beam asymmetry $\Sigma$ are compared to the GRAAL
data. One can see that even close to the $\eta N$ threshold, where the calculations
exhibit a dominant $S_{11}$ production mechanism, the beam asymmetry is non-vanishing for angles  $\cos(\theta)\ge-0.2$. These results demonstrate that this observable is highly sensitive to very small contributions from higher partial waves. At $W=1.68~\mathrm{GeV}$  and forward angles, the GRAAL measurements show a rapid change in asymmetry 
behavior. In~\cite{Shklyar:2006xw} this effect was explained by destructive interference between the  $S_{11}(1535)$ and 
$S_{11}(1650)$
resonances, which induce the  dip at $W \simeq 1.68~\mathrm{GeV}$ in the $S_{11}$ partial wave.
Note that the interference between $S_{11}(1535)$ and $S_{11}(1650)$ and
the interference between different partial waves are of a different nature.
The overlapping of the
$S_{11}(1535)$ and  $S_{11}(1650)$ resonances do not simply mean a coherent sum of two independent
contributions, but also include rescattering, \textit{i.e.}, coupled-channel effects. Such interplay can hardly be simulated by a simple sum of two Breit-Wigner distributions because this approach does not comply with unitarity.

\subsection{Strangeness Production on the Nucleon by Kaon Photoproduction}\label{sec:KaonProd}
\begin{figure}[htb!]
\centering
{
    \includegraphics[width=0.68\textwidth,keepaspectratio]{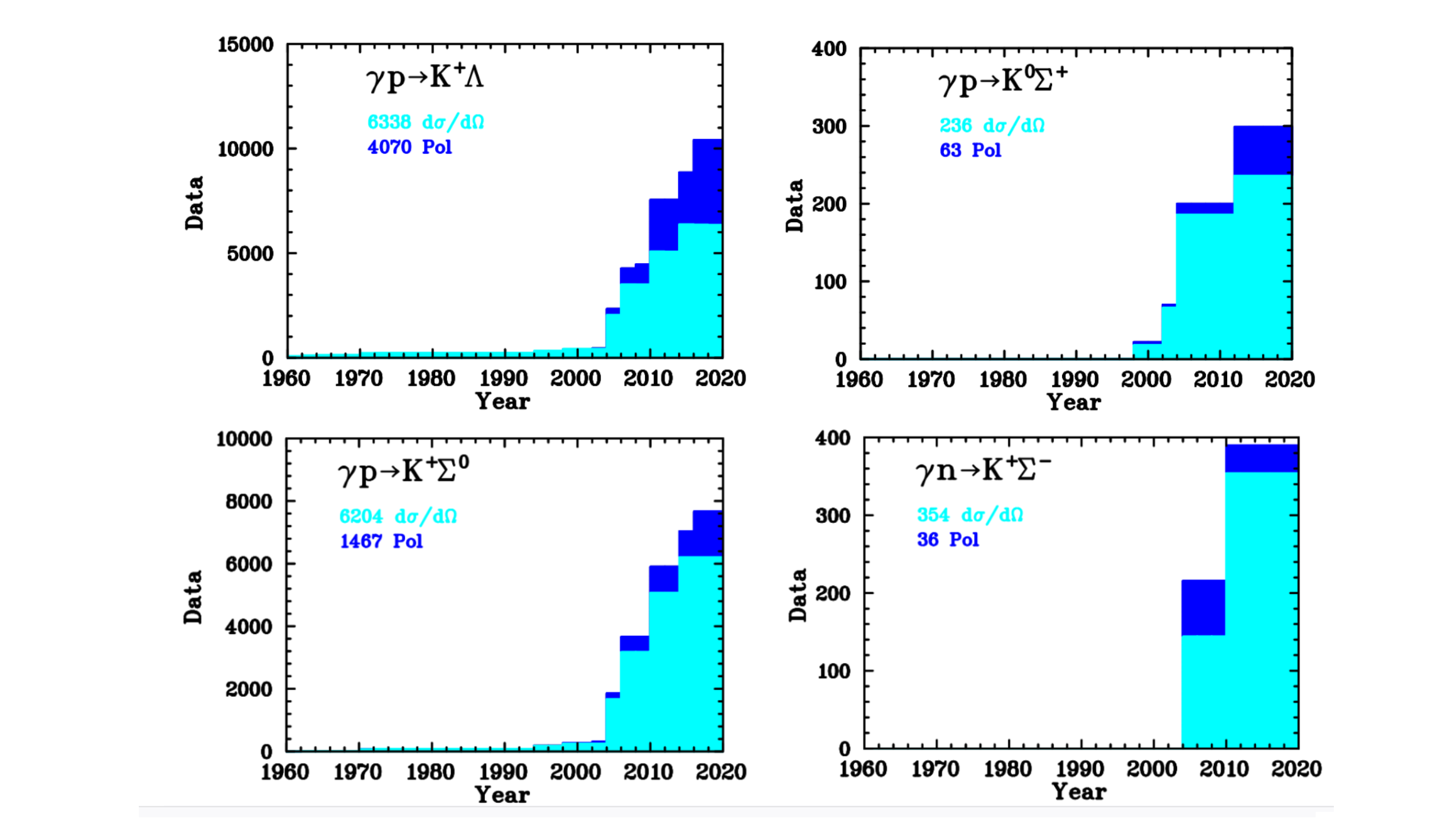}    
}

\centerline{\parbox{\textwidth}{
\caption[] {\protect\small
Database for the Kaon photoproduction data on both proton and neutron targets. Experimental data from the SAID database~\cite{SAID} selected for 1996 through 2018. Right: Amount of data as a
function of time. Full SAID database. The data is shown as a stacked histogram. Light shaded – cross sections, dark shaded – polarization data. The figure is
        adapted from Ref.~\cite{Ireland:2019uwn}.
}
\label{fig:kaon} } }
\end{figure}

As mentioned before, kaon physics was the primer for realizing that CP symmetry is violated, with far-reaching implications for elementary particle physics. The existence of two separate kaon decay branches was the first definite signal of competing and interfering processes. $K^0$ and its antiparticle $\bar{K}^0$ are degenerate mass eigenstates which, however, are physically observed in two distinct \textit{flavor} CP eigenstates, 
$K^0_{1,2} = \frac{1}{\sqrt{2}} (K^0\mp\bar{K}^0)$, 
which agree up to a CP violating mixing amplitude $\sim 10^-3$ with $K_{S, L}$ \cite{Christenson:1964fg}. This special feature enforced the concept of superpositions for neutral elementary particles by necessitating the introduction of the $K_L$ and $K_S$ states of the $K^0/\bar{K}^0$ system. 
The kaon database is shown in Fig.~\ref{fig:kaon}. A similar mixing phenomenon is known for neutral leptons, \textit{i.e.}, neutrinos and antineutrinos.

\begin{figure}
	\centering
	\includegraphics[width=6cm,clip]{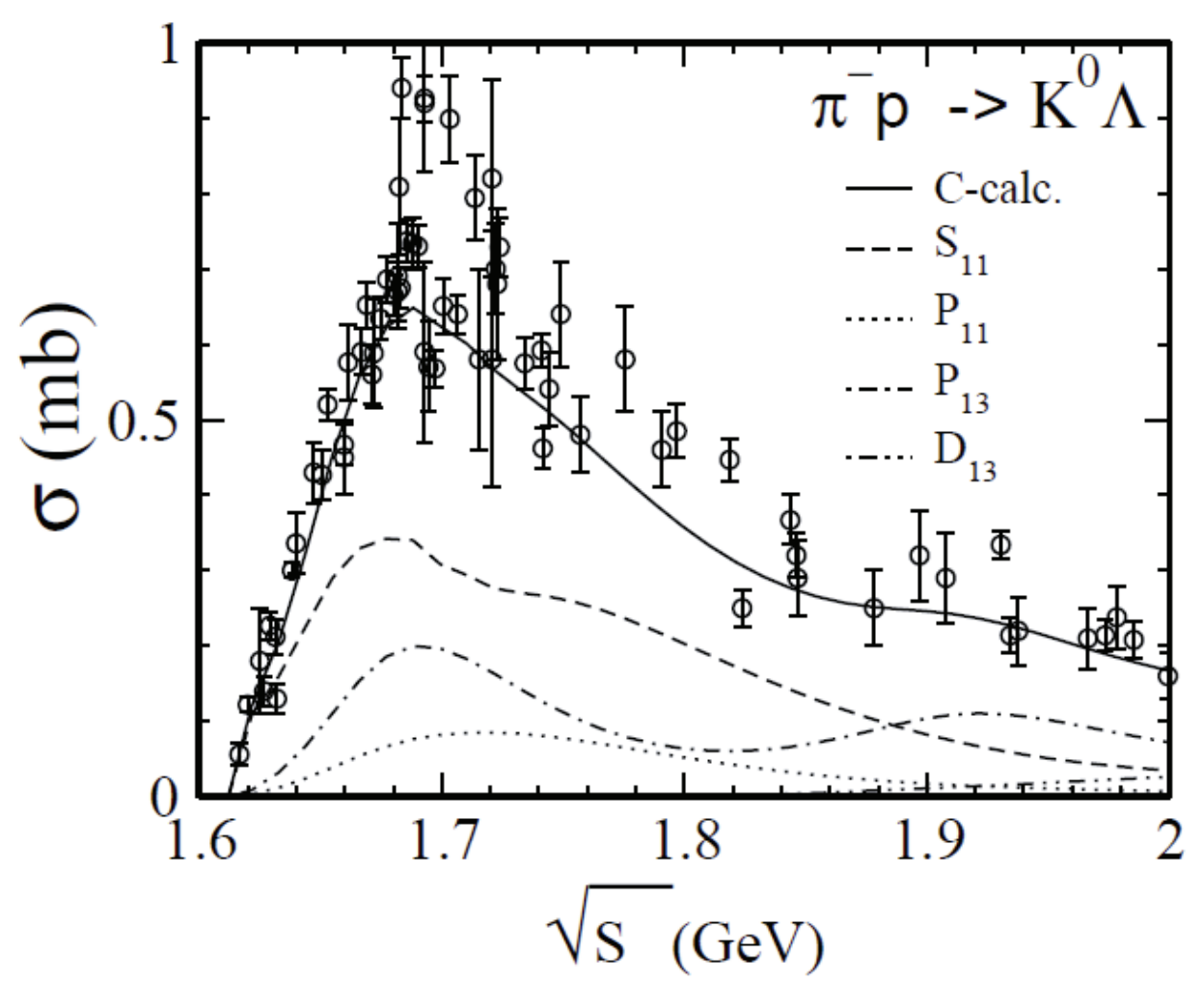} 
     \caption{$\pi^- p\to K^0\Lambda$
partial wave cross sections, predicted by parameter set \emph{C} of 
Ref.~\cite{Shklyar:2005xg}, obtained from a fit to the CLAS data~\cite{CLAS:2004}.
The experimental cross section data are taken from~\cite{Baker:1978, Saxon:1979, Knasel:1975}.
The figure is adapted from Ref.~\cite{Shklyar:2005xg}.
	\label{fig:KLambda}}       
\end{figure}

Strangeness production on the nucleon through the excitation of resonances that decay into kaon-hyperon channels is an important spectroscopic tool, providing access to the SU(3) flavor structure of baryons. Moreover, such \emph{exotic} channels, like the kaon-hyperon final states, are expected to play a key role in identifying hitherto undetected excited states of the nucleon, thus addressing the notorious problem of \emph{missing resonances}. 
In~\cite{Shklyar:2005xg},  pion- and photon-induced $K\Lambda$ reactions were studied using the unitary coupled-channel effective Lagrangian approach. Data on the photoproduction of kaons on the nucleon from the SAPHIR, CLAS, and CBELSA experiments were described by the GiM coupled channels K-matrix approach, also taking into account the full set of all other meson-baryon channels. Thus, a major revision of the complete parameter set was performed. A major goal of those investigations was to address the still open question regarding the major contributions to the associated strangeness production channels. Since  $K\Lambda$ photoproduction data~\cite{SAPH:2005, CLAS:2004} indicated \emph{missing
resonance}  contributions, a combined analysis of the $\pi +N \to K\Lambda$ and the $\gamma+N \to K\Lambda$ reactions was expected to identify these
states. Assuming small couplings to $\pi N$, these \emph{hidden} states
should not manifest in the pion-induced reactions and,
consequently, in the $\pi N \to K\Lambda $ reaction. The calculations aimed to explore whether the data available at that time
could be explained by known reaction mechanisms without introducing new resonances.
 The results for total cross sections are displayed in Fig.~\ref{fig:KLambda} and further results on differential cross sections, polarization observables, and angular distributions are found in~\cite{Shklyar:2005xg}. As discussed in~\cite{Shklyar:2005xg}, the 
 SAPHIR~\cite{SAPH:2005} and the CLAS~\cite{CLAS:2004} data sets lead to two slightly different sets of interaction parameters, reflecting and emphasizing the differences between the two measurements. Below, that point is discussed again.

CLAS-data on $K\Sigma$ production by polarized beams initiated an updated large scale coupled-channels analysis of associated strange production on the nucleon. Based on the effective GiM Lagrangian, a combined CC analysis of $(\pi,\gamma) N \to K\Sigma$ hadro- and photo-production reactions was performed. The analysis covered a center-of-mass energy range up to $2~\mathrm{GeV}$. The central aim was to extract the resonance couplings to the $K\Sigma$ state.  Both $s$-channel resonances and $t,~u$-channel background contributions are found to be important for an accurate description of angular distributions and polarization observables, ensuring a high-quality description of the data. The extracted properties of isospin $I = 3/2$ resonances were discussed in detail. In~\cite{Shklyar:2005xg}, it was found that the $I = 1/2$ resonances are largely determined by the non-strangeness channels.

\begin{figure}
\begin{center}
    \includegraphics[width=10cm]{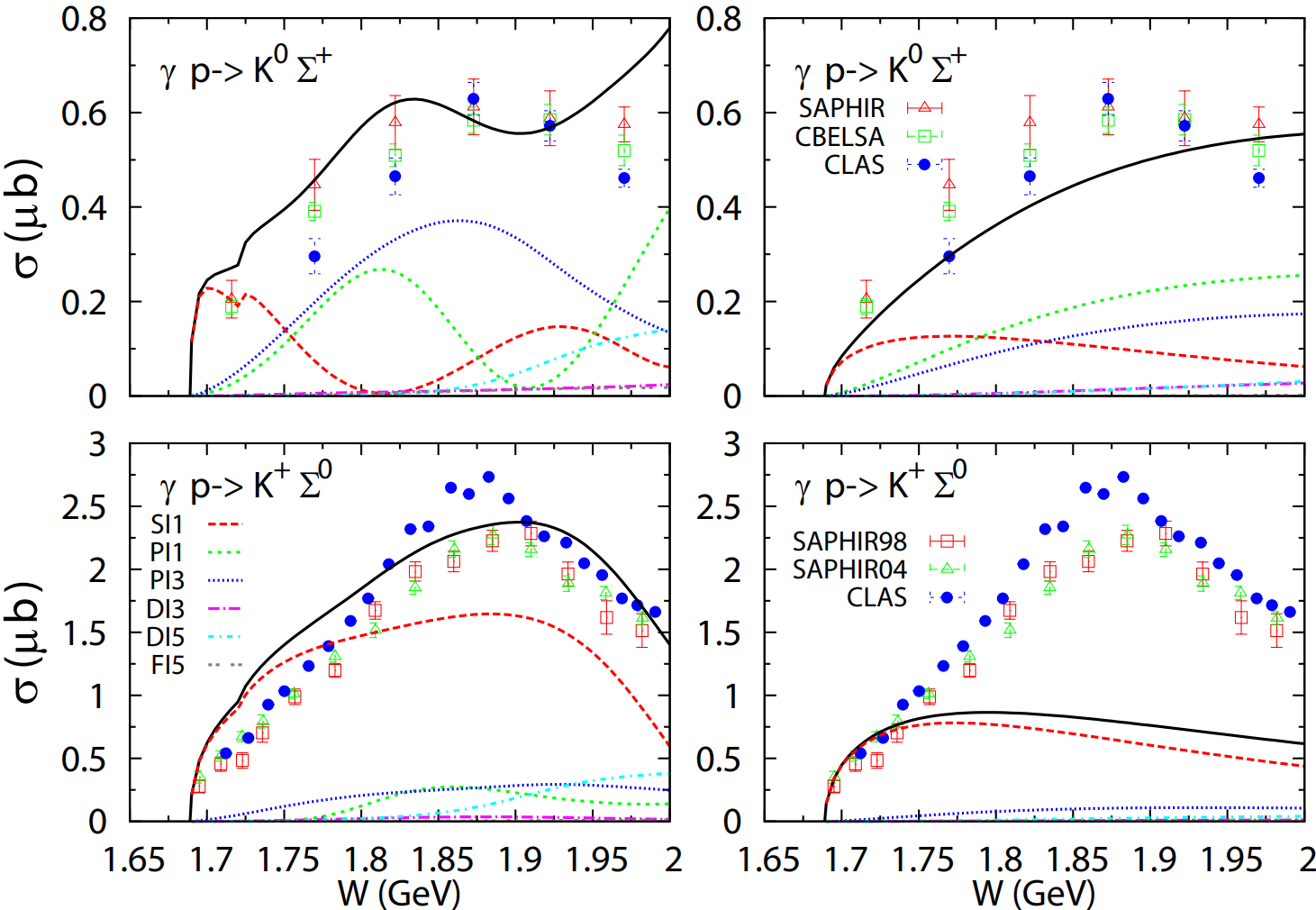}
\caption{Total cross sections for kaon production on the nucleon. Results of the Giessen 
model~\cite{GiM:2013} are compared to CLAS, CBELSA, and SAPHIR data. Results of the full model calculation are shown in the left panel. Results using only the Born amplitudes and t-channel meson exchange are displayed in the right panel. 
The figure is adapted from Ref.~\cite{GiM:2013}.}
\label{fig:Fig1}
\end{center}
\end{figure}

The calculations included 11 isospin $I = 1/2$ resonances and 9 isospin $I = 3/2$ resonances, respectively. The investigations were extended to the $I = 1/2$ and $3/2$ sectors, with the parameters fitted to newly published $K\Sigma$ photoproduction data, together with the previous $\pi N \to K\Sigma$ measurements in the energy region $\sqrt{s} \leq 2.0~\mathrm{GeV}$. The included $K\Sigma$ photoproduction data are those of the $\gamma p \to K^+ \Sigma^0$ published by the LEPS~\cite{LEPS03, Sumihama:2005er, LEPS06Kohri}, CLAS~\cite{CLAS06sigma0, CLAS10Dey}, and GRAAL~\cite{GRAAL07} groups, as well as those of $\gamma p \to K^0 \Sigma^+$ released by the CLAS~\cite{CLASthesis} and CBELSA~\cite{CBELSA08} collaborations, respectively. The SAPHIR data have been omitted here due to the known inconsistencies of the $K^+ \Sigma^0$ data~\cite{SAPH:2005} with the corresponding CLAS and GRAAL data (for details, see Ref.~\cite{CLAS10Dey}). Also, the $K^0 \Sigma^+$ SAPHIR data~\cite{SAPH:2005} have much larger error bars than those of the CBELSA and CLAS groups. The data before 2002 are also no longer used. Results for total cross sections are shown in Fig.~\ref{fig:Fig1}. Up to a total center-of-mass energy of about $\sqrt{s}=2~\mathrm{GeV}$, the data are well described.

\begin{figure}
  \begin{center}
    {\includegraphics*[width=10cm]{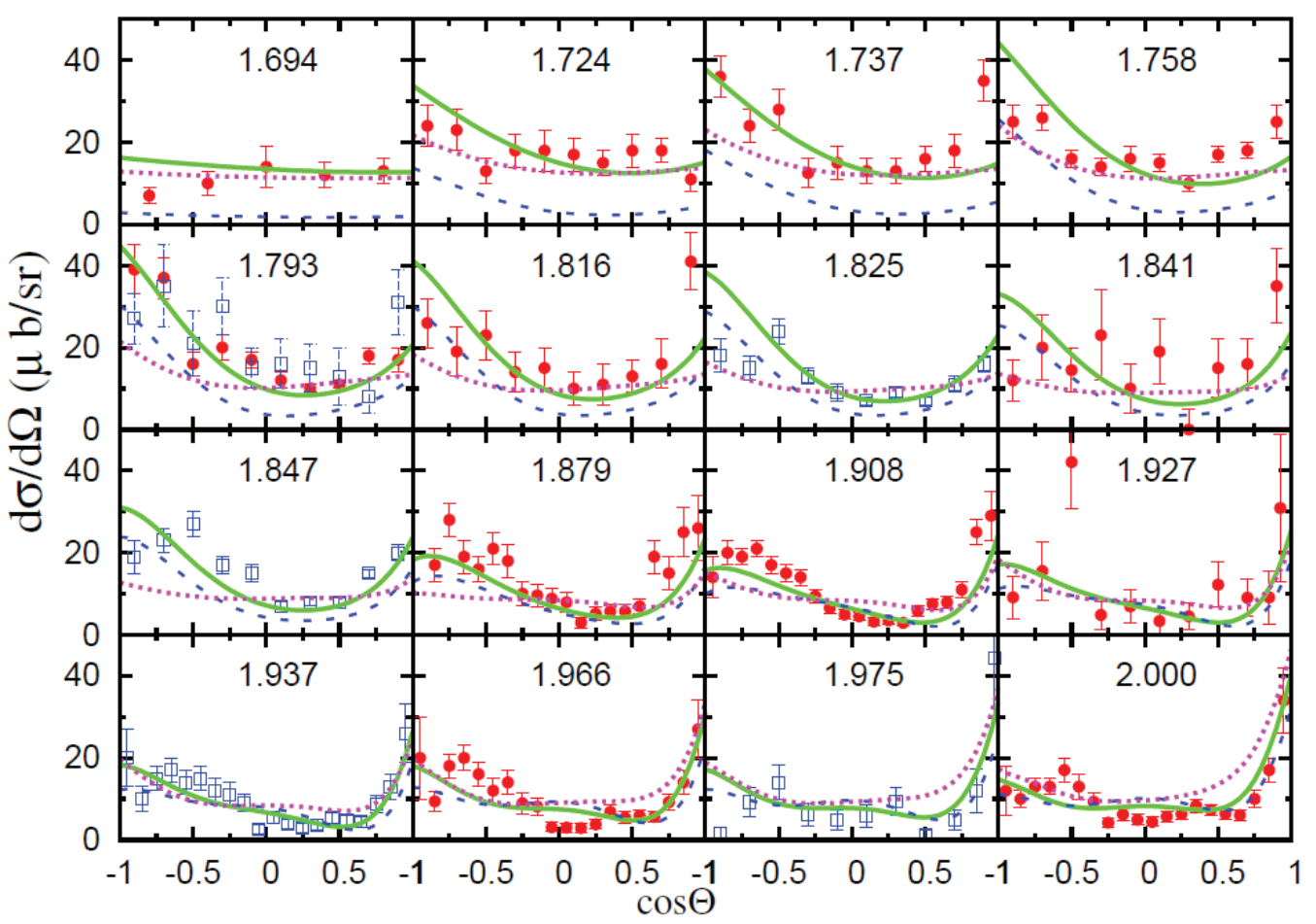}}
       \caption{
Differential cross section of $\pi^- p \to K^0 \Sigma^0$ reaction. The solid (green), dashed (blue), and dotted (magenta) lines are the full model calculation, the model calculation with the $S_{11}(1650)$ and $F_{15}(1680)$ turned off, respectively. The numerical labels denote the center of mass energies in units of GeV. The figure is
        adapted from Ref.~\cite{GiM:2013}.
      \label{fig:ps0dif}}
  \end{center}
\end{figure}

The analysis included all charge channels, $K^0\Sigma^\mp$ and $K^\pm\Sigma^0$.  A quite satisfactory description was achieved  of the $\gamma p \to K^+ \Sigma^0$ data ($\chi^2 = 1.8$) and the $\gamma p \to K^0 \Sigma^+$ data ($\chi^2 = 2.0$). However, the pion-induced strangeness production reactions are described slightly less accurately, as indicated by the corresponding $\chi^2$ values of $\chi^2 =$ 4.1, 3.2, and 2.8 for the $\pi^+ p \to K^+ \Sigma^+$, $\pi^- p \to K^0 \Sigma^0$, and $\pi^- p \to K^+ \Sigma^-$ reactions, respectively. The parameters have been varied in the fit simultaneously to the $I = 1/2$ and $3/2$ sectors. Although the new data are available with reduced total uncertainties, the refitted model parameters changed only very slightly. A typical result is displayed in Fig.~\ref{fig:ps0dif}, illustrating the quality of the description using the example of the $\pi^- p \to K^0 \Sigma^0$ reaction. The complete set of results, including partial wave cross sections, angular distributions of cross sections, and polarization observables for the full set of $K\Sigma$ exit channels, is found in~\cite{GiM:2013}.

\subsection{Vector Meson Photoproduction on the Proton}
\begin{figure}[htb!]
\centering
{
    \includegraphics[width=0.3\textwidth,keepaspectratio]{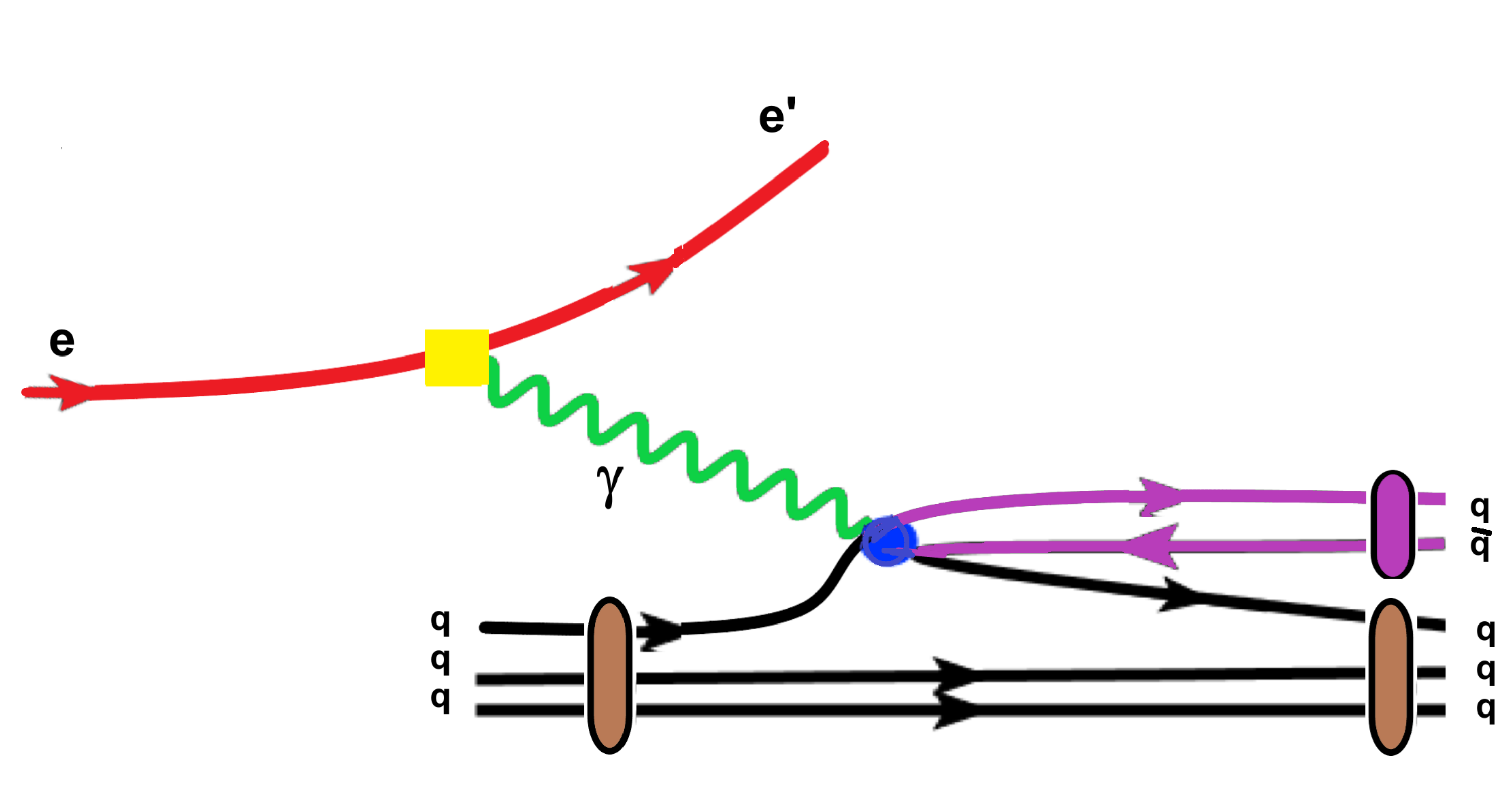}
}

\centerline{\parbox{\textwidth}{
\caption[] {\protect\small
Feynman graph for vector meson photoproduction on a nucleon by bremsstrahlung. A so-called quark transfer process is shown, which is the most likely mechanism for the light vector meson production. The nucleon is denoted as a ($qqq$) state, the $\omega$ meson as a $q\bar q$ state.
}
\label{fig:BremsRad}}}
\end{figure}

For obvious reasons, vector meson beams do not exist, but they can be created \textit{in situ} by taking advantage of \textit{bremsstrahlung}, \textit{e.g.}, by scattering a high-energy electron beam off a proton or a nuclear target, as practiced at all major electron facilities. Physically, such experiments require creating $q\bar q$ pairs through the interaction of bremsstrahlung photons. A generic Feynman diagram is shown in Fig.~\ref{fig:BremsRad}. 

Seen from the outside, the process can be explained by Sakata's VMD hypothesis  (illustrated in Fig.~\ref{fig:fig0})~\cite{Sakurai:1960ju} 
which in this case applies because massless photons - one of the electro-weak gauge bosons - and light massive vector mesons share not only the same intrinsic quantum numbers $I^G(J^{PC}) = 0^-(1^{-~-})$ but also allow the exchange of valence quarks between nucleon and the meson. Thus, at sufficiently high energy, photons may reveal their hadronic content, either in the form of vector mesons, as assumed in VMD, or by coupling to $q\bar q$ pairs in a vector meson configuration. The hit proton is in an excited state, which subsequently decays by $\omega$ emission. 
Examples of $\omega$-production in bremsstrahlung experiments at CLAS@JLab  and MAMI@Mainz are found in~\cite{CLAS:2002cdi} 
and in~\cite{Strakovsky:2014wja}, respectively. The JLab measurement was performed  with a bremsstrahlung photon beam produced by a
continuous electron beam of energy 4.1+~GeV hitting a gold
foil, combined with a bremsstrahlung tagging
system. $\omega$ production  was identified from the $\gamma p\to p +X$ missing mass spectrum. 
For the Mainz experiment, the energy-tagged bremsstrahlung-photon beam
produced by the MAMI electron beam was used, with the Crystal Ball (CB) serving as a central spectrometer, and
TAPS installed as a forward spectrometer for measuring the chain of 
processes 
$\gamma p \to \omega p \to \pi^0 \gamma p \to 3\gamma p$. The 
CBELSA/TAPS experiment at Bonn  
uses similar techniques, see, \textit{e.g.}, \cite{Seifen:2024pxk}. 

\begin{figure}[htb!]
\centering
{
    \includegraphics[width=0.4\textwidth,keepaspectratio]{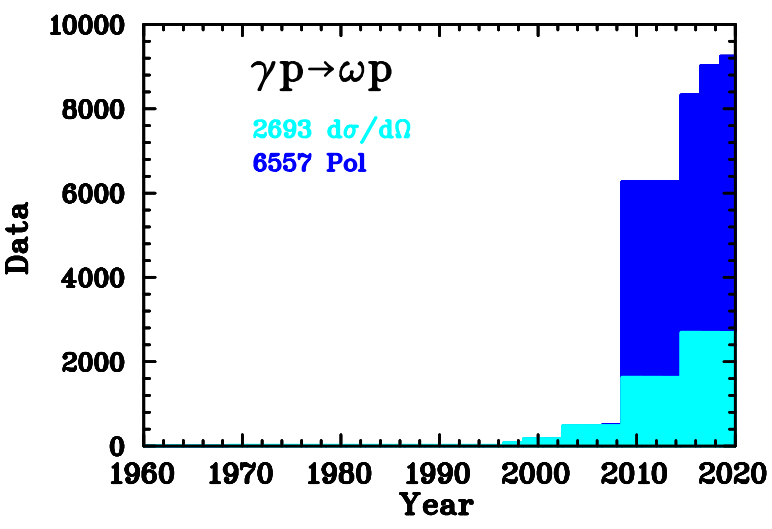}
}

\centerline{\parbox{\textwidth}{
\caption[] {\protect\small
Database for $\gamma p\to\omega p$. Experimental data from the SAID database~\cite{SAID} selected for 1996 through 2018. Right: Amount of data as a function of time. Full SAID database. The data is shown as a stacked histogram for cross sections data (Light shaded ) and polarization data (dark shaded). The figure is adapted from Ref.~\cite{Ireland:2019uwn}.
}
\label{fig:omega}}}
\end{figure}

 In Fig.~\ref{fig:omega}, the data available from JLab, MAMI, ELSA, and GRAAL for the $\omega$ case are shown.
\begin{figure*}[htb!]
\centering
{
    \includegraphics[width=0.4\textwidth,keepaspectratio]{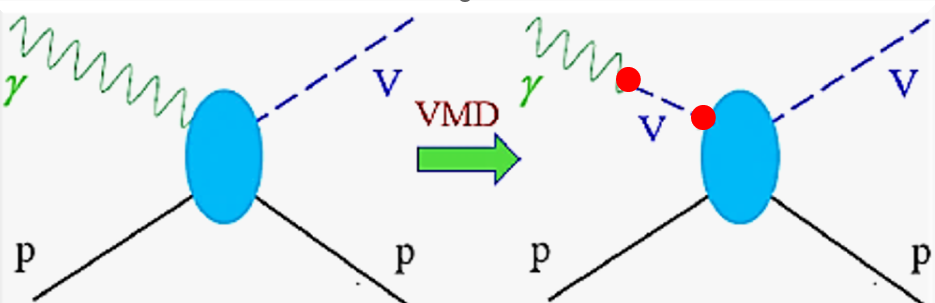}
}

\centerline{\parbox{\textwidth}{
\caption[] {\protect\small
 Schematic diagrams of vector-meson photoproduction (left) and the VMD model (right) in the energy region of threshold experiments. $V$ means vector meson.
} 
\label{fig:fig0} }}
\end{figure*}

High-statistics total cross sections for vector meson photoproduction at the threshold: $\gamma p\to\omega p$ (from A2 at MAMI~\cite{Strakovsky:2014wja}, ELPH~\cite{Ishikawa:2019rvz}, and CBELSA/TAPS)~\cite{CBELSATAPS:2015wwn}, $\gamma p\to \phi p$ (from CLAS~\cite{Dey:2014tfa} and LEPS~\cite{LEPS:2005hax, Chang:2007fc}), and 
$\gamma p\to J/\psi p$ (from GlueX)~\cite{GlueX:2019mkq} allow us to constrain the modulus of the vector meson nucleon scattering length (SL) by using the VMD model to extrapolate to the on-shell point. The extended analysis of $\Upsilon$-meson photoproduction using QCD quasi-data is in perfect agreement with the light-meson findings using experimental 
data~\cite{Guo:2021ibg}. Interestingly, similar to the high-energy limit, the near-threshold production amplitude was found theoretically to be factorizable into gluonic generalized parton distributions (GPD) and a quarkonium distribution amplitude. 

Let us focus on four vector mesons ($\omega$, $\phi$, $J/\psi(1S)$, and $\Upsilon(1S)$) from the $q\bar{q}$ nonet, the widths of which are narrow enough to study meson photoproduction at threshold, and for which data and quasi-data are available (Table~\ref{tbl:VM}). To avoid a broad width problem at the threshold, the $\rho$-meson is not considered a meaningful candidate for the determination of the vector meson nucleon SL. Furthermore, 
$\psi^\prime(2S)$ is not considered because of the disturbing effects from the additional node in the $2S$ radial wave function (WFs). Until recently, it was expected that investigations could not be extended beyond the $\Upsilon$ $b\bar{b}$ region because the $t\bar{t}$ state 
$T(1S)$ seemed not to exist. The situation changed with recent CMS observations of a near-threshold enhancement in top quark pair production~\cite{CMS:2025kzt}.
\begin{table}[htb!]

\centering \caption{List of the vector mesons including quark contents and their widths~\cite {PDG:2024cfk}. The four vector mesons, used for the determination of the scattering length as discussed in the text, are marked in blue.
}

\vspace{2mm}
{%
\begin{tabular}{|c|c|c|}
\hline
Vector      & Quark      & $\Gamma$ \tabularnewline
Meson       & Content    & [MeV] \tabularnewline
\hline
$\rho^+(770)$   & $u\bar{d}$ & 148 \tabularnewline
$\rho^0(770)$   & $\frac{(u\bar{u}-d\bar{d})}{\sqrt{2}}$ & 149 \tabularnewline
\textcolor{blue}{$\omega(782)$} & \textcolor{blue}{$\frac{(u\bar{u}+d\bar{d})}{\sqrt{2}}$} & \textcolor{blue}{8.5} \tabularnewline
$K^{\ast +}(892)$ & $u\bar{s}$ & 51 \tabularnewline
$K^{\ast 0}(892)$ & $d\bar{s}$ & 47 \tabularnewline
\textcolor{blue}{$\phi(1020)$}  & \textcolor{blue}{$s\bar{s}$} & \textcolor{blue}{4.3} \tabularnewline
$D^{\ast +}(2010)$ & $c\bar{d}$ & 0.083 \tabularnewline
$D^{\ast 0}(2007)$ & $c\bar{u}$ & $<$2.1 \tabularnewline
\textcolor{blue}{$J/\psi(1S)(3097)$} & \textcolor{blue}{$c\bar{c}$} & \textcolor{blue}{0.093} \tabularnewline
$\psi'(2S)(3686)$ & $c\bar{c}$ & 0.284 \tabularnewline
\textcolor{blue}{$\Upsilon(1S)(9460)$} & \textcolor{blue}{$b\bar{b}$} & \textcolor{blue}{0.052} \tabularnewline
\hline
\end{tabular}} \label{tbl:VM}
\end{table}
\begin{figure*}[htb!]
\centering
{
    \includegraphics[width=0.4\textwidth,keepaspectratio]{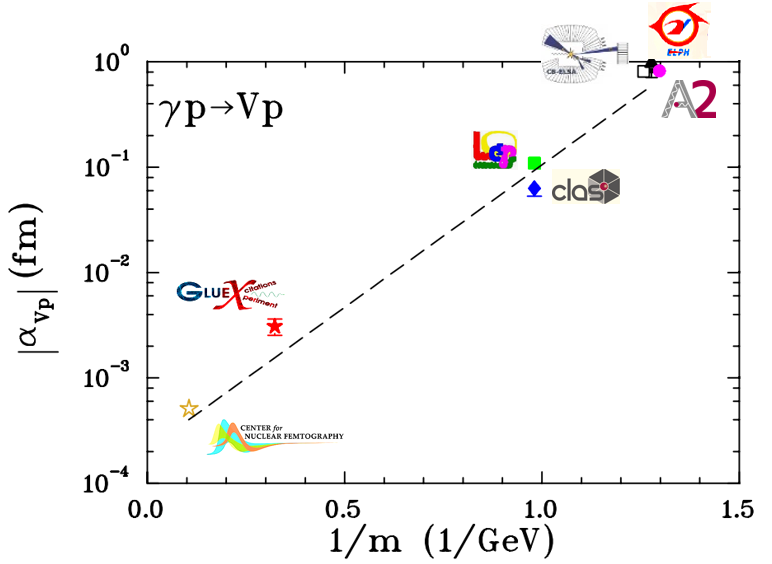}
}

\centerline{\parbox{\textwidth}{
\caption[] {\protect\small
Comparison of the $|\alpha_{Vp}|$ SLs estimated from threshold vector meson photoproduction on the proton target with VMD model contribution vs the inverse mass of the vector mesons. 
Input data for phenomenological analyses came from 
A2 at MAMI (magenta filled circle)~\cite{Strakovsky:2014wja}, 
ELPH (black filled triangle)~\cite{Ishikawa:2019rvz}, and 
CBELSA/TAPS (black open square)~\cite{CBELSATAPS:2015wwn} Collaborations for the $\omega$-meson;
CLAS (blue filled diamond)~\cite{Dey:2014tfa} and 
LEPS (green filled square)~\cite{LEPS:2005hax, Chang:2007fc} Collaborations for the $\phi$-meson; and
GlueX (red filled star)~\cite{GlueX:2019mkq} Collaboration for the $J/\psi$-meson; and quasi-data from Center for Nuclear Femtography (brown open star)~\cite{Guo:2021ibg} for the $\Upsilon$-meson.
Analyses results for
$\omega$-meson is given at Refs.~\cite{Strakovsky:2014wja, Ishikawa:2019rvz, Han:2022khg};
for the $\phi$-meson is given at Refs.~\cite{Strakovsky:2020uqs, Han:2022khg};
for $J/\psi$-meson is given at Ref.~\cite{Strakovsky:2019bev}; and
for $\Upsilon$-meson is given at Ref.~\cite{Strakovsky:2021vyk}.
The black dashed line is hypothetical following $|\alpha_{Vp}| \propto 1/m_V$.
$V$ means vector meson.} 
\label{fig:figSL} } }
\end{figure*}

The suggested approach can be applied to evaluate the $J/\psi$-nucleon SLs, replacing the photon with a $J/\psi$-meson (Fig.~\ref{fig:fig0}). The results then appear to have the order of several units of $10^{-3}~\mathrm{fm}$, $\alpha_{J\psi N}^{(J=1/2)} = (0.2...3.1)\times 10^{-3}~\mathrm{fm}$ and $\alpha_{J\psi N}^{(J=3/2)} = (0.2...3.0)\times 10^{-3}~\mathrm{fm}$, where $J$ corresponds to the total angular momentum of the $J/\psi$-nucleon system~\cite{Du:2020bqj}.

Future high-quality experiments by EIC and EicC will have the opportunity 
to evaluate cases for $J/\psi$- and $\Upsilon$-mesons. It allows one 
to understand the dynamics of $c\bar{c}$ and $b\bar{b}$ production at 
the threshold.  The ability of J-PARC to measure $\pi^- p\to \phi n$ 
and $\pi^-p\to J/\psi n$, independent of the VMD model, is considered.
A new flavor is introduced into these investigations by lattice QCD 
studies of the HAL-QCD collaboration. In \cite{Lyu:2024ttm}, low-energy
interactions $N-J/\psi$ and $N-\eta_c$ are based on (2+1) flavor 
configurations with a nearly physical pion mass, $m_\pi = 146$~MeV. 
The interactions are found to be attractive in all distances and 
possess a characteristic long-range tail consistent with the two-pion 
exchange potential.

\subsection{Electroproduction of Single Pseudoscalar Mesons}
\begin{figure}[ht]
	\centering
	\includegraphics[width=3.0cm]{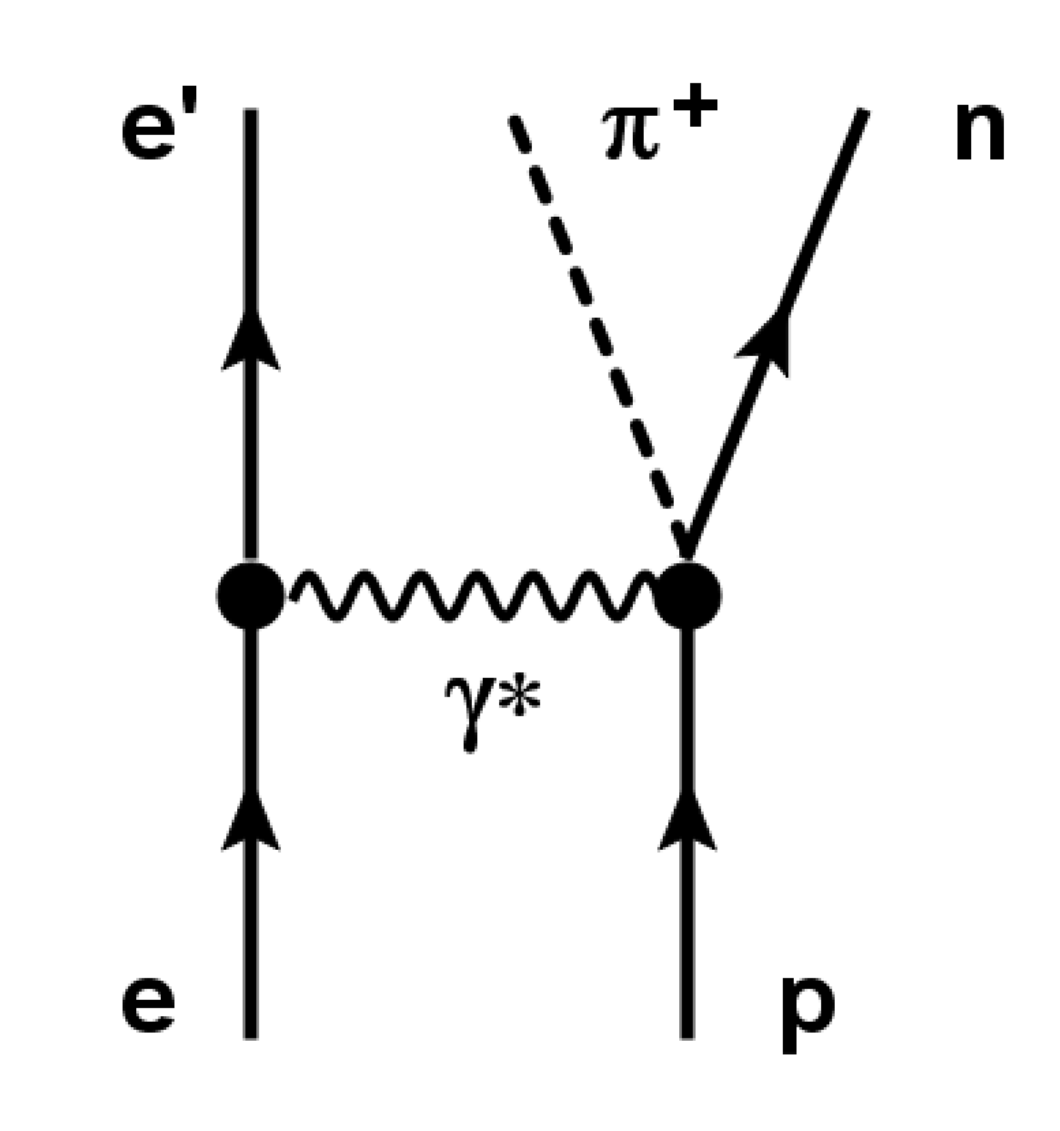}
	\caption{Feynman diagram illustrating electroproduction of a $\pi^+$ meson on a proton by an incident electron through exchange of a virtual (off-shell) photon $\gamma^*$ transferring the off-shell four-momentum $q=(\w,\mathbf{q})^T$.}
	\label{fig:electroProd}
\end{figure}

 An important spectroscopic tool, complementary to photoproduction, is the electroproduction of mesons on the nucleon. As indicated in Fig.~\ref{fig:electroProd}, electron-induced meson production proceeds by the exchange of a virtual photon, which allows for the transfer of energy and momentum beyond the on-shell constraints inherent to photoproduction.  

\begin{figure}[htb!]
\centering
{
    \includegraphics[width=0.225\textwidth,keepaspectratio]{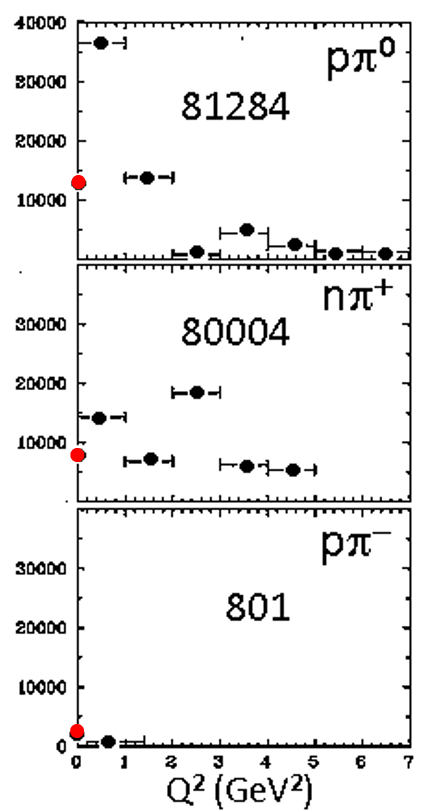} 
}

\centerline{\parbox{\textwidth}{
\caption[] {\protect\small
$Q^2$ distribution of pion electroproduction data, which are now available~\cite{Arndt:2009nv}.
Numbers in the middle of each plot corresponded to the number of events. Red-filled circles show the number of pion photoproduction events.
}
\label{fig:epr} } }
\end{figure}

Thus, electroproduction data are an important tool for studying the properties of non-strange baryons simultaneously in energy and three-momentum transfer as independent variables. Therefore, ongoing PWA fits incorporate the available electroproduction data. The map of $Q^2$ dependence of the pion electroproduction data ($\gamma^\ast p\to \pi^0 p$, $\gamma^\ast p \to \pi^+ n$, and $\gamma^\ast n\to \pi^-p$, no data for $\gamma^\ast n \to \pi^0n$) before 2009 is shown in Fig.~\ref{fig:epr}. One notes that the CLAS Collaboration produced more than 85\% of the world's pion electroproduction data (this database is still growing), much of which was focused on mapping the properties of the $\Delta(1232)$ and higher resonances. Useful comparisons, for instance, will require those involved in this effort to make available all amplitudes obtained in any new determination of ratios $R_{EM}$ and $R_{SM}$ for the transition $N\to\Delta(1232)$, which may be compared with LQCD calculations. Their values are far from those expected in the perturbative regime, $R_{EM} = 1$, and $R_{SM}$ are $Q^2$ independent~\cite{CLAS:2019cpp}.


%% file: DoublePion.tex
\section{Double-Pion Production on the Nucleon} \label{sec:TwoPion}
\subsection{Overview on Double-Pion Production by Pion Beams}
\begin{figure}[htb!]
\centering
{
    \includegraphics[width=0.3\textwidth,keepaspectratio]{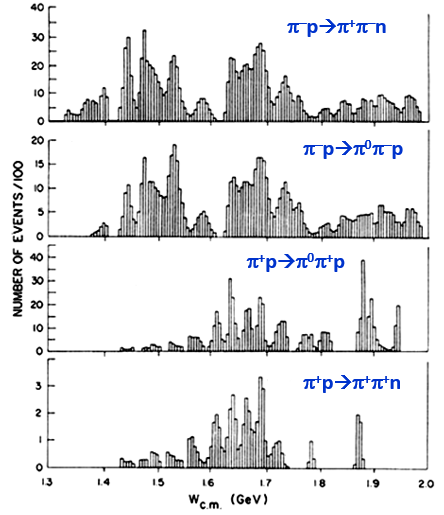} 
}

\centerline{\parbox{\textwidth}{
\caption[] {\protect\small
Summary of the number of Bubble Chamber events for reactions $\pi p\to \pi\pi N$ analyzed at each energy.
}
\label{fig:pi2pi} } }
\end{figure}

Early studies of the dynamics of the $N\pi\pi$ final state were based on bubble chamber data on the reaction $\pi p\to \pi\pi N$ collected in non-polarized experiments by the Berkeley, Saclay, and Rutherford laboratories (Fig.~\ref{fig:pi2pi}). 241,214 Bubble Chamber events have been analyzed in Isobar-model PWA at $W = 1320-1930~\mathrm{MeV}$ by the Virginia Tech group led by Richard Arndt~\cite{Manley:1984jz}. A summary of the number of Bubble Chamber events is given in Fig.~\ref{fig:pi2pi}.

Partial-wave inelasticities for elastic pion-nucleon scattering were determined with the aid of experimental data on $\pi N\to \pi \pi N$ processes in the beam-momentum range $300~\mathrm{MeV/c} < \mathrm{P_{beam}} < 500~\mathrm{MeV/c}$~\cite{Kozhevnikov:2008zz}. Respective partial wave results are displayed in Fig.~32.

However,  an obstacle for a much-desired joint PWA analysis is the inconsistencies that, at present, inhibit merging $\pi N\to \pi N$ and $\pi N\to \pi \pi N$ databases. Problems exist specifically for small inelasticities (Fig.~32)
While the largest inelastic cross section related to the $P_{11}$ Roper resonance is in excellent agreement, it is well described in SAID-SP06 for $\pi N\to \pi N$~\cite{Arndt:2006bf}, see Fig.~32.
\begin{figure}[htb!]
\centering
{
    \includegraphics[width=8cm]{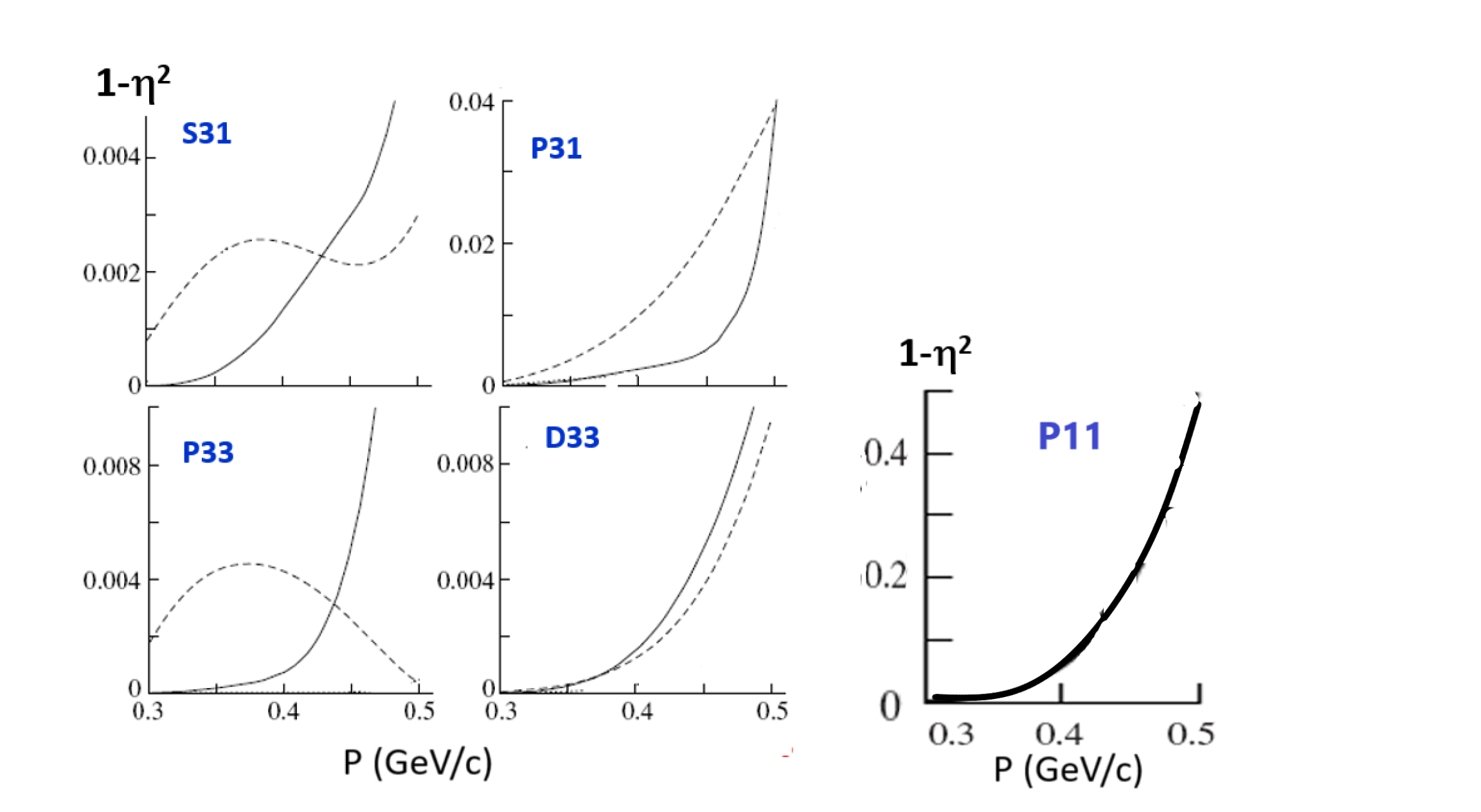}
}
\centerline{\parbox{\textwidth}{
\caption[] {\protect\small
Partial-wave inelasticities $(1 - \eta^2)$ from the $\pi\to 2\pi$ analysis~\cite{Kozhevnikov:2008zz} (dotted curves) and $\pi N$ elastic scattering~\cite{Arndt:2006bf} (solid curves).
\underline{Left}: $S_{31}$, $P_{31}$, $P_{33}$, and $D_{33}$. \underline{Right}: $P_{11}$.}
}
}\label{fig:Etasq}  
\end{figure}

Besides photo-production and  electro-production on the nucleon \cite{Manley:1984jz, Arndt:2006bf, Arndt:2009nv, Aznauryan:2005tp} 
double-pion production was also measured systematically in NN scattering at WASA@CELSIUS and WASA@COSY~\cite{CELSIUSWASA:2008dnf, Bashkanov:2009zz, WASA-at-COSY:2011bjg}.

Theoretically, double-pion channels are of high interest for $\chi$EFT. In~\cite{Fettes:1999wp}, the J\"ulich group has been studying pion production off nucleons in heavy baryon chiral perturbation theory to third order in the chiral expansion, aiming at the determination of the low–energy constants. Most of the available 
differential cross sections and angular correlation functions at low pion incident energies could be described together with
total cross sections at higher energies. The contributions from the one-loop graphs were found to be essentially negligible once the dominant terms 
related to pion–nucleon scattering graphs with one pion
were added at the second and third orders. An interesting aspect is that the $\pi \pi N$ channels provide the possibility of extracting the pion–pion $S$–wave scattering
lengths, which are otherwise hard to access.

Present coupled-channel models for double-meson production involve approximations, \textit{e.g.}, describing three-body interactions by effective two-body configurations. The caveats are well understood on a formal level, but their numerical implementation is a highly non-trivial task. Improvements and optimization strategies are ongoing tasks in coupled-channel research, as is a deeper look into the developments documented in the discussions of competing approaches, \textit{e.g.},~\cite{Dosch:1968mv, Fettes:1999wp, Fix:2005if, Huang:2011as, Haberzettl:2011zr, Shklyar:2014kra, Anisovich:2010mks}. A practical approach is discussed in the following paragraph. 

\subsection{Double-Pion Production in Pion-Proton Reactions}
\begin{figure}
  \begin{center}
\includegraphics[width=11cm]{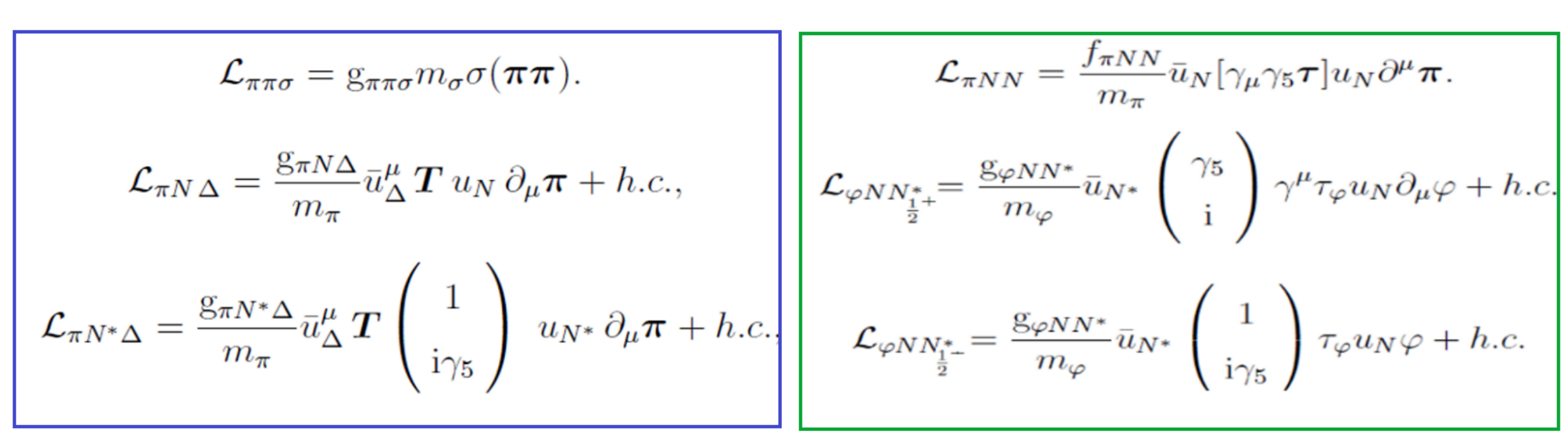}
       \caption{
The GiM Lagrangians underlying two-pion production on the nucleon. In the left column, the Lagrangian for
$\pi\pi\leftrightarrow \sigma$, $\pi N \Delta$, and $\pi N^*\Delta$ interactions are shown, where in the bottom row the upper and lower choice of operator relates to $N^*$ of $J^\pi=\frac{1}{2}^\pm$. In the right column, the $\pi NN$ Lagrangian and the Lagrangians of mesons 
$\varphi \neq \pi$ interacting with a nucleon and $J^\pi=\frac{1}{2}^\pm$ resonances are displayed. The upper and lower choice of operator relates to mesons of negative and positive intrinsic parity, respectively. }
      \label{fig:2piLagrange}
  \end{center}
\end{figure}

\begin{figure}
  \begin{center}
    \includegraphics[width=8cm]{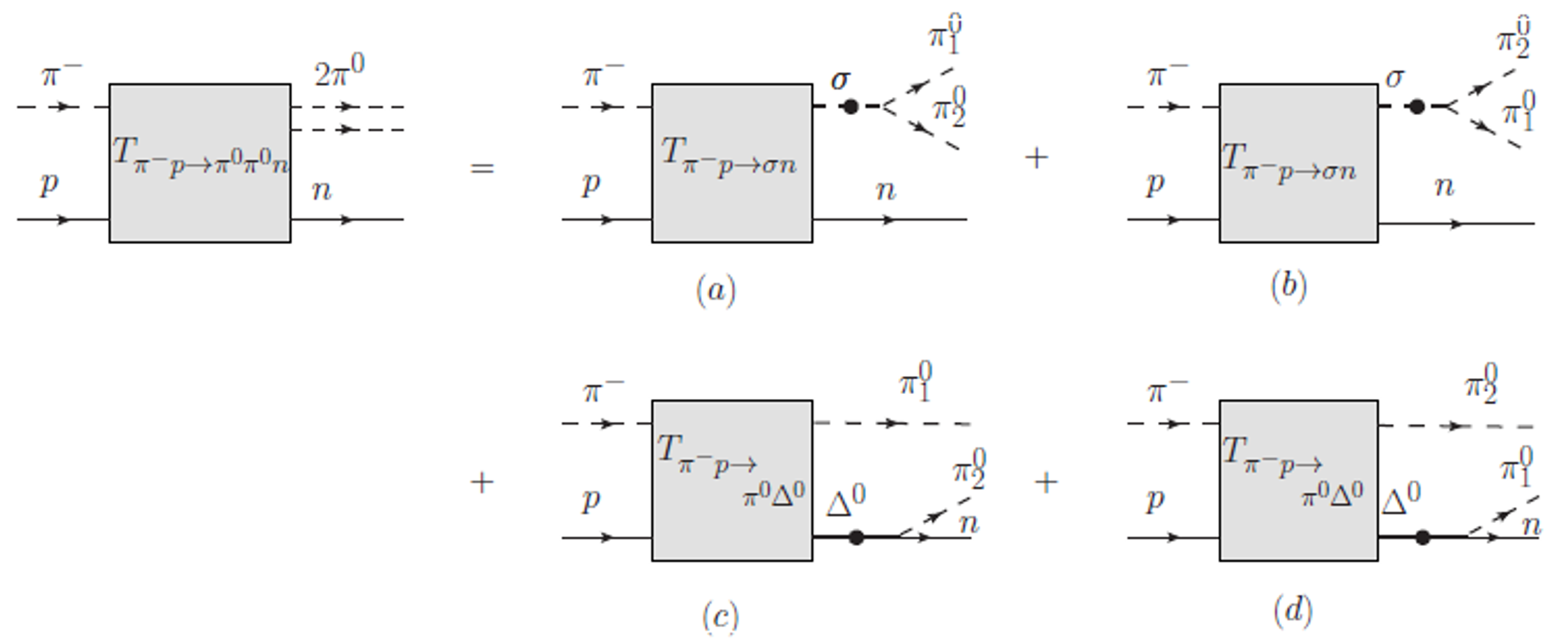}
       \caption{
The processes contributing to double-pion production T-matrix are depicted diagrammatically: (a) and (b) production through the $\sigma$-isobar, (c) and (d) production through the $\Delta^0$-isobar. Symmetrization of the exit channels is indicated in, see (b) and (d).}
      \label{fig:2piD}
  \end{center}
\end{figure}

In certain energy regions, the $\pi N \to 2\pi N$ reaction accounts for up to 50\% of the $\pi N$ inelasticity. Therefore, this production channel must be included in any CC-PWA approach, as \textit{e.g.}, practiced in GiM Ref.~\cite{GiM:2016, Lenske:2018bgr}. The Lagrangians of the interactions considered in double-$\pi^0$ production 
are displayed in Fig.~\ref{fig:2piLagrange}. In view of the foregoing complexities, physically meaningful approximations are necessary. The guiding principle is to retain the essential dynamical aspects while making numerical calculations feasible. This goal is reached by an isobar description of intermediate two-pion configurations and their decay into the final double-pion states on the mass shell. The processes contributing to the T-matrix of 
$\pi^0\pi^0$ pair production on the nucleon in that energy region are depicted in Fig.~\ref{fig:2piD}.

This approach allows for the direct analysis of the $2\pi N$ experimental data. Since the corresponding Dalitz plots have a strongly non-uniform structure, it is natural to assume that the main effect of the reaction comes from the resonance decays into isobar sub-channels~\cite{Manley:1984jz}.
The most important contributions are expected to be from the intermediate $\sigma N$, $\pi \Delta$,  and $\rho N$ states. The analysis of the $\pi N\to 2\pi N$ reaction would, therefore, provide very important information about the resonance
decay modes into different isobar final states. The much richer baryon spectrum found in LQCD simulations~\cite{Edwards:2011jj, Duerr:2008}, the functional DSE
approaches~\cite{SanchisAlepuz:2011jn}, and the CQM results~\cite{Koniuk:1979vy, Capstick:1998uh} that are observed in scattering experiments indicate the necessity for broader investigations, including a larger class of reaction channels. Experimentally,  most of the non-strange baryonic states have been identified from the analysis of the elastic $\pi N$ data~\cite{Arndt:2006bf, Cutkosky:1979fy}.
As pointed out in~\cite{Koniuk:1979vy}, the signal of excited states with a small  $\pi N$ coupling will be suppressed in the elastic  $\pi N$ scattering.
As a solution to this problem, a series of photoproduction experiments has been conducted to accumulate enough data for the study of the nucleon excitation spectra.
However, the results from the photoproduction reactions are still controversial. While recent investigations of the photoproduction reactions presented by the BnGa
group~\cite{Anisovich:2011fc} reported indications of some new resonances, not all of these states are found in other calculations~\cite{PDG:2016}.
This raises a question about the independent confirmation of such states from the investigations of other reactions.

Because of the smallness of the electromagnetic couplings,   the largest contribution to the resonance self-energy comes from the hadronic decays.
If the $N^\ast\to \pi N$ transition is small, one can expect a sizable resonance contribution to the remaining hadronic decay channels.
As  a result, the effect of the resonance with a small
$\pi N$ coupling could still be significant in inelastic pion-nucleon scattering. here, the smallness of the resonance coupling to the
initial $\pi N$ states could be compensated for by the potentially large decay branching ratio to other inelastic final states. Such a scenario is realized, \textit{e.g.}, in the case of the well-known $N^\ast(1535)$
state. While the effect of this resonance on the elastic $\pi N$ scattering is only moderate at the level of the total cross section, its contribution to the $\pi N \to\eta N$ channel turns out to be
dominant~\cite{Shklyar:2012js}. Since the $\pi N \to 2\pi N$ reaction could account for up to 50\% of the total  $\pi N$ inelasticity, this channel becomes
very important not only for the investigation of the properties of already known resonances, but also for the search for the signals of possibly
unresolved states.

\begin{figure}
  \begin{center}
    \includegraphics[width=9cm]{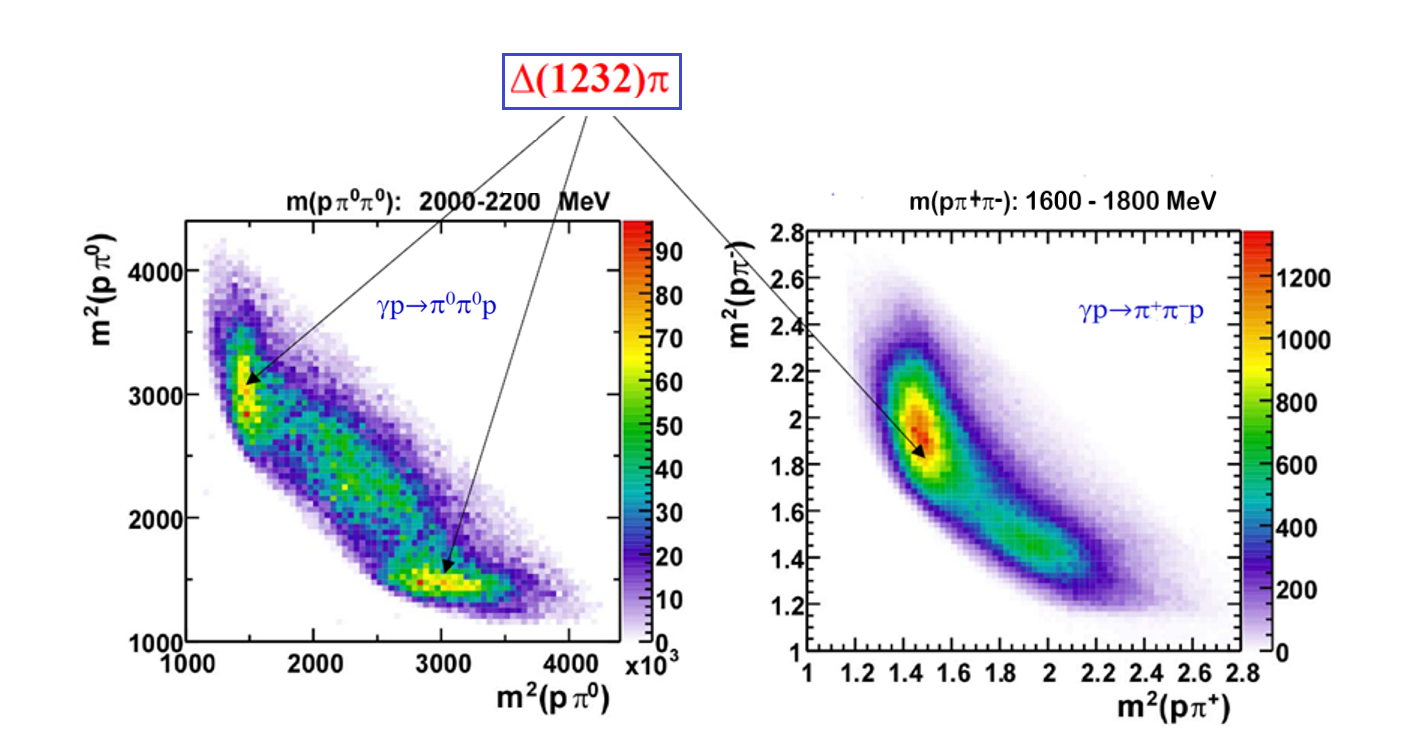}
       \caption{Double-pion production through excitation of the $P_{33}(1232)$ Delta-resonance. The Dalitz plots were generated from CBELSA (left) and CLAS (right) photo-production data .}
      \label{fig:2piDalitz}
  \end{center}
\end{figure}

Another important issue in studies of the $2\pi N$  channel is the possibility of investigating
cascade transitions like ${N^\ast}'\to\pi N^\ast\to \pi\pi N$, where a massive state ${N^\ast}^\prime$ decays via intermediate  excited
$N^\ast$ or $\Delta^\ast$. The Lagrangians shown in Fig. \ref{fig:2piLagrange} contain the proper interactions. Experimentally, such processes are clearly observed, as seen in Fig.~\ref{fig:2piDalitz} where Dalitz plots are shown for two-pion photo-production through the excitation of the $P_{33}(1232)$ Delta-resonance. It is interesting to check whether such decay modes are responsible for the large decay width of higher-lying mass states. So far, only the $\pi N^\ast(1440)$ isobar channel has been considered in a partial wave analysis (PWA) of the $\pi N\to 2\pi N$  experimental data~\cite{Manley:1984jz}.

The coupled-channel analysis of $2\to 3$ transitions is confronted with several complications.
The first one is the difficulty of performing the partial-wave decomposition
of the three-particle state.
The second complication is related to the issue of three-body unitarity.
For a full dynamical treatment of the $2\to 3$ reaction, the  Faddeev equations must be solved.
Although appropriate theoretical and numerical methods are known, the effort inhibits practical implementation.
Here, a coupled-channel approach for solving the $\pi N \to 2\pi N$ scattering problem in the isobar approximation is used, as is widely practiced. In this formulation, the $(\pi/\pi\pi) N\to  (\pi /\pi\pi) N$  coupled-channel equations are reduced to effective two-body scattering equations, taking advantage of intermediate two-body isobar production.  Such a description accounts, by construction, for the full spectroscopic strength of intermediate channels and, in addition, provides a considerable
numerical simplification. 

\begin{figure}
  \begin{center}
    \includegraphics[width=7cm]{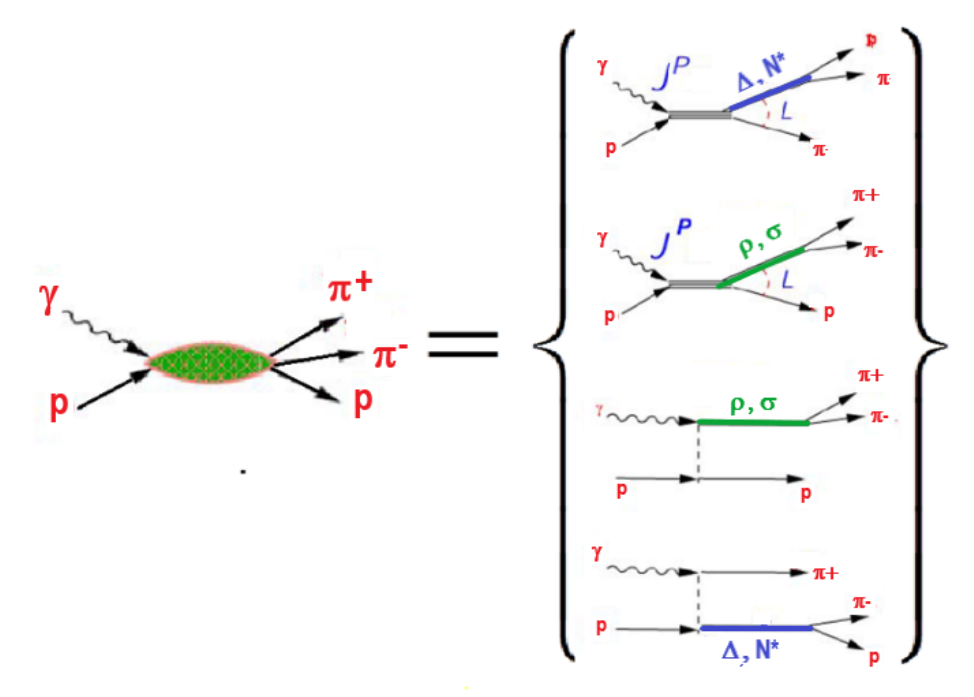}
       \caption{
Overview of elemental Feynman diagrams considered in the calculation of the 2-to-3-body amplitudes in double-pion production reactions. The incident photon may be replaced by an incoming hadron. }
      \label{fig:2piS}
  \end{center}
\end{figure}

Three-body unitarity leads to a relation between the imaginary part of the
elastic scattering amplitude and the sum of the total elastic and inelastic cross sections, as described by the well-known optical theorem. Since, in the isobar approximation, the pions in the $\pi\pi N$  channel are produced from 
isobar sub-channels, all contributions to the total $\pi N \to \pi\pi N$ cross section are driven by isobar 
production. The optical theorem can be fulfilled if all discontinuities in isobar sub-channels are taken into
account. In the present work, three-body unitarity is maintained up to
interference terms between the isobar sub-channels.  
The calculation of the 2-to-3-body amplitude within this scheme is illustrated in Fig. \ref{fig:2piS}.

The first resonance energy region is of particular interest because of the sizable effect from $N^\ast(1440)$.  The dynamics of the Roper resonance turn out to be rich because of the two-pole structure reported in earlier studies~\cite{Arndt:1985vj, Cutkosky:1990zh}, (see~\cite{Arndt:2006bf,  Doring:2009yv, Suzuki:2009nj} for the recent status of the problem).
The origin of the Roper resonance is also a matter of controversy. For example, the calculations
in the J\"ulich model, explain this state as a dynamically generated pole due to the strong attraction in the $\sigma N$ sub-channel.
At the same time, the Crystal Ball Collaboration finds no evidence of strong $t$-channel
sigma-meson production in their $\pi^0\pi^0$ data~\cite{Craig:2003yd}.
From the further analysis of the $\pi^0\pi^0$ production, the effect of the sigma meson was found to be small~\cite{Prakhov:2004zv}.
On the other hand, the $pp\to pp\pi^0 \pi^0$ scattering experiment by the CELSIUS-WASA
collaboration~\cite{Skorodko:2008zzb} finds the $\sigma N$ decay mode of the Roper resonance to be dominant.
\begin{figure}
  \begin{center}
        {\includegraphics*[width=7cm]{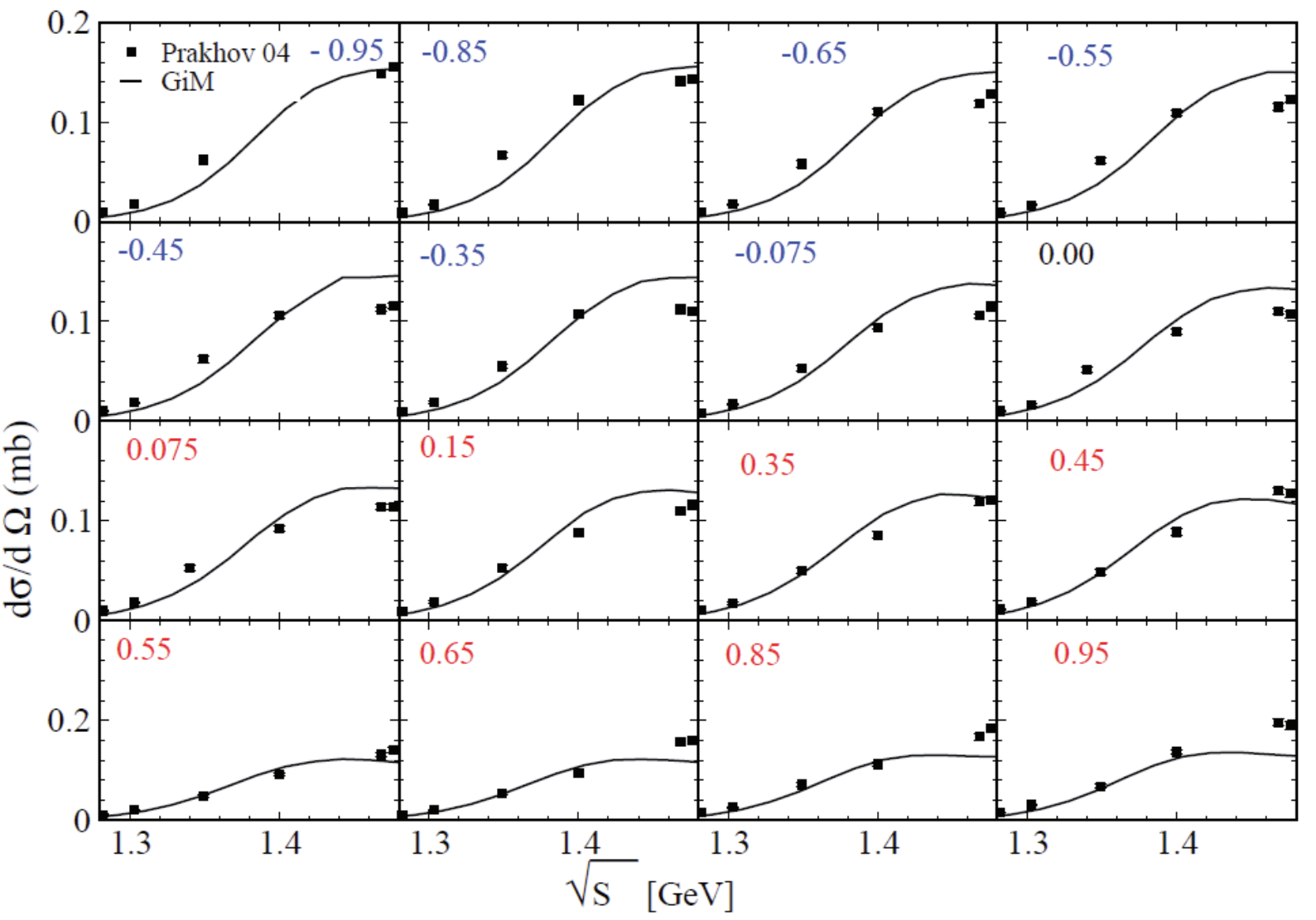}}
       \caption{$\pi^0\pi^0$ differential cross sections for the reaction $\pi^-p \to\pi^0\pi^0n$ at fixed $t_{\pi^0\pi^0}=\cos{\theta_{\pi^0\pi^0}}$, shown in the upper left corner of the panels. Energy distributions for $-0.95\leq t_{\pi^0\pi^0} \leq +0.95$ are shown. The experimental data are from~\cite{Prakhov:2004zv} The figure is
        adapted from Ref.~\cite{GiM:2016}.}
      \label{fig:2pidsig}
  \end{center}
\end{figure}

In the region of the Roper resonance, the calculations are able to describe the mass distributions quite satisfactorily.
Also, in this region, the production strength shifts to higher invariant masses $m_{\pi^0\pi^0}^2$.
At the same time, a peak at small  $m_{\pi^0\pi^0}^2$
also becomes visible. In these calculations, the fit tends to decrease.
The magnitude of the $ \pi\Delta$ production compensates for it by enhancing the strength to $\sigma N$.
The obtained decay branching ratio of $N^\ast(1440)$   for the $\sigma N$ channel is about twice as large as that for the $\pi\Delta$.

\begin{figure}
  \begin{center}
        {\includegraphics*[width=7cm]{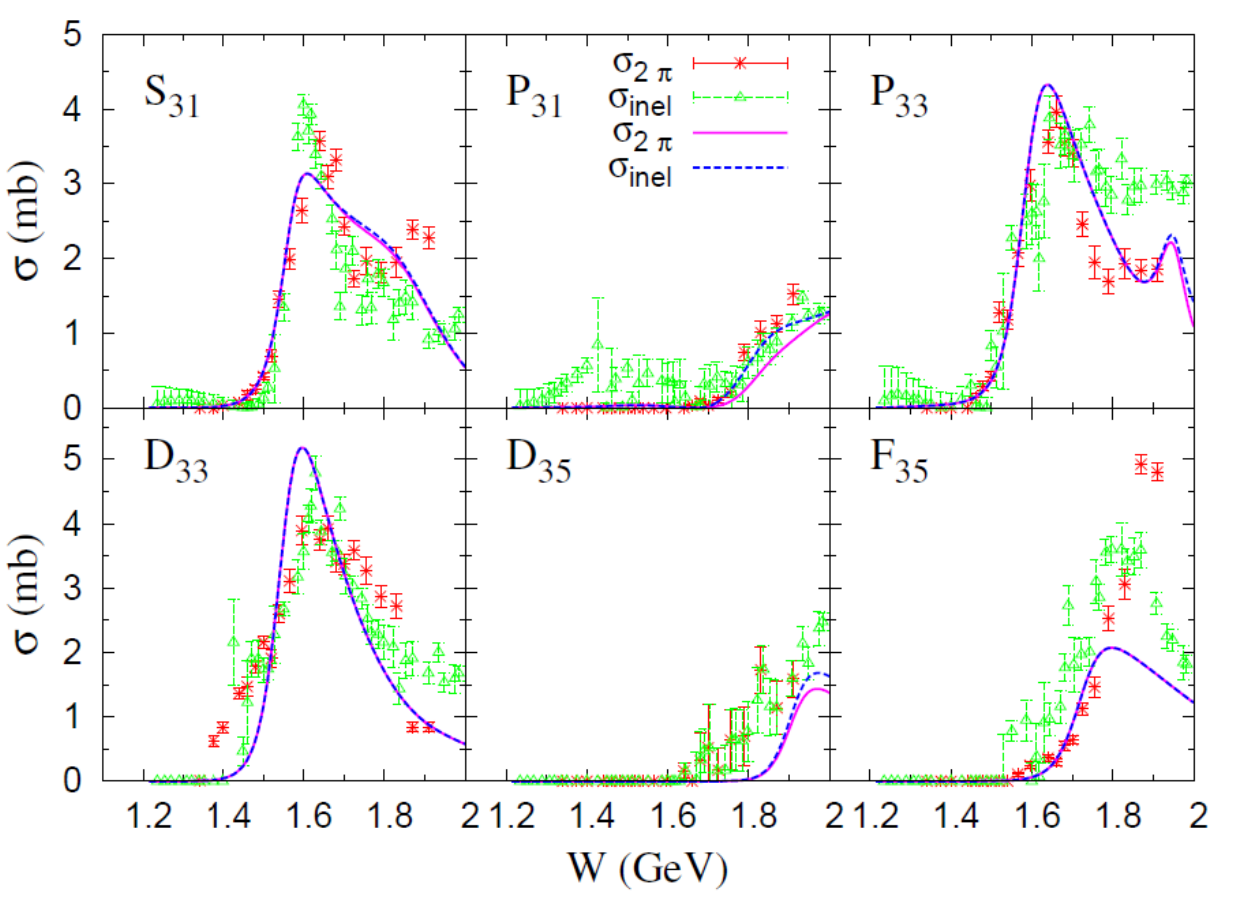}}
       \caption{Two-pion production partial wave cross sections comparing GiM results to the Manley and GWU analyses as indicated.
        Adapted from Ref.~\cite{GiM:2016}.}
      \label{fig:2piPW}
  \end{center}
\end{figure}

Both the small peak at small invariant masses and the broad structure at large invariant masses are well reproduced, indicating an important interplay between the $\sigma N$ and $\pi \Delta$ production mechanisms.
Interestingly, the isoscalar correlations in the $\pi\pi$ rescattering are also found
to be necessary in order to reproduce the asymmetric shape of the mass distributions.
Though the $\pi\Delta$ production gives rise to a two-peak structure, only the first one at small  $m_{\pi^0\pi^0}^2$ is visible at energies $1.4-1.468~\mathrm{GeV}$. The second peak at high  $m_{\pi^0\pi^0}^2$ is not seen because of the
large $\sigma N$ contributions. In the current approaches
$\pi^0 \pi^0 n$ production is calculated as a coherent sum of isobar contributions. Though the interference effects are important,
they are found to be very small at the level of the total cross sections.

To simplify the analysis, the  $S_{11}$ and $P_{11}$  $\pi N$ partial waves  are directly constrained by the single energy solutions (SES)
derived by GWU(SAID)~\cite{Arndt:2006bf}. The experimental data on the $\pi^- p \to\pi^0\pi^0 n$ reaction are taken from~\cite{Prakhov:2004zv}. These measurements provide high-statistics data on  the angular distributions ${\rm d} \sigma 
/{d\Omega_{\pi\pi}}$, where $\Omega_{\pi\pi}$ is the scattering angle of the $\pi\pi$ pair (or the  final nucleon in c.m.).

The calculated  $\pi^0\pi^0$ differential cross sections are shown in Fig.~\ref{fig:2pidsig} and compared to the Crystal Ball data as a function of the c.m. energy. The measurements demonstrate a rapid rise in the cross sections at the energies $1.3-1.46~\mathrm{GeV}$, indicating a strong contribution from the Roper resonance, as also found in the GWU(SAID)~\cite{Arndt:2006bf} analysis.

The invariant $\pi^0\pi^0$ mass distributions play a crucial role in the separation of the isobar contributions.
The  $\pi^- p \to \pi^0\pi^0 n$ reaction close to the threshold is dominated by the
$\sigma N$ production due to the $t$ channel pion exchange. The nucleon Born term contribution to the $\pi\Delta$ channel is found to be less significant. The decay branching ratios of $N^\ast(1440)$ are obtained as  $R_{\sigma N}^{N(1440)}=27^{+4}_{-9}\,\%$ and  $R_{\pi \Delta}^{N(1440)}=12^{+5}_{-3}\,\%$.

The parameters extracted independently using the different approach of the BnGa group~\cite{Sarantsev:2007aa} are $R_{\sigma N}^{N(1440)}=17^{+7}_{-7}\%$ and  $R_{\pi \Delta}^{N(1440)}=21^{+8}_{-8}\%$, demonstrating that, despite the visible differences in the central values, these quantities still coincide within the error bars.

In Fig.\ref{fig:2piPW}, two-pion partial wave cross sections are compared for three
different CC approaches. Although the GiM approach differs in detail considerably from the Manley and GWU-SAID descriptions, the results are in favorable overall agreement, thus confirming that the essential physics features are covered in the CC scheme. 

\begin{figure}
  \begin{center}
    {\includegraphics*[width=7cm]{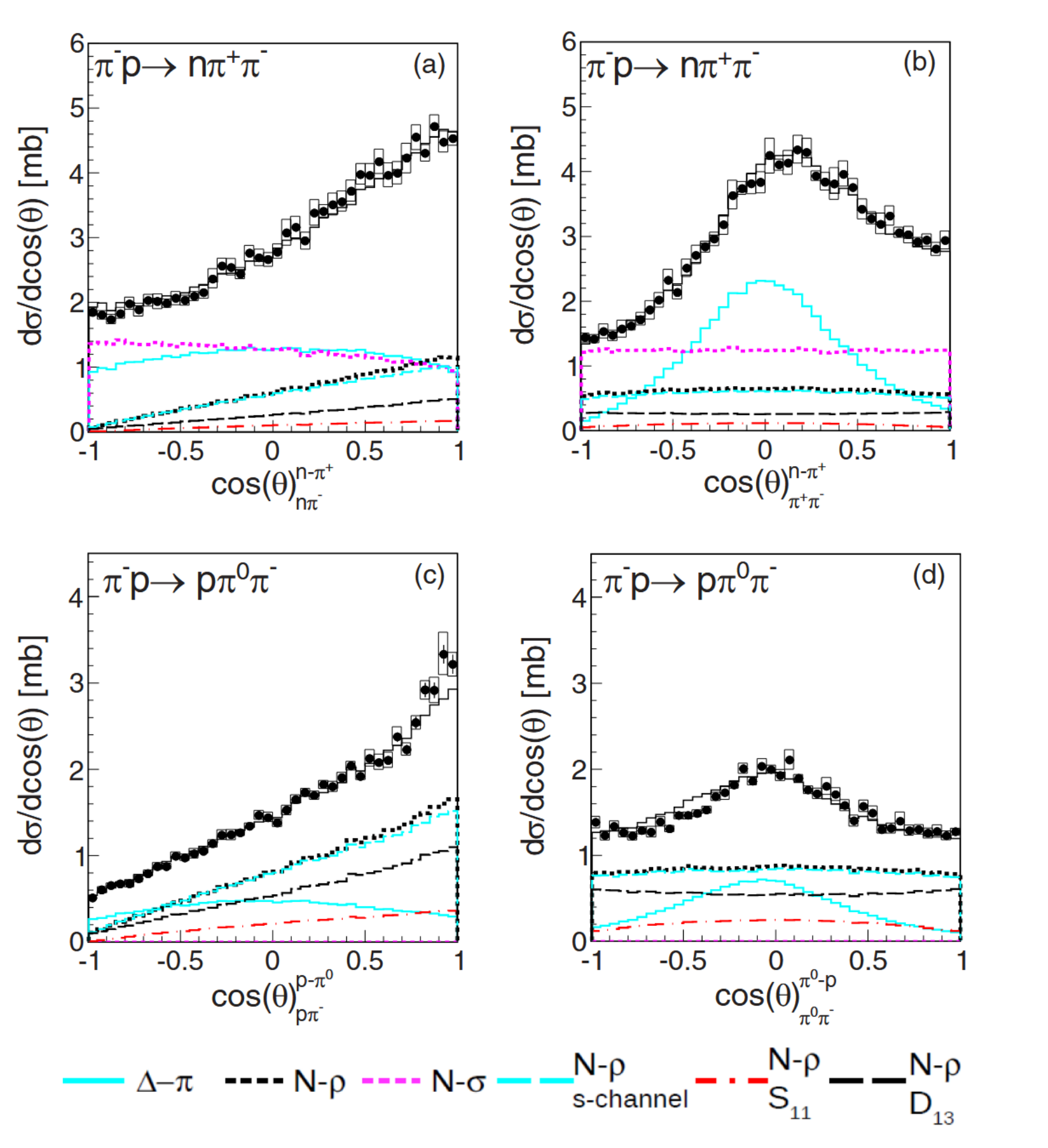}}
\caption{Angular distributions of pions in the nucleon-pion (left) and nucleons in the pion-pion (right) helicity frames for the 
$\pi^-p \to h\pi^+\pi^-$
(upper) and $\pi^-p \to p\pi^0\pi^-$
 (lower) reaction channels. The helicity frame is indicated by the subscript labels, while the angle
between the given particles in that frame is denoted by superscript labels. The z-axis of the helicity frame is chosen opposite to the neutron (upper panel) and the proton (lower
panel) directions. Color (gray) curves display various final-state contributions (indicated in the legend).
 The figure is
        adapted from Ref.~\cite{HADES:2020kce}.}
      \label{fig:HADES}
  \end{center}
\end{figure}
     
Double-pion production has been part of the research program of the HADES@GSI experiment. 
The GSI pion beam facility was used to measure excitation functions and angular distributions for $\pi^-p \to h\pi^+\pi^-$
and $\pi^-p \to h\pi^0\pi^-$ reactions  at four different pion beam momenta:
0.650, 0.685, 0.733, and $0.786~\mathrm{GeV/c}$~\cite{HADES:2020kce}. The primary goal was to study the very specific question of which of the resonances in the second resonance region around $1600~\mathrm{MeV}$ are dominant in $N\rho$ interactions.  In Fig.~\ref{fig:HADES}, angular distributions in the $n\pi^+\pi-$ and $p\pi^0\pi^-$ final channels are shown. The BnGa framework was used to study the partial wave and the isotopic channel composition of the measured angular distributions. Two major conclusions are drawn: First, the  double-pion cross section  is influenced by interference
effects between isospin $I = 1/2$ and $I = 3/2$ amplitudes, which are
constructive for $n\rho^0$ and destructive in the $p\rho^-$
case. Second, identify $N(1520)\frac{3}{2}^-$ with a branching ratio 
of $12,2\%$ as the dominant state for $N\rho$ self-energies.    

\subsection{Double-Pion Production and Polarization}
An important set of observables on hadron structure and production dynamics is obtained from polarization measurements; see, \textit{e.g.}, the discussion around Fig.~\ref{fig-3}. For single-pion production, such measurements are conducted at practically all hadron facilities and have become routine work in most charged current (CC) approaches.   

The situation is different, however, for double-pion production. The probably only double-pion production experiment, including polarization measurements and subsequent theoretical analysis, was conducted some time ago by Alekseev \textit{et al.}~\cite{Alekseev:1998tm} at ITEP (Moscow). Data were taken for the reaction $\pi^- \vec{p}\to \pi^+ \pi^-+p$ on polarized targets and at a beam momentum of $1.78~\mathrm{Gev/c}$. By means of the SPIN spectrometer, designed especially for measurements of polarization observables in reactions leading to charged two-
and three-particle final states, the full set of 14 spin observables could be measured. The primary aim of the experiment was to study specifically pion-pion interactions in the mass region of the $\rho(770)$ meson, while aiming to narrow down the mass of the - until today - heavily disputed mass (and width) of the iso-scalar scalar $\sigma/f_0(500)$-meson (see Fig.~\ref{fig:Nonets} and the listings in the 2024 issue of the Review of Particle Physics~\cite{PDG:2024cfk}). 

The authors of Ref.~\cite{Alekseev:1998tm} could indeed extract highly valuable information from the data on the $\pi^+\pi^-$ S-wave scattering phase shift in a PWA by including the polarization observables. However, the data obtained from that single measurement were not sufficient to determine the desired mass parameter unambiguously. Still, that - until today - single case experiment showed the important gain in information on the dynamics and spectroscopy of a two-body sub-system within the three-hadron final state, populated in double-pion production on the nucleon.    

\subsection{Double Pseudoscalar Meson Production Beyond Pions}
Double-kaon or $\eta$-meson production, as well as vector meson (photo) production, is a task that is highly demanding, both experimentally and theoretically. In a recent note, the ALICE collaboration announced the first data on the pion and kaon pair 
production~\cite{Schicker:2024dga}. From their mass spectra, they concluded that the states observed in the mass region of $2~\mathrm{GeV}$ are well described by CQM calculations in the tradition of Karl and Isgur.  

There is also ongoing research activity on the theoretical side. 
In~\cite{Kang:2024fsf}, the authors presented a new approach describing double pion photoproduction off the nucleon in covariant chiral perturbation theory, thus connecting two important concepts of hadron physics.  

The recent BnGa paper reported a combined analysis of photo- and pion-induced double pion production~\cite{Sarantsev:2025lik}. Recent CBELSA/TAPS and CLAS (photo case) and Crystal Ball at BNL and HADES at CERN (pion case) collaboration data, plus $\pi N$ PWA amplitudes from the SAID and Karlsruhe-Helsinki groups, were involved in the analysis.  The critical motivation of this study is the separation of $\pi\Delta$, $\rho N$, and $f^0(500)N$ final states, because the experimental state for each case is the same: namely, two pions and a nucleon. 
The CLAS experiment recently reported double-pion electroproduction on the proton and on deuterium~\cite{CLAS:2023mfc}, thus expanding the research into the domain of finite momentum transfer.


%% file: GiMStrangeEta.tex
\subsection{Pion Production Spectra and Resonance Spectrum of the Nucleon}\label{sec:YRDiscussion}
A natural and perhaps the most important question in hadron physics is how well the experimental spectrum is described by theory. 

Here, we address this issue in the context of the GiM coupled channels approach, which may serve as a representative example for this class of models. As mentioned before, the GiM is based on a phenomenological covariant field theory of baryons, mesons, and their interactions. By construction, the fundamental symmetries of QCD, including chiral symmetry, are conserved. Resonance and background contributions are generated consistently from the tree-level interactions as defined by the underlying Lagrangian. 

\begin{figure}
  \begin{center}
    \includegraphics*[width=9cm]{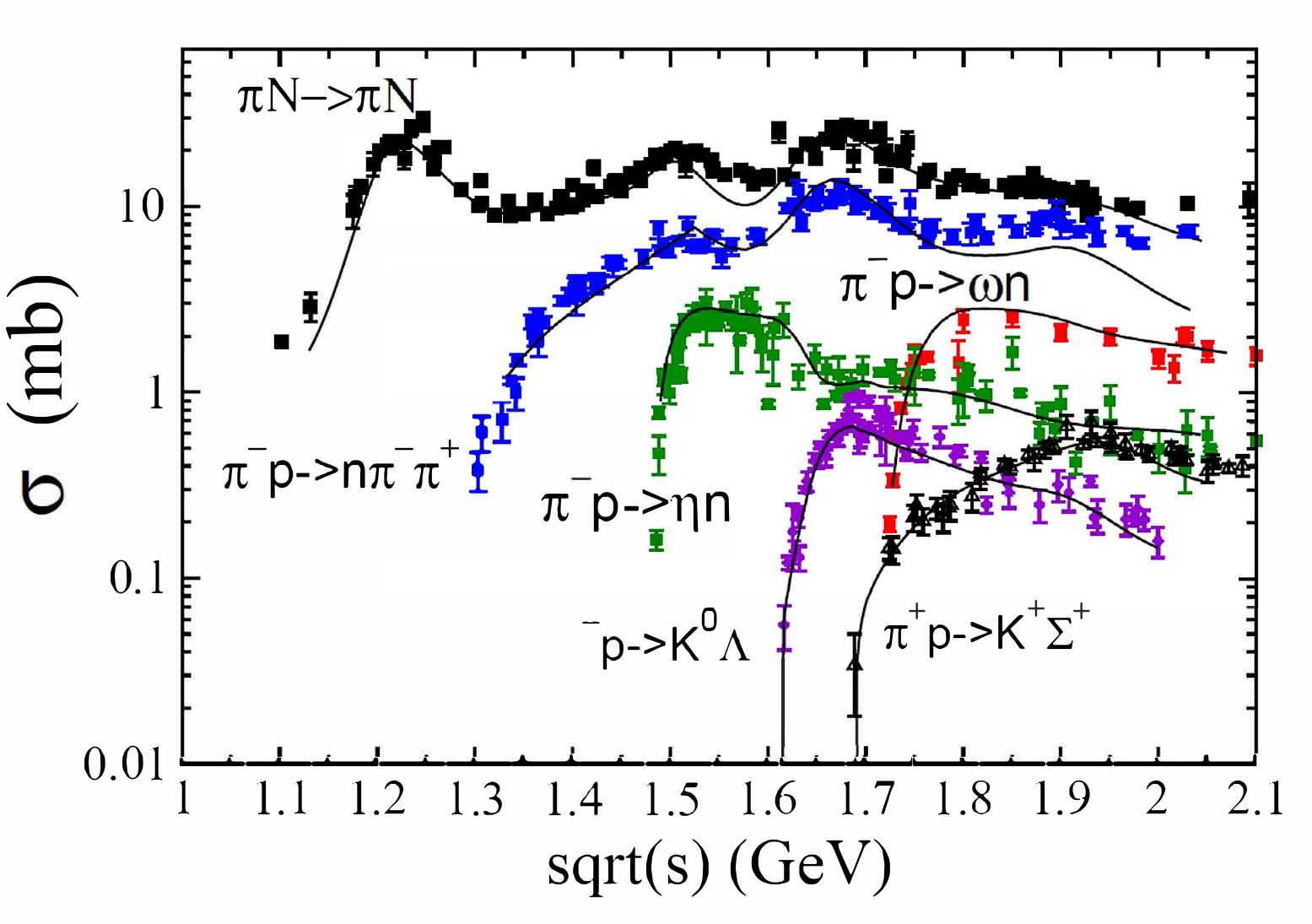}
       \caption{Comparison of GiM CC-results (lines) to the data set of measured meson production spectra (symbols) observed in $\pi + p$ reactions. 
      \label{fig:PionProd_Data}}
  \end{center}
\end{figure}

The scattering amplitudes are determined by a linear system of coupled equations, which is solved numerically in the K-matrix approximation and partial wave representation. High-spin resonances are incorporated in a gauge-invariant manner. Applications to selected reaction channels have been presented, ranging from single pion, eta,  and kaon production to double-pion production to investigations of omega and $K^\ast$ vector meson channels.

Hence, CC approaches like the GiM and the other aforementioned projects incorporate the defining elements of hadron physics at both the theoretical and numerical levels. The approaches use comparable numerical methods for solving the CC problem, where attention is paid to the fact that the numerical procedures conserve the basic symmetries and conservation laws.  

A direct reliability test of CC models is the comparison of numerical results to data recording the spectra in the various channels under theoretical scrutiny. In Fig.\ref{fig:PionProd_Data} a variety of total meson production cross sections observed in pion-proton reactions  are compared to results of GiM CC calculations. Overall the large set of divers data is quite well described, hence confirming convincingly the validity of CC approaches like the GiM to hadron production.  

\begin{figure}
  \begin{center}
    \includegraphics*[width=9cm]{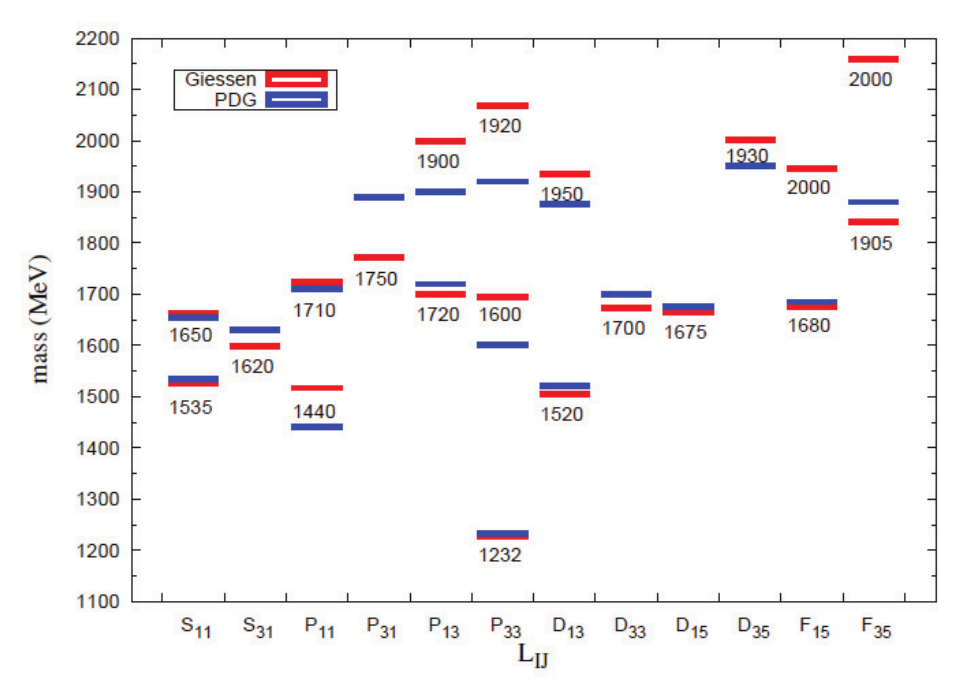}
       \caption{Comparison of the GiM $N^\ast$ resonance level scheme (red bars) in various partial waves to the PDG compilation~\cite{PDG:2016} (blue bars). 
      \label{fig:GiM_Levels}}
  \end{center}
\end{figure}

The GiM baryon level scheme resulting from the investigations of meson production channels is summarized in Fig.~\ref{fig:GiM_Levels} and compared to the observed spectrum, as found in the  PDG resonance compilations~\cite{PDG:2016, PDG:2024cfk}. It is noteworthy that the GiM spectrum is the result of a network of calculations in which resonances of one kind serve as intermediate states in the analyses of resonances in reactions of another kind, \textit{e.g.}, the $\pi \Delta$ channels in double-pion production. The rather satisfactory agreement is a strong and encouraging confirmation of CC approaches as a tool for spectroscopic research in hadron production reactions.  

\section{Hadron Production Induced by Nucleons, Nuclei, and Lepton Beams}
Past measurements involving pion and kaon scattering were made at a variety of laboratories, mainly in the 1970s and 1980s, when experimental techniques were far inferior to the standards of today. In the US, pion beams in the momentum range $190~\mathrm{MeV/c}$ to $730~\mathrm{MeV/c}$ were available at the ``Meson Factory'' LAMPF in Los Alamos. This means that the maximum c.m. energy for baryon spectroscopy measurements at LAMPF was only $W \approx 1500~\mathrm{MeV}$. LAMPF was a linear accelerator for $1000~\mathrm{\mu A}$ protons at $800~\mathrm{MeV}$. The meson factory PSI (formerly SIN) near Zurich was a sector-focused cyclotron capable of $100~\mathrm{\mu A}$ of protons at $600~\mathrm{MeV}$, and the Meson Factory TRIUMF in Vancouver was a sector-focused cyclotron for negative hydrogen ions up to $100~\mathrm{\mu A}$ at $500~\mathrm{MeV}$~\cite{Ericson:1991ae, Strakovsky:2023}. Single-pion production was also part of the WASA@COSY research program~\cite{WASA-at-COSY:2018vjy}.

Important work is being conducted through experiments at other hadron facilities that were not mentioned in detail. Historically, SATURN at Saclay and the Synchro-Phasotron in Dubna (JINR), for example, provided important insights into meson production processes on nuclei. The STAR@RHIC (BNL) experiment and the HADES@GSI experiment produced - and are producing - a wealth of data on meson and kaon, and partially also vector meson production in heavy ion collisions at relativistic energies. The meson yields of such processes, occurring in compressed and/or heated nuclear matter, are essential for understanding the dynamics of compressed and heated baryon matter.

Practically all currently operating LHC experiments are searching for exotic hadrons and nuclei, exploring, \textit{e.g.}, flavored hypernuclei and antimatter-nuclei, the latter used for searching signals of CP and CPT violation. COMPASS and AMBER Collaborations are devoted to meson physics; both are also located at CERN. J-PARC at Tokai has a strong program in hadron spectroscopy and strangeness studies \cite{Nagamiya:2012zza}.

A different class of facilities is the electron-positron colliders BESIII and Belle~II, utilizing electro-weak interactions for hadron production in exit channel configurations of total $J^\pi=1^-$ as dictated by the producing annihilation reaction. The experiments are primarily focused on flavor physics, see \textit{e.g.},~\cite{RevModPhys:2011bes, Liu:2025puu}.

\input{SumOut.tex}


%% file: SumOut.tex
\section{Model-Independent Analysis and Optimized Treatment of Hadronic Reaction Data}
The previous section was devoted to an involved multi-channel description of experimental data. For that purpose, models of the reaction mechanism and the interactions had to be developed, mostly on phenomenological grounds. Hence, model parameters were adjusted by fitting them to the data. As seen, those approaches are successful in the sense of reproducing data in a consistent manner, which, in view of the complexity of the task, is a remarkable achievement.

However, modern accelerator facilities, in combination with detectors, are capable of providing a tremendous amount of experimental data. Here, the problem arises: how to present those numerous detailed data in a model-independent manner and make them available - and manageable - for further investigations. The standard approach is a graphical representation, displaying various cuts through the multidimensional volume of measured data, resulting in a multitude of graphs. That approach is necessarily selective and lacks complete coverage. 

\begin{figure}[htb!]
\centering
{~~~~
    \includegraphics[width=0.7\textwidth,keepaspectratio]{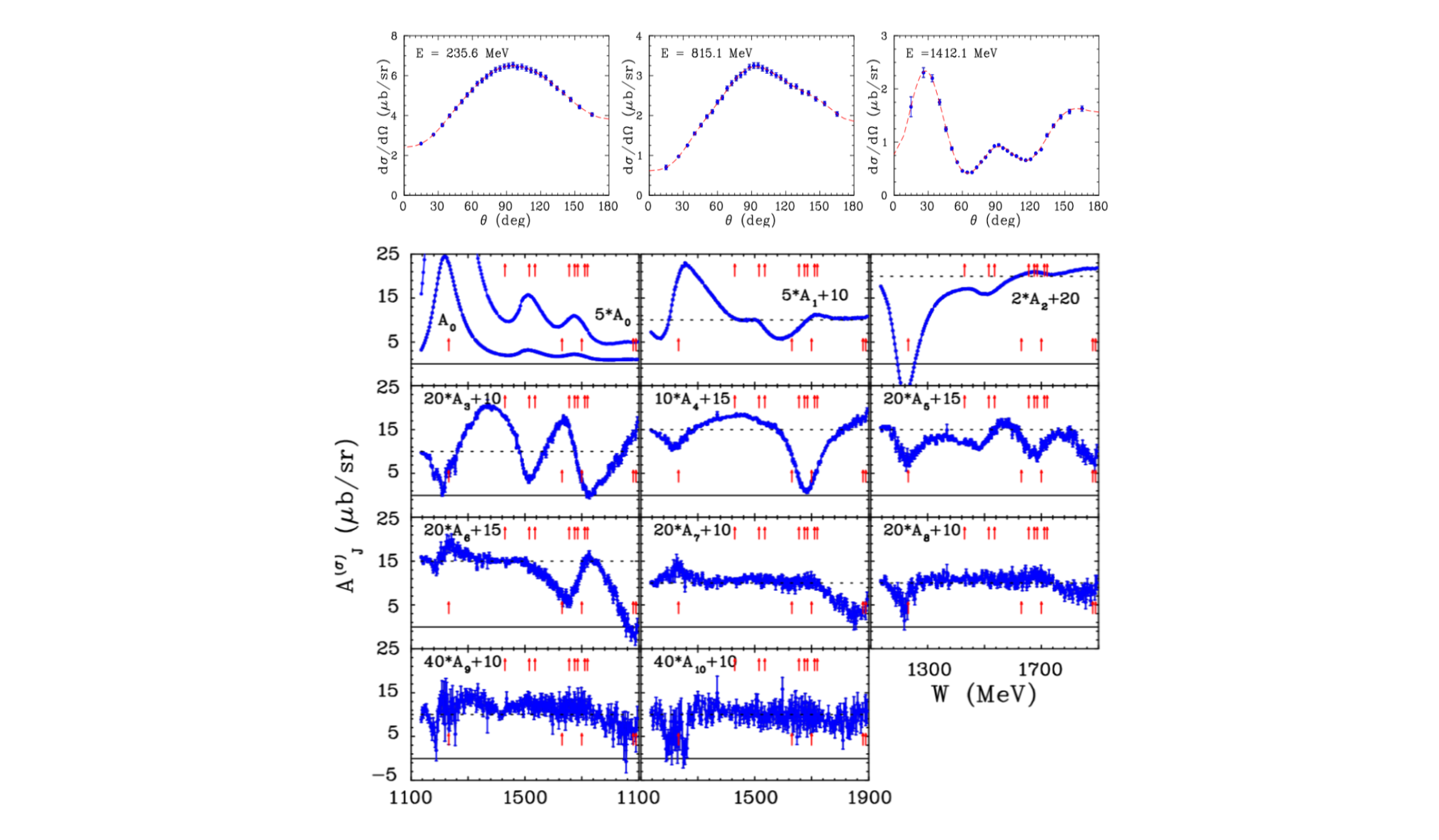}
 
}

\centerline{\parbox{\textwidth}{
\caption[] {\protect\small
\textbf{Top panel}: Samples of the $\gamma p \to \pi^0p$ differential cross 
sections, $d\sigma/d\Omega$, from A2 Collaboration at MAMI measurements (blue
filled circles)~\cite{A2:2015mhs} with the best fit results using Legendre polynomials (red dashed lines). The error bars on all data points represent statistical uncertainties only. Values of $E$ in each plot indicate the laboratory photon energies. 
\textbf{Bottom panel}: Coefficients of Legendre polynomials (blue filled circles). The error bars 
of all values represent $A^{(\sigma)}_J(W)$ uncertainties from the fits in which only the statistical 
uncertainties were used. Solid lines are plotted to help guide the eye. Red vertical arrows indicate 
masses of the four-star resonances (BW masses) known in this energy
range~\cite{PDG:2024cfk}. The upper row of arrows corresponds to $N^\ast$ states with isospin $I = 1/2$ and the lower row corresponds to $\Delta^\ast$ with $I = 3/2$.  Adopted from Ref.~\cite{Azimov:2016djk}.
}
\label{fig:leg} } }
\end{figure}

As an alternative, Azimov \textit{et al.}~\cite{Azimov:2016djk} have suggested expanding experimental data into a series of a complete set of orthogonal functions, \textit{e.g.}, Legendre polynomials $P_J(z)$. For unpolarized differential cross sections, measured angular distributions, for example, may be decomposed as
\begin{equation}
    \frac{d\sigma}{dz}(W, z) = \sum_{J = 0}^{\infty} A^{(\sigma)}_J(W) P_J(z)
    \>,
\label{eq:leg}
\end{equation}
where $W$ is the c.m. energy, $z = \cos \theta$, and $\theta$ is the polar c.m. angle. Formally, this series is infinite. However, in real situations, only a finite number of Legendre coefficients $A^{(\sigma)}_J(W)$ are effectively necessary. As a cut-off criterion, one may impose that only those are considered with coefficients whose central values are larger by a given limit than the estimated error. The method was illustrated, for instance, in 
Ref.~\cite{A2:2015mhs} using photoproduction data of the reaction $\gamma p \to \pi^0p$. As a result, the entire set of data appears to be represented in a compact form of energy-dependent form. 
Legendre coefficients (Fig.~\ref{fig:leg}), easy to use in later theoretical re-analyses or any other application of the data. The approach may be optimized for other kinds of data by choosing function systems that are well adapted to the case. Such compression strategies are widely used in modern data management and \textit{data mining}. 

The Legendre coefficients also have other, less evident properties. For parity-conserving reactions, some of these properties are related to partial-wave amplitudes for states of definite parity. In the unpolarized cross section, the positive- and negative-parity amplitudes always appear symmetrically (that is why the unpolarized cross section by itself does not allow for determining the parity of a particular state). It is not quite the same for $A^{(\sigma)}_J$. One can show that the Legendre coefficients provide specific discrimination of parities: $A^{(\sigma)}_J$ with odd $J$ contains only interferences of states with opposite parities, while $A^{(\sigma)}_J$ with even $J$ contains only interferences
of states with the same parities, positive or negative. And, of course, only the even $J$ coefficients may contain squares of the absolute values of various partial-wave amplitudes.

The above statements can be summarized as follows:
\begin{itemize}
\item Legendre expansions provide a model-independent approach suitable for the presentation of modern detailed (high-precision and high-statistics) data for two-hadron reactions.
\item This approach applies to both cross sections and polarization variables; it is much more compact than traditional methods, at least at not very high energies.
\item The Legendre coefficients reveal specific correlations and interferences between states of definite parities.
\item Due to interference with resonances, high-momentum Legendre coefficients open a window to study higher partial-wave amplitudes, which are beyond reach in any other way.
\end{itemize}

In conclusion, it is worth emphasizing that direct interference has become a useful instrument for searching for and studying rare decays of well-established resonances. However, its possibilities are limited by restrictions on the resonance quantum numbers. Rescattering interference is not limited by such requirements and, therefore, may provide effective methods to search for and study new resonances with arbitrary quantum numbers. Data on multi-hadron decays of heavy particles also present a rapidly expanding area for applications of different kinds of interference, both to study the spectroscopy of resonances and to establish their characteristics.

A special kind of challenge is the treatment and analysis of large data sets, both from large scale experiments and numerical studies. Graph neural networks, especially designed and trained artificial intelligence (AI) systems, are indispensable tools for handling the formidable tasks of data evaluation, pattern recognition, and archiving. In \cite{DeZoort:2023vrm}, the status and perspectives of using graph neural networks, for example, at LHC experiments, are discussed as a flexible and efficient approach to encoding heterogeneous information. 
In~\cite{Caron:2025rir}, an initiative on AI infrastructure for
particle, nuclear, and astroparticle physics was started recently. The authors strongly emphasize that Artificial Intelligence (AI) is transforming scientific research with deep learning
methods. Applications will play a central role in data analysis, simulations, and signal detection across the fields of physics.

\section{Summary and Outlook}\label{SumOut}
The long journey from the first, at that time not understood, signals observed by Victor Hess in high-altitude balloon campaigns, which were increasingly confirmed in different independent experiments in the following years, to modern hadron physics was reviewed. The improving experimental techniques were accompanied by epochal developments in theory, changing our understanding of the material world and the origin of matter. That exciting journey is by no means coming to an end  - it is ongoing, and many surprises can be expected in the future.

\begin{figure}
\begin{center}
\includegraphics[width = 5cm]{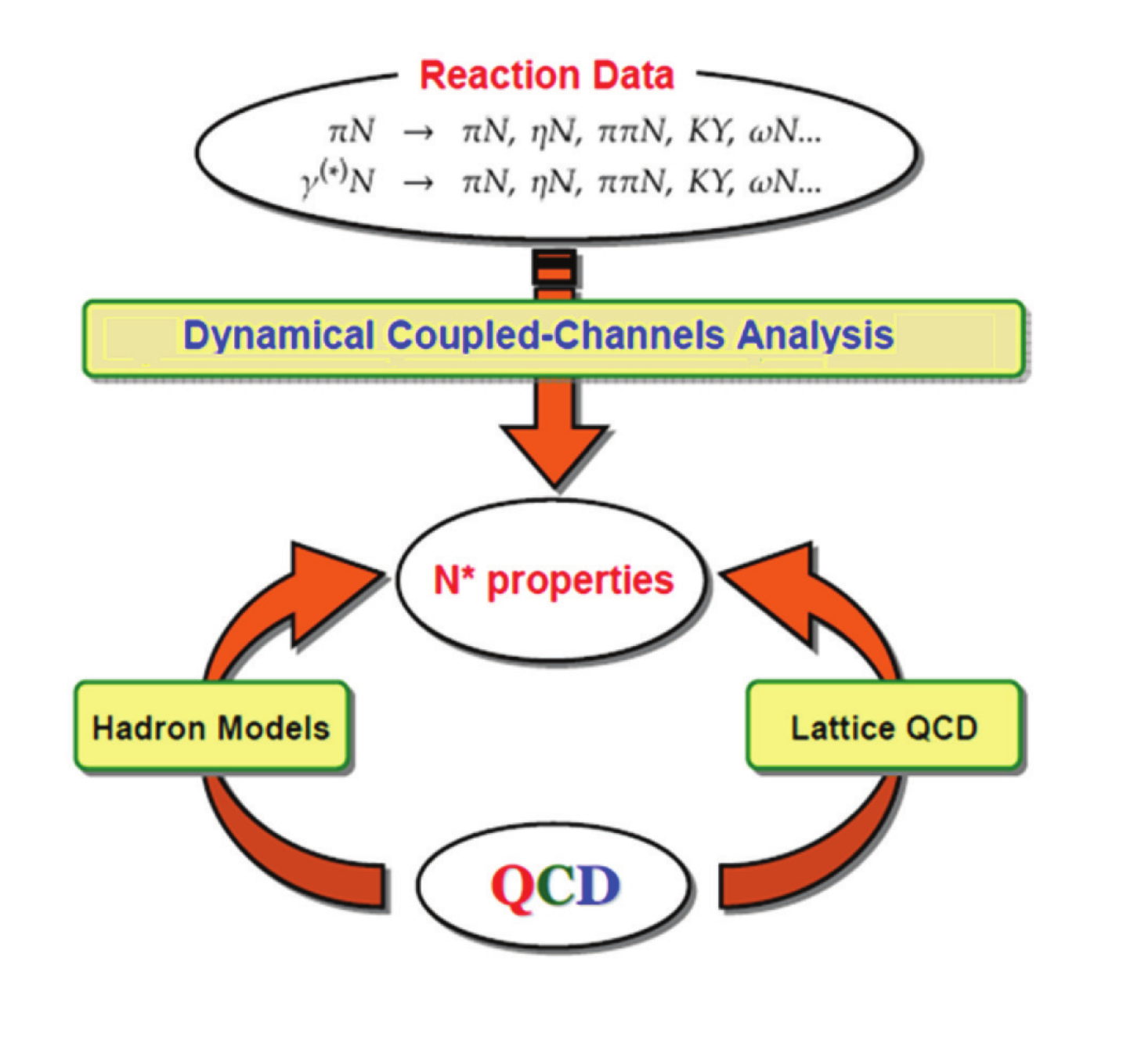}
\caption{Illustration of the long-term goal of hadron physics, aiming at connecting hadron phenomenology to QCD.}
\label{fig:Nstar_Ident}
\end{center}
\end{figure}

The final goal - yet waiting to be achieved - is to connect the results derived by CC methods from data to the QCD predictions obtained by LQCD and the respective functional approaches discussed in this Encyclopedia. The iterative approach, eventually converging to a (self-)~consistent picture, is indicated in Fig.~\ref{fig:Nstar_Ident}.

